\documentclass[final,twocolumn,preprintnumbers,superscriptaddress,nofootinbib,amssymb,aps,prb]{revtex4-2}
\usepackage{soul}
\usepackage[colorlinks = true, linkcolor = blue, urlcolor  = blue, citecolor = blue, anchorcolor = blue]{hyperref}

\usepackage{graphicx} 
\usepackage{booktabs}
\usepackage{amsmath}
\usepackage{braket}
\usepackage{amsbsy}
\usepackage{float}
\usepackage{hyperref}
\hypersetup{colorlinks,urlcolor=blue,citecolor=blue,linkcolor=blue}
\usepackage{dcolumn}
\usepackage{multirow}
\usepackage{enumitem,booktabs,cfr-lm}
\usepackage[referable]{threeparttablex}
\renewlist{tablenotes}{enumerate}{1}
\makeatletter
\setlist[tablenotes]{label=\tnote{\alph*},ref=\alph*,itemsep=\z@,topsep=\z@skip,partopsep=\z@skip,parsep=\z@,itemindent=\z@,labelindent=\tabcolsep,labelsep=.2em,leftmargin=*,align=left,before={\footnotesize}}
\makeatother
\usepackage{epsfig}
\usepackage{lmodern}
\usepackage[T1]{fontenc}
\usepackage[utf8]{inputenc}
\usepackage[english]{babel}
\usepackage{bm}
\usepackage{xfrac}
\usepackage[version=3]{mhchem}
\usepackage[usenames,dvipsnames,svgnames,table]{xcolor}
\usepackage{todonotes}

\begin{document}

\title{Band filling effects on the emergence of magnetic skyrmions: Pd/Fe and Pd/Co bilayers on Ir(111)}

\author{I. P. Miranda}
\email{Corresponding author: ivan.miranda@alumni.usp.br}
\affiliation{Universidade de S\~ao Paulo, Instituto de F\'isica, Rua do Mat\~ao, 1371, 05508-090 S\~ao Paulo, S\~ao Paulo, Brazil}
\affiliation{Department of Physics and Astronomy, Uppsala University, 75120 Box 516 Sweden}
\author{A. B. Klautau}
\affiliation{Faculdade de F\'isica, Universidade Federal do Par\'a, CEP 66075-110, Bel\'em, PA, Brazil}
\author{A. Bergman}
\affiliation{Department of Physics and Astronomy, Uppsala University, 75120 Box 516 Sweden}
\author{H. M. Petrilli}
 \affiliation{Universidade de S\~ao Paulo, Instituto de F\'isica, Rua do Mat\~ao, 1371, 05508-090 S\~ao Paulo, S\~ao Paulo, Brazil}
\date{\today}

\begin{abstract}
Structurally similar transition metal systems can have widely differing magnetic properties. A prime example of this is found for bilayers on Ir(111), where Pd/Fe/Ir(111) ground state  
  has a well-established noncollinear spin texture, while Pd/Co/Ir(111) present a ferromagnetic (FM) single-domain. To unravel the origins of these
different magnetic behaviors, an investigation of Pd/Fe and Pd/Co bilayers on Ir(111), is here performed. Based on the obtained \textit{ab-initio} electronic structure, exchange coupling parameters ($J_{ij}$) and Dzyaloshinskii-Moriya interactions (DMI), we demonstrate that, although in Pd/Co/Ir(111) the DMI is significant, 
two ingredients play a key role on the origin of noncollinearity in Pd/Fe/Ir(111): the presence of magnetic frustrations and a much more in-plane DMI, both with a long-range influence. The $J_{ij}$ and DMI behaviors in both systems can be explained in terms of a simple rigid-band-like model. Also, by performing 
spin-dynamics simulations with the magnetic parameters tuned from the  Pd/Co/Ir(111) to Pd/Fe/Ir(111) \textit{ab-initio} values, we could find conditions for the emergence of skyrmionic phases in the originally FM Pd/Co/Ir(111).
\end{abstract}

\maketitle
 
\section{Introduction}
\label{sec:introduction}
Among the manifestations of the nanoscale magnetism, one of high interest  is the phenomenon of noncollinear magnetism, characterized by 
magnetic moments of neighboring atoms with no unique quantization axis.
Although the nature of this kind of magnetic configuration in solid-state systems is not completely understood on a microscopic level, many different expressions of noncollinear magnetism have been found by the joint effort of experimental and theoretical investigations: parity effects \cite{Holzberger2013,Lounis2008}, vortex-like magnetic structures \cite{Igarashi2012b,Ribeiro2011}, complex spin textures \cite{Igarashi2016,Lounis2014,Bezerra-Neto2013,Phark2014,Cardias2016} and skyrmions \cite{Fert2017,Wiesendanger2016,Romming2013,Mulhbauer2009,Sampaio2013}. The observation of skyrmions \cite{Mulhbauer2009} in magnetic materials has attracted attention, due to their potential applications in novel spintronic technologies, such as logical \cite{Zhang2015a} and non-volatile \cite{Kiselev2011} data storage devices. 

Different materials have been  investigated inspecting the possibility of exhibiting noncollinear configurations and, in particular, skyrmions. Those which are constituted by magnetic overlayers onto high spin-orbit coupling (SOC) substrates are often good candidates as skyrmionic systems \cite{Romming2013,Heinze2011,Herve2018}, due to the combination of different exchange mechanisms, such as the Dzyaloshinskii-Moriya interaction (DMI) \cite{Dzyaloshinskii1958,Moriya1960}, magnetic frustrations \cite{Bergman2006,Lounis2014,Okubo2012,Rozsa2016a}, substrate and overlayers hybridizations \cite{Jadaun2020} and even higher-order interactions (\textit{e.g.}, four-spin) \cite{Heinze2011}.

 Examples of complex magnetism in atomically thin layers and at interfaces can be found by theoretical studies and measurements, {\it e.g.} the extensively investigated Pd/Fe bilayer on  Ir(111) \cite{Romming2013,Romming2015}, as well as by decorating the film edge with Fe/Co spots \cite{Spethmann2022}. In this system, the ground state is characterized by a spin-spiral configuration \cite{Rozsa2016,Simon2014} and, when subjected to the action of an external perpendicular magnetic field
  \cite{Romming2013,Hanneken2015,Bottcher2018,Romming2015,Dupe2014} (of $\sim 1-1.5$ T and at low temperatures), the emergence of a skyrmion lattice can be observed.
As a substrate, the $5d$ element Ir is particularly interesting
since the SOC should be significant what, in conjunction with a low symmetry situation provided by the surface, leads  to an important DMI. In addition, $4d$ elements like Pd are particularly attractive to be used as overlayers, since 
 large magnetic moments can be induced, due to their high magnetic susceptibility. However, if the Fe layer is replaced by Co (periodic table Fe row nearest neighbor element) to generate Pd/Co/Ir(111), the system presents a ferromagnetic (FM) single-domain state \cite{Dzemiantsova2012}.

On one hand, the local magnetic properties of the Pd/Co bilayer on Ir(111) have not yet been deeply investigated  by theoretical approaches.
In order to 
understand the microscopic origins and favorable conditions for the emergence of noncollinear magnetism, in special  skyrmions \cite{Back2020}, 
a full comparison of Pd/Fe/Ir(111) and Pd/Co/Ir(111) offers a special opportunity, through the analysis of the exchange coupling parameter ($J_{ij}$) and the DMI, discussing the role of each type of interaction. On the other hand, recent studies \cite{Mankovsky2021,Yang2018,Gusev2020,Desplat2021,Srivastava2018} have shown that, even in the absence of favorable intrinsic DMI and frustrated exchange interactions, such properties can be tuned by external/internal modifications in the systems, such as the application of long-term \cite{Mankovsky2021,Srivastava2018,Yang2018,Paul2021} or short-term \cite{Desplat2021} electric fields, strain \cite{Gusev2020}, or even by introducing defects with strong spin-orbit coupling \cite{Bezerra-Neto2013}. In view of that electronic structure-level controlling of magnetic interactions, one can conjecture that, for given appropriate conditions, skyrmionic phases can be found in the originally FM Pd/Co/Ir(111). 

Here we first perform a first-principles investigation of the electronic structure and magnetic properties of 
Pd/Fe and Pd/Co bilayers on an Ir(111) substrate. The analysis is focused on the conditions that lead to the emergence of noncollinear spin textures in Pd/Fe/Ir(111), in particular skyrmionic phases (SkPs): in this work, we consider a SkP the metastable state which present skyrmions, which can be mixed with other magnetic structures (\textit{e.g.}, spin-spirals) or not. 
Last, we explore through atomistic spin dynamics simulations, how the exchange interactions in the pristine Pd/Co/Ir(111) would have to be altered in order to support the creation of such SkPs, showing the resulting phase diagrams. 

\section{Computational Details}
\label{sec:methodology}
The electronic structure calculations  were performed in the framework of the Density Functional Theory (DFT) using the self-consistent real-space linear muffin-tin orbital method within the atomic sphere approximation (RS-LMTO-ASA)  \cite{Peduto1991,Frota-Pessoa1992}, which has
been generalized to describe noncollinear magnetism \cite{Bergman2006a,Bergman2007,BERGMAN20064838,PhysRevB.93.014438,PhysRevB.83.014406}. 

The RS-LMTO-ASA is based on the LMTO-ASA formalism \cite{Andersen1975} and solves the eigenvalue problem directly in real space using the recursion method \cite{Haydock1980}. All calculations were performed within the local spin density  approximation (LSDA) \cite{Barth1972} and including 
SOC, $\vec{l}\cdot\vec{s}$, in each variational step \cite{Andersen1975}. In the recursion method, the continued fractions have been terminated with the Beer-Pettifor approach \cite{Beer1984}  after  $LL=22$ recursion levels.
After the self-consistent procedure for  the collinear solution,
the 
isotropic exchange parameters $J_{ij}$ are calculated with the Liechtenstein-Katsnelson-Antropov-Gubanov (LKAG) formula \cite{Liechtenstein1987} as implemented in the RS-LMTO-ASA method \cite{Frota-Pessoa2000,PhysRevB.96.144413}, as well as the anisotropic exchange, DMI \cite{Cardias2020}, $\vec{D}_{ij}$. 
We note that higher-order interactions (HOI) can have an influence on the magnetic ordering. However, recent investigations \cite{Gutzeit2021} have shown that, for Co-based bilayers, the HOI are relatively small; in the Pd/Fe/Ir(111) case, the four-spin interaction can lead to an enhancement of skyrmion stability. 

For the calculated magnetic interactions here, we also considered both $J_{ij}(E)$, i.e.
 $J_{ij}$ as a function of the integration energy limit
(in the expression \cite{Liechtenstein1987} for the exchange interactions), and the $(s,p,d)$ orbital contributions to $J_{ij}$
(calculated including only the corresponding orbital indices in the integration \cite{Frota-Pessoa2000}). An analogous approach (replacing the integration limit to a given energy $E$) is here used to obtain the DMI vector components as a function of 
energy.

Pd/Fe and Pd/Co bilayers on  Ir(111) substrates
were simulated in real space by  semi-infinite clusters  with $\sim 16$,000 atoms 
 generated using the \textit{fcc} Ir experimental 
lattice constant ($a=3.84$ \AA \cite{Ashcroft1976}).
One layer of empty spheres was also included to simulate the vacuum region
 above the Pd layer. In all cases, the magnetic Fe (or Co) atoms are placed on the surface in the fcc-stacking. 
 Relaxations towards the surface in the real-space calculations were neglected, 
 since our results presented already a good agreement with experimental observations \cite{Romming2013,Dzemiantsova2012}.

After the electronic structure calculations, 
atomistic spin dynamics (ASD) were performed using the UppASD package \cite{Skubic2008,Eriksson2017}. Using the \textit{ab-inito} calculated  
$J_{ij}$ and $\vec{D}_{ij}$ the spin Hamiltonian $\mathcal{H}_{i}$ at a given site $i$ can be obtained from

\begin{equation}
\label{eq:spinhamiltonian}
\begin{split}
    \mathcal{H}_{i}=-\sum_{i\neq j}J_{ij}(\hat{e}_{i}\cdot\hat{e}_{j})-\sum_{i\neq j}\vec{D}_{ij}\cdot(\hat{e}_{i}\times\hat{e}_{j})-\mu_B\sum_{i}\vec{B}\cdot\hat{e}_{i}\\
    +\sum_{i}K_{\textnormal{eff}}^{i}\left(\hat{e}_{i}\cdot\hat{e}_{z}\right)^{2},
\end{split}
\end{equation}

\noindent where $\vec{B}$ is the external applied magnetic field (optionally included in the Hamiltonian), $\hat{e}_{i}$ is the direction of the local $i$-th magnetic moment, $\mu_B$ is the Bohr magneton, and $K_{\textnormal{eff}}^{i}$ are the anisotropy strengths in the $\hat{e}_{z}$ direction (where $K_{\textnormal{eff}}^{i}<0$ indicates an out-of-plane easy axis). From this spin Hamiltonian $\mathcal{B}^{\textnormal{eff}}_{i}=-\frac{\partial\mathcal{H}_{i}}{\partial\hat{e}_{i}}$ of the Landau-Lifshitz-Gilbert (LLG) equation \cite{Eriksson2017} can be obtained, which is then used 
in the ASD simulations. 

In the ultrathin films investigated here, the dipole-dipole interaction energy term is small (of the order of $10^{-1}$ meV/atom) when compared to the DMI, and its effects can be reproduced by considering it as part of an effective anisotropy. \cite{Rohart2016,Lobanov2016}. Therefore, the  $K_{\textnormal{eff}}^{i}$ parameters were split in two parts: (\textit{i}) the SOC-induced magnetocrystalline anisotropy ($K_{\textnormal{SOC}}^{i}$), and the so-called shape anisotropy ($K_{dd}^{i}$), related to the dipole-dipole interactions. The $K_{dd}^{i}$ values can be obtained by the Ewald's summation technique \cite{Guo1991}. In turn, $K_{\textnormal{SOC}}^{i}$ was obtained using the Quantum ESPRESSO (QE) package \cite{Giannozzi2017,Giannozzi2009}, using a fully relativistic scheme based on ultrasoft pseudopotentials \cite{Urru2019}. In these QE calculations, the plane wave basis-set was considered withing a cutoff of 110 Ry and the Brillouin zones were sampled by a $12\times12\times2$ $\vec{k}$-point mesh, following the Monkhorst-Pack method \cite{Monkhorst1976}.  The Pd/Co/Ir(111) system was modelled with 8 Ir layers and $\sim10\,$\AA$\,$ of vaccuum, while the first three layers were relaxed to achieve the optimal geometry. For Pd/Fe/Ir(111), we used the experimental value for the effective anisotropy, of 0.4 meV/Fe \cite{Romming2015,Spethmann2022}.

The simulation of spin systems with a Hamiltonian given by Eq. \ref{eq:spinhamiltonian} and subjected to periodic boundary conditions, can exhibit an energy penalty if the cells are incommensurate with the magnetic textures. Here, we found that a square lattice with $250\times250$ atomic spins is suitable for simulating the Fe and Co layers of Pd/Fe/Ir(111) and Pd/Co/Ir(111). The convergence to an optimum state with ASD followed the simulated annealing (SA) method \cite{Kirkpatrick1983}, gradually decreasing the temperature and equilibrating the spin configuration with the heat-bath Monte-Carlo algorithm in every temperature step. Each simulation was performed two times, starting from a FM and a random (paramagnetic) spin configurations. The two total energies obtained from the SA were then compared, finally allowing the choice of the lower-energy state. For the Pd/Fe/Ir(111) phase diagram, we also started from the lowest-energy magnetic configuration in each point \cite{Bottcher2018}.
The presence of  skyrmionic structures  was analyzed through the topological charges $Q$ of the entire spin lattice \cite{Berg1981}.

\color{black}\section{Results}
\label{sec:results-discussion}
\subsection{Magnetic moments and anisotropy}
First, we performed calculations in a collinear approach for Pd/Fe and Pd/Co bilayers on Ir(111). The obtained spin magnetic moment ($m_{s}$) was $2.78\,\mu_{B}$/atom at the Fe sites in Pd/Fe/Ir(111) and $1.85\,\mu_{B}$/atom at the Co sites in Pd/Co/Ir(111).
The Fe (Co) atoms induced spin polarizations on Pd sites 
at the outermost layer, and the magnetizations 
were not small
: $m_s = 0.32\mu_{B}$ ($0.30\,\mu_{B}$)/atom. 
On the other hand, the induced spin moments on Ir atoms at the first layer (denoted herein by Ir$_{1}$), nearest to the Fe (or Co) layer, 
$-0.03\,\mu_{B}$/atom, are small. The $- (+)$ sign indicates that the induced moments at Ir sites 
are aligned antiparallel (parallel) to the spin moments of the $3d$ metals. These Ir $m_{s}$ quickly vanishes with increasing distance from the magnetic layer, in an Friedel-like oscillatory pattern (such as the charge density). The orbital contributions to the magnetic moments ($m_o$) were all found to be relatively small: $m_o \sim 0.1\;\mu_B$/atom ($\sim 0.09\,\mu_B$/atom) for Fe (Co), and $m_o \sim 0.03\; \mu_B$/atom  for Pd.
All 
 magnetic moments 
obtained here are in agreement with previous results \cite{Weber1991,Dzemiantsova2012,Dupe2014}. With these \textit{ab-initio} magnetic moments, the calculated $K_{dd}^{i}$ parameters (see Eq. \ref{eq:spinhamiltonian}) for the fcc(111) monolayer of  Co  is $K_{dd}^{i}=0.16$ meV/atom,  with an in-plane contribution, as expected. In turn, using the $K_{dd}^{i}$ and the obtained $K_{\textnormal{SOC}}^{i}$ values from $\vec{k}$-space calculations, we find that the effective anisotropy strength (see Eq. \ref{eq:spinhamiltonian}) is $K_{\textnormal{eff}}^{i}=-0.38$ meV/atom  for Pd/Co/Ir(111). This $K_{\textnormal{eff}}^{i}$ was used in all spin-dynamics simulations reported later.

\begin{figure}[htb]
\begin{center}
\hspace*{-1.3cm}
\includegraphics[width=1.3\columnwidth]{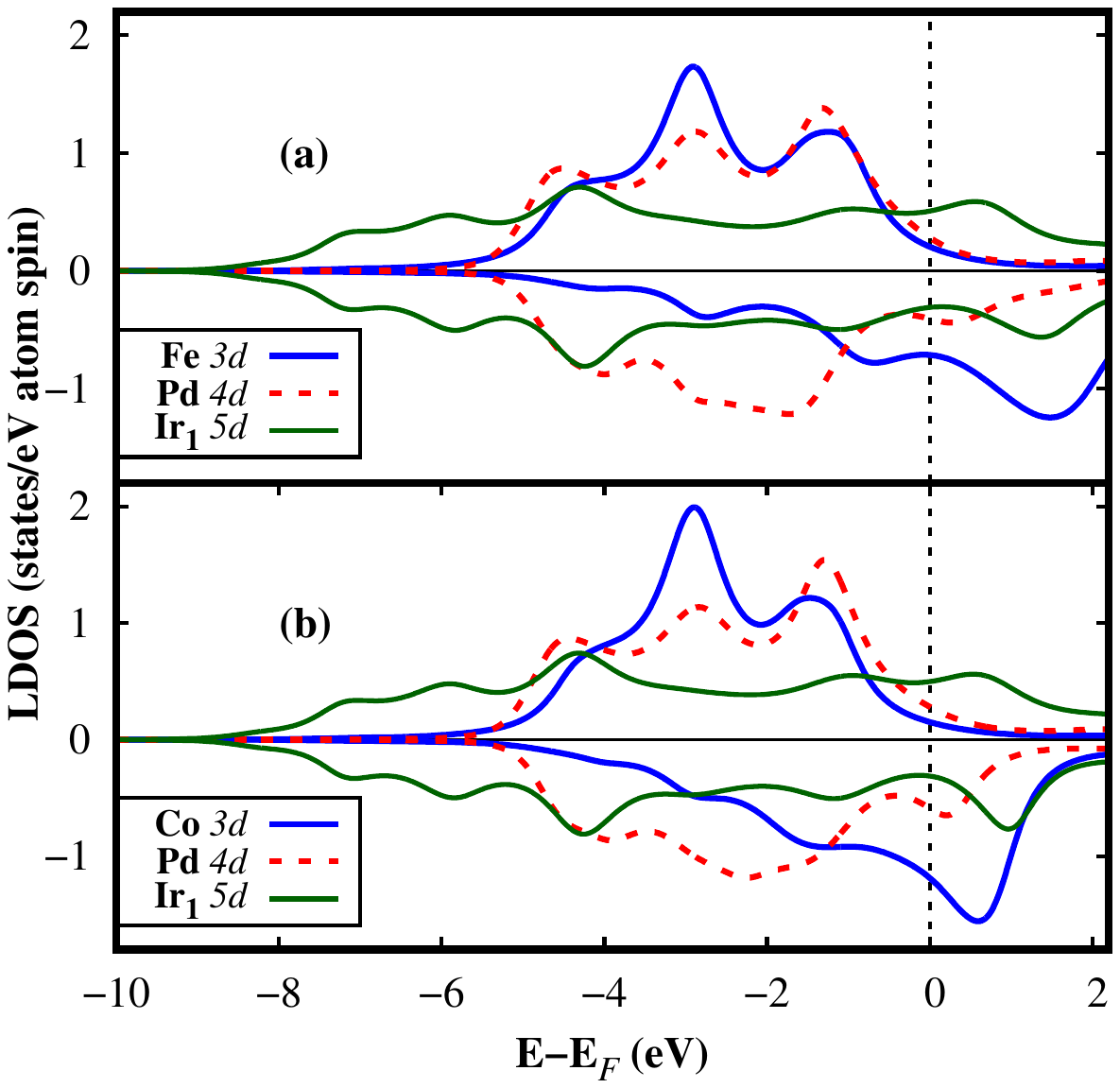}
\vspace*{-0.4cm}
	\caption{ 
	Spin-polarized $d$ orbitals projected local density of states (LDOS) 
	at Pd, Fe and the first Ir layer (Ir$_1$) 
	in (a) Pd/Fe/Ir(111) and (b) Pd/Co/Ir(111). The minority density of states are shown as negative.} 
	\label{fig:ldos-pd-fe-ir}
\end{center}
\end{figure}

The spin-polarized local density of states (LDOS) for the $d$ orbitals of representative layers among the systems studied here are shown in Figure~\ref{fig:ldos-pd-fe-ir}; $s$ and $p$ orbitals are not shown since they are very small.
It can be noticed the similarity of the Fe (Figure~\ref{fig:ldos-pd-fe-ir}a) and Co (Figure~\ref{fig:ldos-pd-fe-ir}b) systems, where an almost rigid band model can be applied (roughly, it is the Fermi level position what is changed by adding one electron when going from Fe to Co). 
There is a relevant overlap energy interval of the 
$3d$, $4d$ and $5d$ 
states.
 Due to this large $3d$-$4d$ hybridization and the high Pd polarizability \cite{Weber1991}, there is an electron charge transfer from Pd $4d$ minority states to Fe $3d$ majority states, 
 what explains the calculated non-negligible induced magnetic moment 
 at the Pd surface layer. 
The 
Ir$_1$ 
LDOS are more extended 
than the $3d$ and $4d$ elements. 

\subsection{Heisenberg exchange interactions}

\begin{figure}[htb]
	\centering
	\hspace*{-0.8cm}
\includegraphics[width=1.20\columnwidth]{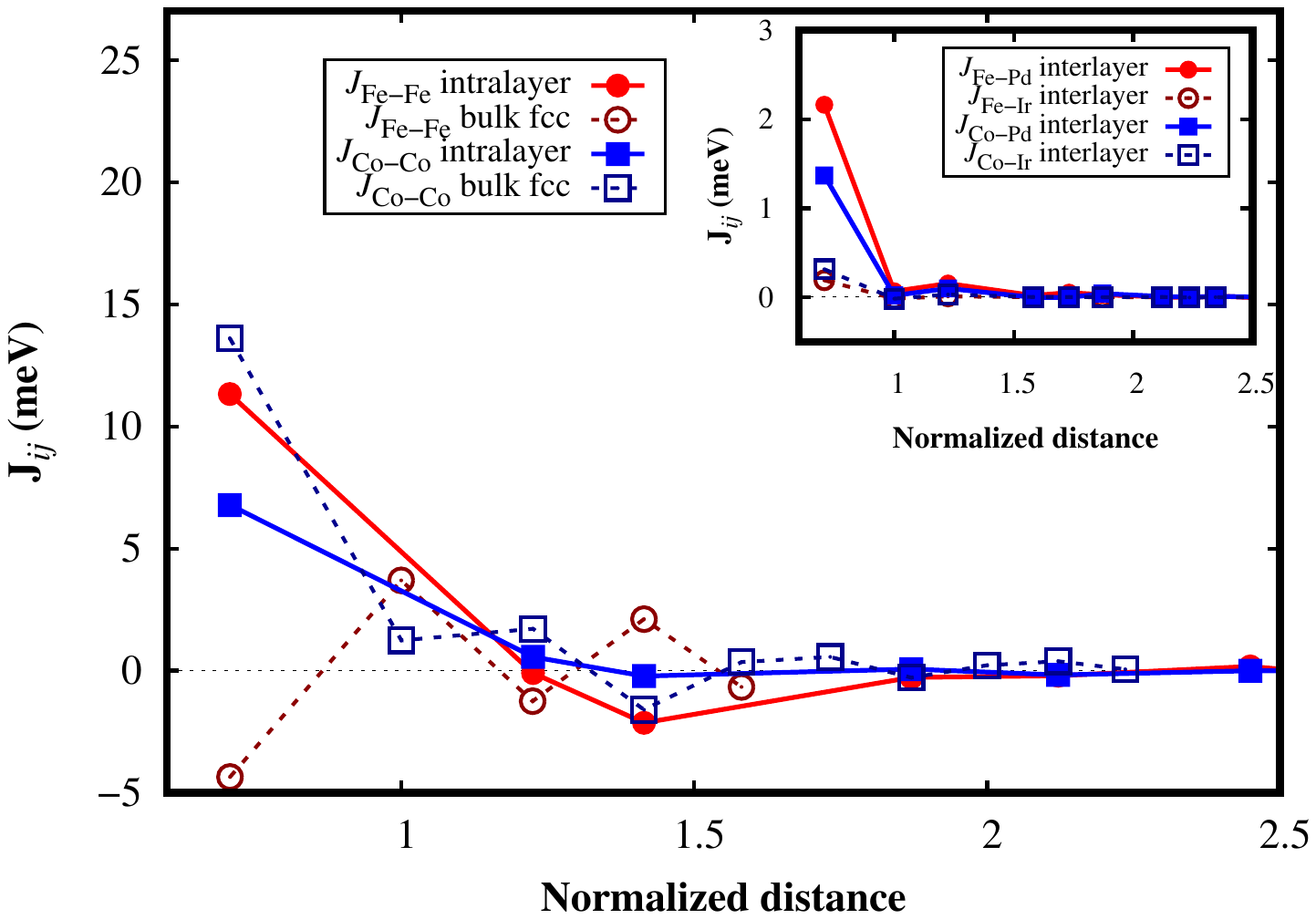}
\vspace*{-0.8cm}
	\caption{(Color online) Isotropic exchange coupling parameter ($J_{ij}$) between $3d$-$3d$ atoms as function of the normalized interatomic distance (units of lattice parameter), for Pd/Fe/Ir(111) (solid red circles), Pd/Co/Ir(111) (solid blue squares), and the correspondent fcc bulks of Fe (dark-red open circles) and Co (dark-blue open squares). Data for fcc Co and fcc Fe extracted from Refs. \cite{Frota-Pessoa2000} and \cite{Cardias2017}, respectively.
	\textit{Inset}: exchange coupling between $3d$-$4d$ atoms and $3d$-$5d$ atoms (interlayer) in Pd/Fe/Ir(111) (full and open red circles) and Pd/Co/Ir(111) (full and open blue squares). The lines are guide for the eyes.
	}	
	\label{fig:exchange-fe-fe}
\end{figure}

The calculation of the Heisenberg (isotropic), $J_{ij}$, and the DMI (anisotropic) exchange parameters were done for a collinear spin configuration, and we here consider an approximation \cite{Polesya2010} in which the Pd and Ir moments are treated as independent spin degrees of freedom. Figure \ref{fig:exchange-fe-fe} shows the intralayer $3d$-$3d$, as well as the interlayer $3d$-$4d$ and $3d$-$5d$ $J_{ij}$ parameters in Pd/X/Ir(111) (X = Fe, Co), including a comparison with Fe and Co fcc bulks exchange couplings, both with the same lattice parameter as Ir bulk. 

Concerning the Pd/Fe bilayer on Ir(111), the intralayer isotropic exchange coupling 
($J_{\textnormal{Fe}-\textnormal{Fe}}$) is FM between spin moments of first neighbors Fe atoms ($J^{1st}_{\textnormal{Fe}-\textnormal{Fe}}= 11.3$ meV) 
and its strength  decreases with inter atomic distance, oscillating between 
positive and negative values for more distant atoms. 
 In particular, from the second neighbors  to the fifth neighbors, the  interactions favors the AFM coupling (with emphasis on the third neighbors' interactions, of $J^{3rd}_{\textnormal{Fe}-\textnormal{Fe}}= -2.1\,$ meV), 
as shown in Fig.~\ref{fig:exchange-fe-fe}, which characterizes a long-range magnetic frustration effect. This remark is particularly relevant, given that frustrated isotropic exchange interactions have shown to play a key role on the stabilization of complex noncollinear (spin spiral, skyrmionic) states in transition metal 2D systems \cite{Rozsa2016a,Okubo2012}, as well as effective $J_{ij}$'s are closely related to the formation of lower-energy spin spiral states \cite{Dupe2016}. Fe bulk fcc ($\gamma$-Fe) is known to exhibit a ground state noncollinear magnetic ordering, very sensible to the lattice parameter \cite{Sjostedt2002}, but the fcc structure is not the only cause \cite{Liz??rraga2004} for the Fe noncollinearity in Pd/Fe/Ir(111): the $J^{1st}_{\textnormal{Fe}-\textnormal{Fe}}$ is drastically modified when Fe changes from the fcc bulk structure to a monolayer (see Fig. \ref{fig:exchange-fe-fe}).
Beyond $\sim2a$, the coupling $J_{\textnormal{Fe}-\textnormal{Fe}}$ becomes very small. These values are in agreement with earlier results for Pd/Fe bilayer on Ir(111) \cite{Bottcher2018,Simon2014,Dupe2014}. The coupling between a typical Fe atom and its first neighbor Pd is FM, to be precise  $J^{1st}_{\textnormal{Fe}-\textnormal{Pd}}=2.2$ meV, while between Fe and Ir nearest neighbors are much weaker ($\sim9\%$ of the Fe-Pd interaction), as shown as \textit{Inset} in Fig. \ref{fig:exchange-fe-fe}, due to the large Ir bandwidth.

\begin{figure}[htb]
	\centering
	\hspace*{-0.40cm}
\includegraphics[width=1.20\columnwidth]{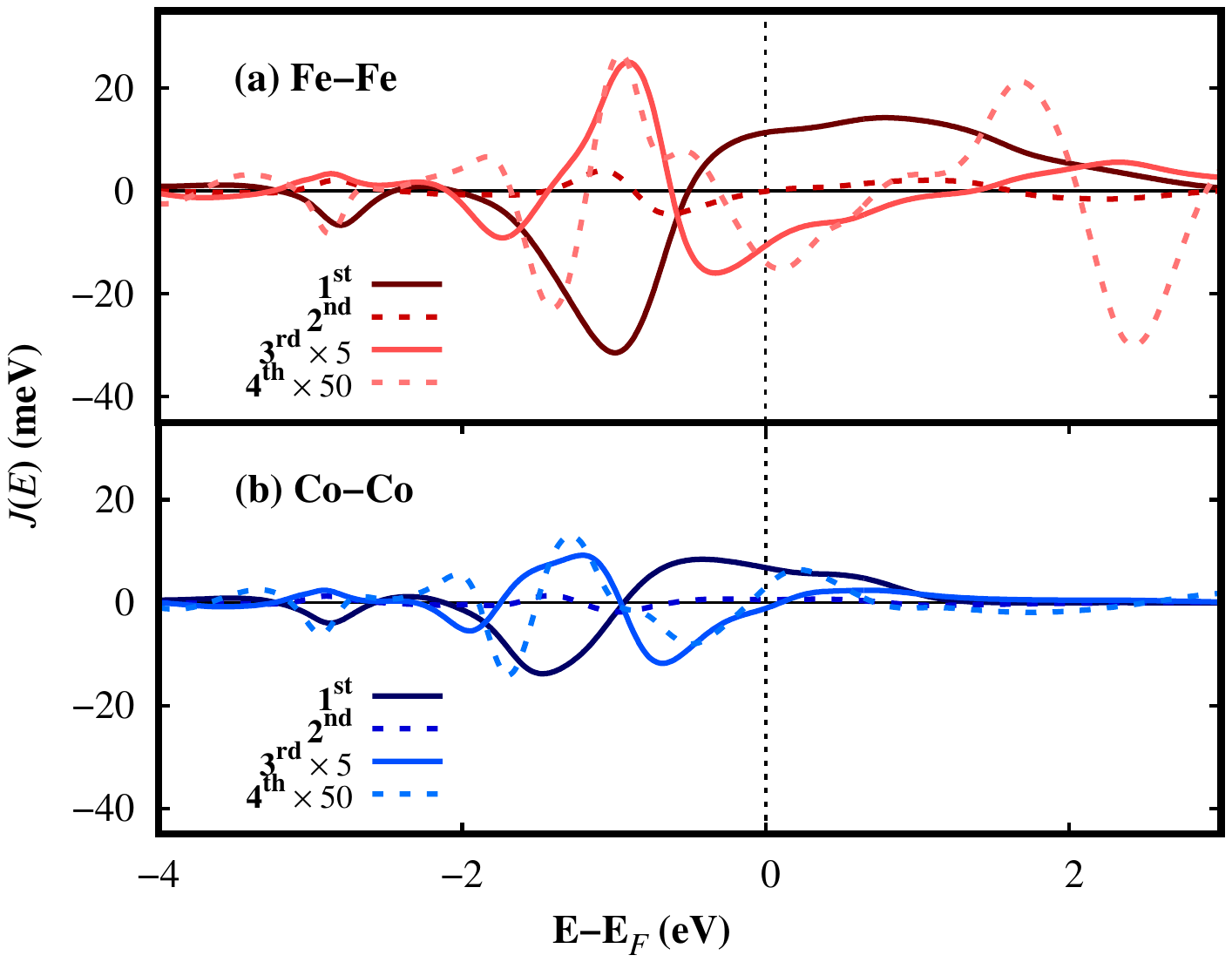}
\vspace*{-0.8cm}
	\caption{(Color online) 
	Isotropic exchange coupling as a function of energy (J($E$)),  between 
	(a) Fe and (b) Co 1$^{st}$, 2$^{nd}$, 3$^{rd}$, and 4$^{th}$ nearest-neighbors for Pd/Fe/Ir(111) and Pd/Co/Ir(111). $E_{F}$ is the Fermi energy.}
	\label{fig:j-energy}
\end{figure}

For the Pd/Co/Ir(111), the intralayer isotropic exchange interaction is FM  between Co first neighbors ($J^{1st}_{\textnormal{Co}-\textnormal{Co}}=6.8$ meV), 
and  monotonically decreases to zero as the Co-Co distance increases. The third and fifth neighbors' interactions favors the AFM coupling ($J^{3rd}_{\textnormal{Co}-\textnormal{Co}}\approx J^{5th}_{\textnormal{Co}-\textnormal{Co}}= -0.2$ meV) but the FM $J^{1st}_{\textnormal{Co}-\textnormal{Co}}$ is the most relevant contribution. Therefore, unlike Pd/Fe/Ir(111) and even Co bulk fcc (see Fig. \ref{fig:exchange-fe-fe}), no relevant AFM-FM competition is obtained for Pd/Co/Ir(111).
The interlayer isotropic coupling between first neighbors Co and Pd atoms is FM, 
$J^{1st}_{\textnormal{Co-Pd}}=1.4$ meV.
In order to analyze the isotropic exchange coupling behavior in more detail for Pd/Fe and Pd/Co on Ir(111), 
we show in Fig. \ref{fig:j-energy} the $J_{ij}$ as a function of energy between Fe-Fe (or Co-Co) atoms, from first up to fourth nearest neighbors ($J_{\textnormal{Fe}-\textnormal{Fe}}(E)$ and $J_{\textnormal{Co-Co}}(E)$).
The actual $J_{ij}$ value (as expressed in Fig. \ref{fig:exchange-fe-fe}) is obtained when 
the $J(E)$ is at the Fermi level ($J(E=E_{F})$). 
From the inspection of the $J(E)$ curves around the Fermi level 
 it can be inferred how structural relaxations as well as effects of increasing or reducing the band filling would affect the isotropic exchange coupling  parameter ($J_{ij}$) values \cite{Cardias2016}. 

 From Figs. \ref{fig:j-energy}(a,b), we can see that the nearest neighbor coupling 
 is by far the most important one.
 The similarity observed between the curves up to and near $E_{F}$ for Fe-Fe and Co-Co exchange couplings suggests that we can assume a rigid-band-like model to explain the obtained difference between the $J_{\textnormal{Fe-Fe}}$ and $J_{\textnormal{Co-Co}}$ characters, attributing it to the extra $3d$ valence electron of Co when compared to Fe. With the displacement of the Fermi level in the Fe $\rightarrow$ Co transition, for instance, third neighbours' $J_{ij}$ diminish the relative AFM strenght and fourth neighbors' exchange interactions change from AFM to FM in Pd/Co/Ir(111) (Fig. \ref{fig:j-energy}b). As $J_{ij}(E)$ is modulated by the magnetic moment, $J_{\textnormal{Co-Co}}$ curves are flatter than $J_{\textnormal{Fe-Fe}}$.

 Fig. \ref{fig:j-energy}(a) shows that it is quite difficult to switch  the interaction between Fe nearest 
 neighbors from FM to AFM just by changing the band filling, since the 
 $J^{1st}_{\textnormal{Fe-Fe}}(E)$ curve has  
 a constant positive level around the $E_{F}$. 
Moreover, as also pointed out in Ref. \cite{Cardias2016}, the  structural relaxation would tend to shift the $J(E)$ curve to the right (higher values of energy). Thus, this indicates that  the FM coupling for this system 
will be, in general,  stable under structural relaxations. 
In turn, Fig. \ref{fig:j-energy}(b) shows a qualitatively similar behavior for the interaction between Co nearest neighbors  by changing the band filling, but with a decreasing $J^{1st}_{\textnormal{Co-Co}}(E)$  curve around $E_{F}$. For both systems, the exchange couplings between more distant Fe (or Co) neighbors are small.

Concerning the orbital projected contributions to the total $J_{\textnormal{Fe-Fe}}$ and $J_{\textnormal{Co-Co}}$ interactions \cite{Cardias2017}, we first point out that $s-$ and $p-$ contributions are very small (always less than $2\%$), and therefore negligible. For both first and fourth neighbors interactions in the Pd/Fe and Pd/Co bilayers, the main contributions are from $3d_{xz}-3d_{xz}$ and $3d_{xy}-3d_{xz}$ orbitals. For second and third neighbors interactions, however, it is interesting to notice that the shift in $E_{F}$ changes the major orbital projected contributions. Regarding the second neighbors interactions, the main contributions for the total $J_{ij}$'s in Pd/Fe/Ir(111) and Pd/Co/Ir(111) come from $3d_{xy}-3d_{z^2}$, and $3d_{xy}-3d_{xy},3d_{z^2}-3d_{z^2},3d_{xz}-3d_{xz}$, respectively. In turn, for third neighbors, the main contributions arise from $3d_{xz}-3d_{xy},3d_{xz}-3d_{xz}$, and $3d_{xz}-3d_{z^2}$ for Pd/Fe/Ir(111), and $3d_{xz}-3d_{z^2}$ for Pd/Co/Ir(111). These changes in second and third neighbors $J_{ij}$ interactions are important and will be explored in Section \ref{sec:spin-dynamics} when analyzing the boundaries for the emergence of noncollinear textures in Pd/Fe/Ir(111).

\subsection{Dzyaloshinskii-Moriya interactions}

\begin{figure}[htb]
	\centering
\begin{tabular}{c}
		\hspace{-1.3cm}
\includegraphics[width=1.3\columnwidth]{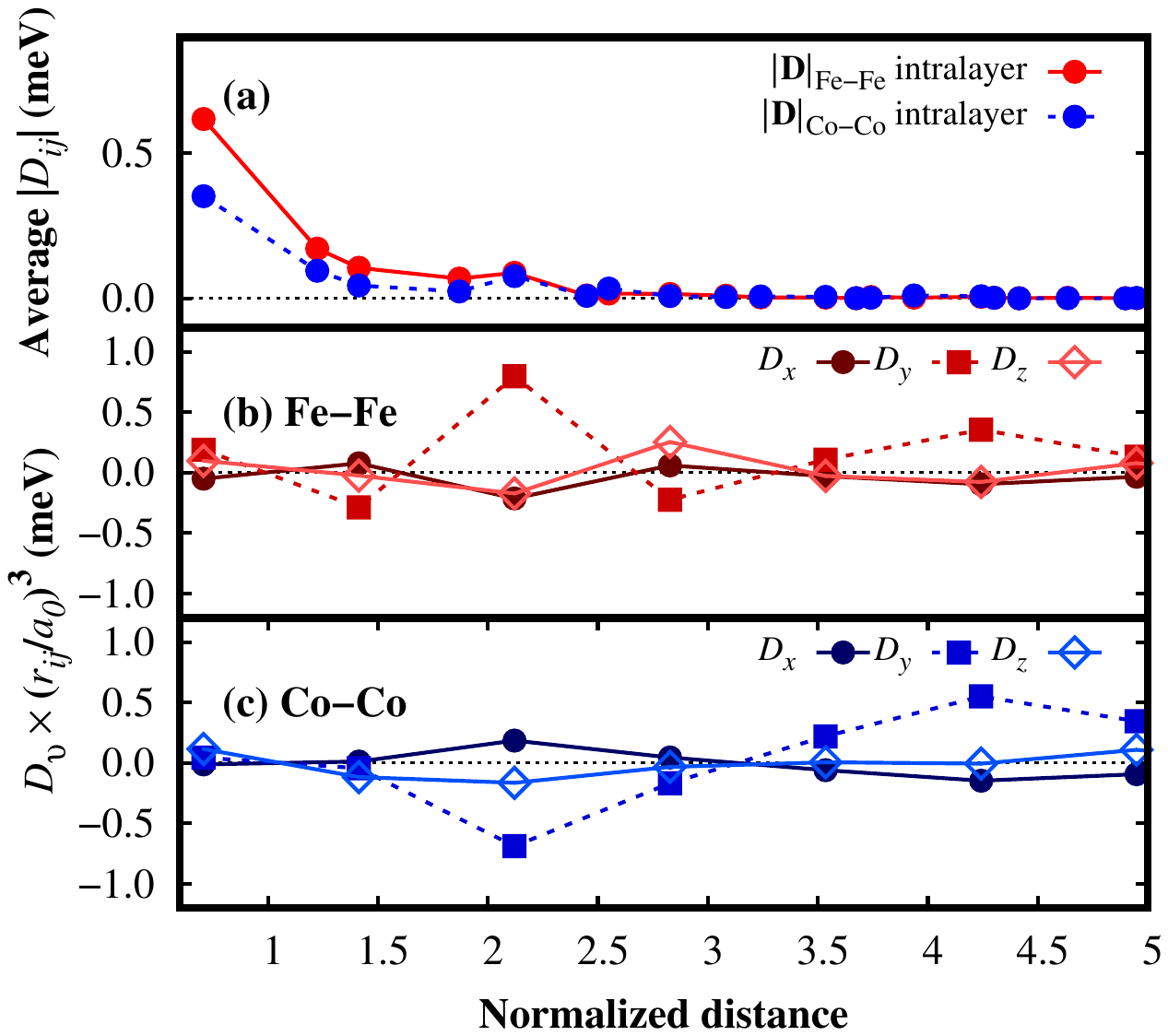}
\end{tabular}
\vspace{-0.4cm}
\caption{(Color online)  Average magnitude of (a) the DMI vectors, $|\vec{D}_{ij}|$, for \textit{all} possible Fe-Fe (or Co-Co) bonds in an fcc(111) layer; and (b,c) 
$D_{x}$, $D_{y}$ and $D_{z}$ components times $\left(\frac{r_{ij}}{a}\right)^{3}$, as a function of the normalized interatomic separation (units of lattice parameter) for (b) Fe-Fe and (c) Co-Co atoms,  considering \textit{only} the in-plane $\left[\bar{1}01\right]$ pairing direction. The DMI calculations were performed with collinear magnetic configuration.
The lines are guides for the eyes.	 	
}
	\label{fig:dm-fe-fe}
\end{figure}

Turning now to the anisotropic DMI, we first stress that this interaction comes as a surface effect. Let us consider the interaction vector $\vec{D}_{ij}=(D_{x},D_{y},D_{z})$ between $i$ and $j$ sites ($i\neq j$), whose components are individually denoted by $D_{\nu}$ ($\nu=\{x,y,z\}$). Herein, the $z$-axis is considered to be out-of-plane ($z\parallel[\bar{1}\bar{1}\bar{1}]$), and $(D_{x},D_{y})$ are in-plane components, where the $x$-axis is parallel to the nearest Fe-Fe (or Co-Co) bond direction.
 Fig. \ref{fig:dm-fe-fe} shows the average magnitudes of $\vec{D}_{ij}$ between the (a) Fe-Fe ((b) Co-Co) atoms as a function of the
interatomic distance for Pd/Fe/Ir(111) (Pd/Co/Ir(111)), as well as 
  the  components, considering the in-plane $\left[\bar{1}01\right]$ pairing direction. It can be seen from Fig. \ref{fig:dm-fe-fe}(a) that, in both cases,  the magnitude of the  DMI vector
between Fe-Fe (or Co-Co) atoms is larger for the first neighbors ($D^{1^{st}}$). Moreover, the magnitude of  
 the DMI vector tends to decrease with respect to the interatomic separation, but not monotonicaly, presenting peaks for enlarged pairwise distances, such as $\frac{3}{2}\sqrt{2}a$ ($\sim 8.2\,$\AA). This demonstrates the long-range nature of the 
  DMI in these systems, as a consequence of electronic hopping mediated mostly by the extended Ir $5d$ states (Fig. \ref{fig:ldos-pd-fe-ir}). 
We obtained $D^{1^{st}}_{\textnormal{Fe-Fe}}=0.62$ meV, for the Fe nearest neighboring interactions, which reveals a good agreement with previous studies ($D^{1^{st}}_{\textnormal{Fe-Fe}}=0.82$ meV for a $5\%$ inward relaxed Fe layer \cite{Simon2014}, and $D^{1^{st}}_{\textnormal{Fe-Fe}}=1.00$ meV for the pristine Pd/Fe/Ir(111) system \cite{Dupe2014}), considering the different theoretical approaches used. Given the simple FM (single-domain) observed for Pd/Co/Ir(111) \cite{Dzemiantsova2012}, it is relevant to notice also the relatively high $D^{1^{st}}_{\textnormal{Co-Co}}$ value, of $0.35$ meV. Concerning the DMI vector components,  we notice that in both Pd/Fe/Ir(111) and Pd/Co/Ir(111) multilayers, 
all $D_{\nu}$ components between Fe-Fe (or Co-Co) atoms 
exhibit an oscillatory character, specially $D_{y}$, as shown in Fig. \ref{fig:dm-fe-fe}(b,c), in a Ruderman-Kittel-Kasuya-Yosida-like (RKKY-like \cite{Ruderman1954,Kasuya1956,Yosida1957}) behavior \cite{Khajetoorians2016}, mediated by the conduction electrons \cite{Fert1980}. These oscillations in $D_{\nu}$ cause the DMI to vary its orientation as the distance from the reference atom changes. 

\begin{figure}[htb]
	\centering
\includegraphics[width=\columnwidth]{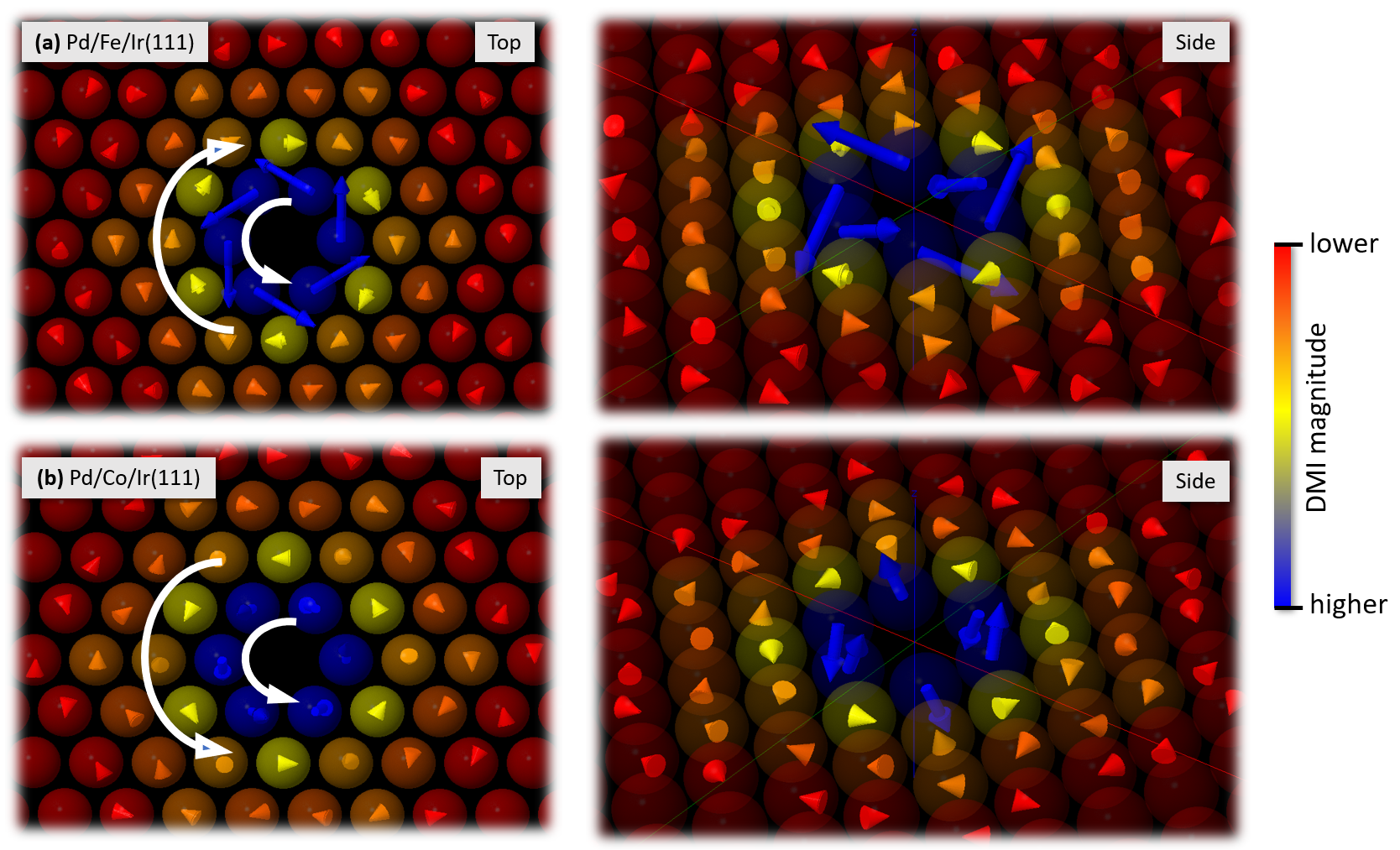}
\vspace*{-0.4cm}
	\caption{
	(Color online) Schematic top and side views of a typical $3d$ atom (in dark gray) and its closer neighbors in: (a) Pd/Fe; and (b) Pd/Co bilayers on Ir(111) surfaces. The arrows indicate the DMI vectors between the central atom and the atom in which the arrow is located, while the DMI magnitude is represented by the color scheme (as indicated in the colorbar). The $z$-axis corresponds to the out-of-plane direction ($[111]$).
	}	
	\label{fig:dm-fe-fe-sketch}
\end{figure}

This can be directly seen in Fig.~\ref{fig:dm-fe-fe-sketch}, where we present the DMI directions between a typical atom at the $3d$ monolayers and its intralayer neighbors and where the colors (arrows) denote, respectively, the modulus (orientation) of $\vec{D}_{ij}$, in case of a collinear magnetic configuration calculation.
For both Pd/Fe and Pd/Co bilayers on Ir(111), a inspection on Fig.~\ref{fig:dm-fe-fe-sketch}
shows that the DMI directions keep the $C_{3v}$ point group symmetry
of the fcc(111) layer, always following the relation $\vec{D}_{ij}\cdot\vec{r}_{ij}\sim0$ ($\vec{r}_{ij}=\vec{r}_{i}-\vec{r}_{j}$ is the position vector), the only constrain for this type of surface expected from the Moriya's symmetry rules \cite{Moriya1960,Crepieux1998}. Similar result was found in Ref. \cite{Simon2014} for the Pd/Fe/Ir(111) system.
In Fig.~\ref{fig:dm-fe-fe-sketch}(a), it is worth noting that 
the average angle between the NN $\vec{D}_{\textnormal{Fe-Fe}}$ and the Ir(111) surface plane, $\bar{\phi}$, is $\bar{\phi} \sim26.5^{\circ}$. This mostly in-plane $\vec{D}^{1^{st}}_{\textnormal{Fe-Fe}}$ (the most proeminent) favors a rotation of the local spin moments towards the Ir(111) surface to minimize the system energy, directly influencing the emergence of noncollinear magnetism. The reasoning is the same for $\vec{D}^{2^{nd}}_{\textnormal{Fe-Fe}}$ and $\vec{D}^{3^{rd}}_{\textnormal{Fe-Fe}}$, specially because of the long-range influence of $D_{y}$ (see Fig.~\ref{fig:dm-fe-fe}b). 
Concerning the $\vec{D}_{\textnormal{Co-Co}}$ interaction vectors, from Fig.~\ref{fig:dm-fe-fe-sketch}b, it is shown an almost out-of-plane nearest neighbors DMI vectors. Indeed, the obtained average angle between the $\vec{D}^{1^{st}}_{\textnormal{Co-Co}}$ and the Ir(111) surface is about $\bar{\phi}\sim 68.6^{\circ}$. When moving to further neighborhoods, and following the long-range $D_{y}$ RKKY-like influence (Fig.~\ref{fig:dm-fe-fe}c), $\vec{D}_{\textnormal{Co-Co}}$ periodically changes from almost in-plane to almost out-of-plane.
 Therefore, an opposite behavior of $\vec{D}^{1^{st}}_{\textnormal{Co-Co}}$ 
 is verified when comparing Pd/Fe/Ir(111) and Pd/Co/Ir(111) systems. 
 It is also worth stressing that the chirality is changed for second neighbors' DMI, but maintained for nearest neighbors $\vec{D}_{ij}$.

\begin{figure}[htb]
	\centering
	\begin{tabular}{c}
	\hspace{-0.55cm}
\includegraphics[width=1.23\columnwidth]{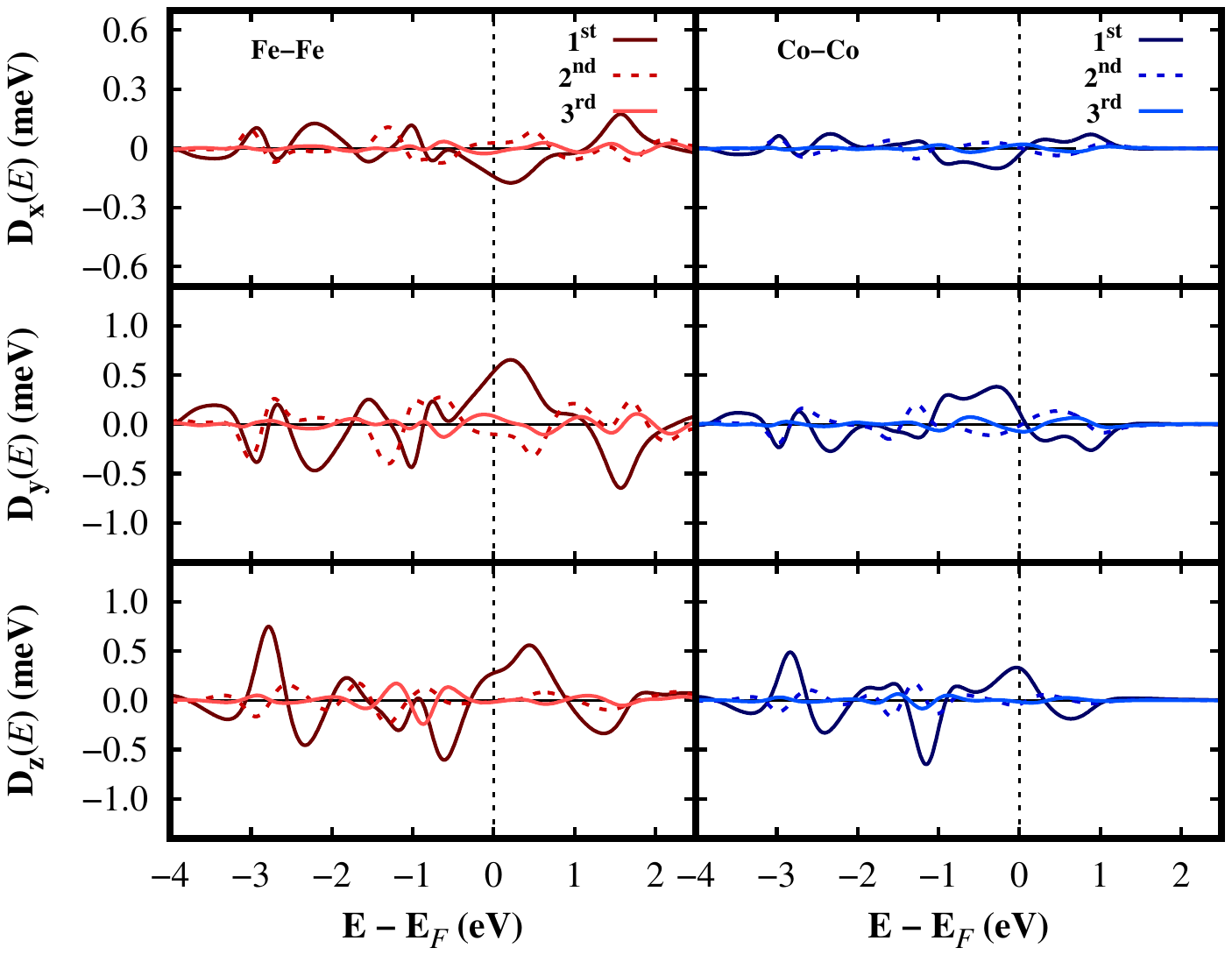}
	\end{tabular}
	\vspace*{-0.4cm}
	\caption{(Color online) 
	Anisotropic DMI vector components $D_{x}, D_{y}$, and $D_{z}$ as a function of the energy,  between 
	Fe (left panel, red curves) and Co (right panel, blue curves) 1$^{st}$, 2$^{nd}$, and 3$^{rd}$ nearest-neighbors for the Pd/Fe/Ir(111) and Pd/Co/Ir(111) in the in-plane $[\bar{1}01]$ pairing direction. $E_{F}$ is the Fermi energy.}
	\label{fig:d-energy}
\end{figure}

Analogous investigation as for $J_{ij}$ in Fig. \ref{fig:j-energy} can be done for the DMI. Let us consider $\vec{D}_{ij}(E)$ the interaction vector as a function of the energy for each component, $D_{\nu}(E)$. The results for Fe-Fe and Co-Co interactions for the in-plane pairing direction $[\bar{1}01]$ are shown in Fig \ref{fig:d-energy}(a,b). The results are similar for the $[\bar{1}10]$ in-plane direction (data not shown).  
For $\vec{D}_{ij}(E)$ we also obtain very similar curves for the Fe-Fe and Co-Co cases, indicating, again, that a simple rigid-band-like model can be used to explain the differences in the DMI directions in both systems. The extra $3d$ valence electron in the minority spin states of Co shifts $E_{F}$ to higher energies, in a region in which $D_{z}$ is more pronounced in Pd/Co/Ir(111), diminishing at the same time the corresponding importance of $D_{x}$ and $D_{y}$ components. In Pd/Fe/Ir(111), however, the case is the inverse: while $D_{z}(E_F)$ becomes relatively smaller, $D_{x}(E_F)$ and $D_{y}(E_F)$ become comparatively relevant, constraining $\vec{D}_{ij}$ to a more in-plane interaction vector ($\vec{D}_{ij}\perp[111]$). Even though the $D_{\nu}(E)$ curves profiles for Pd/X/Ir(111) (X = Fe, Co) are very similar to each other, their magnitudes differ (being smaller for Co-Co interactions), a trend which is indirectly driven by Hund's first rule \cite{Belabbes2016}.
With this simple rigid-band-like model, for example, it is also possible to predict the situation of one $3d$ valence electron \textit{more} in Co (Co $\rightarrow$ Ni), in which $E_{F}$ is shifted to higher energies. The DMI would diminish in magnitude, specially due to the lower peaks in $D_{\nu}(E)$ (in accordance to Ref. \cite{Belabbes2016}), and the interaction vectors would be prominently in-plane \cite{Carvalho2021} (while the $J_{ij}$ would present FM neighbors couplings, see Fig. \ref{fig:j-energy}b and Ref. \cite{Carvalho2021}). 
The analysis of $D_{z}$ on Pd/Fe/Ir(111) and Pd/Co/Ir(111) also suggests that the DMI would not be profoundly affected by small structural relaxations.

In order to compare our results for the DMI with the experimental one concerning the Pd/Fe/Ir(111)  
\cite{Romming2015}, which obtained  $D=\left(3.9\pm0.2\right)$ mJ/m$^2$, expressed as an energy density \cite{Rohart2013}, one can define an effective DMI vector magnitude as $D^{\textnormal{eff}}=\sum_{i\neq1}\left|\vec{D}_{1i}\right|$, where the site "$1$" represents a reference 
(Fe or Co) site (central atom) and the sum goes over its nearest neighbors atoms of the $3d$ intralayer hexagon, presenting an area of $\frac{3l^2\sqrt{3}}{2}$, with $l=a\frac{\sqrt{2}}{2}$. Therefore,
our first principles DMI vector density $D$ for the Fe layer in Pd/Fe/Ir(111) 
 is equal to $D = D_{\textnormal{Fe}} = \frac{D^{\textnormal{eff}}}{\frac{3l^2\sqrt{3}}{2}}=\frac{3.72\,\textnormal{meV}}{19.16\,\textnormal{\AA}^2}\sim 3.1\,\textnormal{mJ/m}^2$, which is 
in good agreement with the experimental result.
For Pd/Co/Ir(111) this quantity is $\sim60\%$ of $D_{\textnormal{Fe}}$ ($D_{\textnormal{Co}}\sim1.8$ mJ/m$^2$). 

Skyrmion formation and shape are often related to 
the ratio between the anisotropic and the isotropic exchange couplings \cite{Polesya2014,Fert2013,Simon2014,Sampaio2013}. 
For the Pd/Fe/Ir(111) system, we obtained $\frac{\left|\vec{D}\right|}{J}\sim 0.055$ for the average nearest  neighbors Fe-Fe interactions, which is in agreement with Ref. \cite{Simon2014}. Interestingly, for the Pd/Co/Ir(111) this value is similar, $\frac{\left|\vec{D}\right|}{J}\sim 0.051$. As skyrmions in Pd/Fe/Ir(111) were experimentally verified \cite{Romming2013}, one would have expected, by looking at the higher $\frac{\left|\vec{D}\right|}{J}$ obtained ratio, that Pd/Co/Ir(111) has a tendency towards the emergence of noncollinear spin configurations or even skyrmions. Nevertheless, this is not the case \cite{Dzemiantsova2012}. 

\subsection{\label{sec:spin-dynamics} Phase diagrams and boundaries for the noncollinear textures}

So far, we have shown that a simple rigid-band-like model can explain the differences in the \textit{ab-initio} $J_{ij}$ and $\vec{D}_{ij}$ parameters of Pd/Fe and Pd/Co bilayers on Ir(111). Also, we have demonstrated the long-range character of $\vec{D}_{ij}$, and the in-plane (or out-of-plane) preference of the DMI vectors. Now, we apply these parameters we have obtained in the DFT calculations in the ASD simulations, constructing the magnetic configurations for several external magnetic fields ($B$) and temperatures ($T$) conditions. Before we discuss our results, however, it is noteworthy that we had to consider not only the first shell of Fe-Fe and Co-Co neighbors in the simulations, but further neighbors to obtain the correct magnetic configurations for Pd/Fe/Ir(111) and Pd/Co/Ir(111), as it will be discussed in Section \ref{sec:boundaries}.  Concerning Pd/Fe/Ir(111), it was experimentally \cite{Romming2013,Romming2015,Lindner2020} and theoretically \cite{Rozsa2016,Bottcher2018,Bessarab2018,Dupe2014,Simon2014} shown that the system presents a spin-spiral (SS) ground state which evolves to a FM skyrmion lattice (SkL) when an external magnetic field is applied (of $\sim1-1.5$ T), reaching a FM (field-polarized) state for $B\sim2.5-3$ T. However, Pd/Co/Ir(111) presents a single-domain FM state in an applied field of $B=\pm1$ T \cite{Dzemiantsova2012}. We explored the ground and excited states (with $B>0$ and $T>0$) of both systems, for external magnetic fields applied in the out-of-plane direction. In all simulations we considered the \textit{ab-initio} Gilbert damping values of $\alpha_{\textnormal{PdFe}}=0.023$ for Pd/Fe/Ir(111) and $\alpha_{\textnormal{PdCo}}=0.040$ for Pd/Co/Ir(111), obtained with the same method for calculating $\alpha$ in real-space as described in Ref. \cite{Miranda2021}. These $\alpha$ values take into account the moment-weighted average of the layer-resolved dampings shown in Table \ref{tab:gilbert-damping} for typical sites in the Fe (or Co) layers \cite{Miranda2021} and typical sites in the correspondent Pd layers. For example, $\alpha_{\textnormal{PdCo}}=\frac{m_{s}^{\textnormal{Pd}}\alpha_{\textnormal{Pd}}+m_{s}^{\textnormal{Co}}\alpha_{\textnormal{Co}}}{m_{s}^{\textnormal{Pd}}+m_{s}^{\textnormal{Co}}}$. The results presented here are not strictly dependent on the damping parameter choice. We note, however, that one should be careful when using the damping values for the investigation of $\alpha$-dependent properties, as the effective $\alpha$ has shown to change when in noncollinear spin environments (such as in Pd/Fe/Ir(111)) \cite{Andre2016,Rozsa2018}.

\begin{table}[h]
	\centering
	\caption{First-principles damping values calculated for a typical atom in the Fe ($\alpha_{\textnormal{Fe}}$), Co ($\alpha_{\textnormal{Co}}$), and Pd ($\alpha_{\textnormal{Pd}}$) layers of Pd/Fe/Ir(111) and Pd/Co/Ir(111).}
	\vspace*{0.3cm}
	\begin{tabular}{c|c}
	\hline\hline
	 \textbf{Pd/Fe/Ir(111)} & 
	 \textbf{Pd/Co/Ir(111)}  \\ \hline
	 $\alpha_{\textnormal{Fe}}=0.004$ \cite{Miranda2021} & $\alpha_{\textnormal{Co}}=0.003$ \cite{Miranda2021} \\ 
 	 $\alpha_{\textnormal{Pd}}=0.192$ &
 	 $\alpha_{\textnormal{Pd}}=0.274$ \\
 	 \hline\hline
	\end{tabular}
	\label{tab:gilbert-damping}
\end{table}

\begin{figure}[htb]
	\centering
\includegraphics[width=\columnwidth]{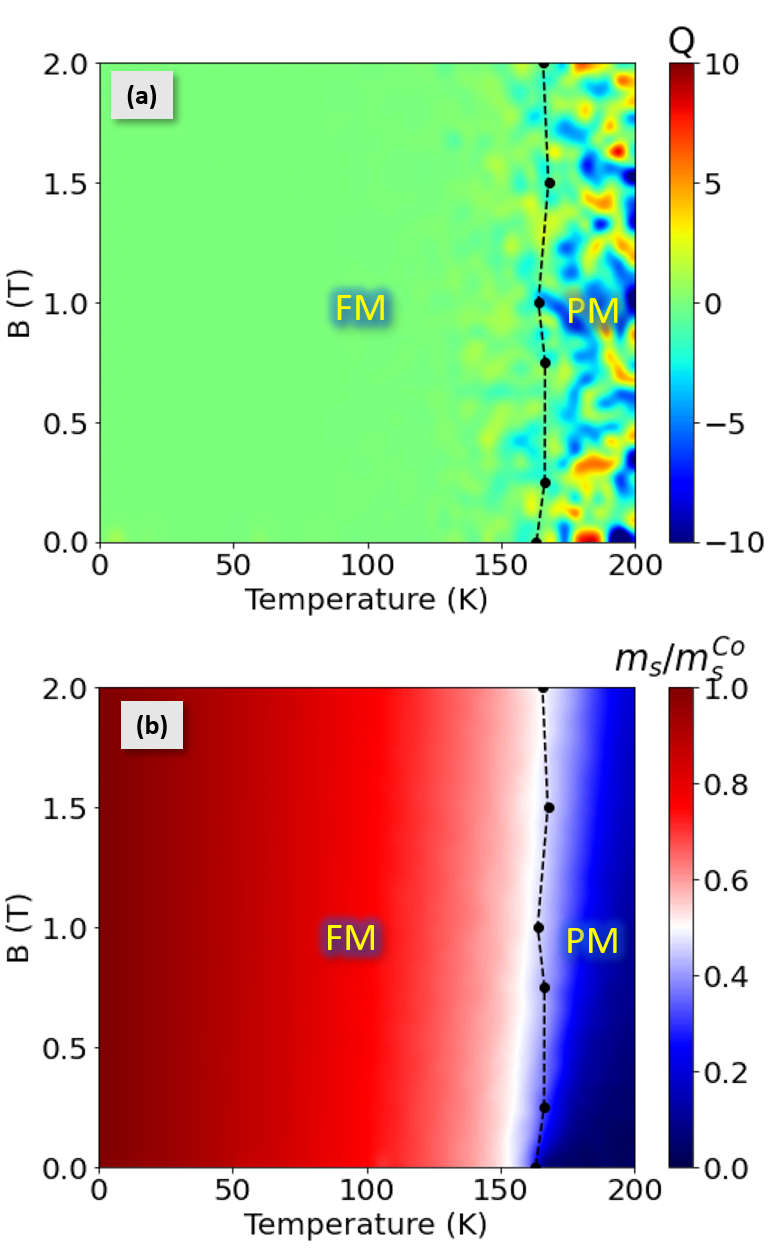}
\vspace*{-0.4cm}
\caption{(Color online) $B-T$ phase diagram of Pd/Co/Ir(111), showing: (a) the time-average total topological number $\bar{Q}$; and (b) the average relative magnetic moment $\left(\frac{\bar{m}_{s}}{m_{s}^{\textnormal{Co}}}\right)$ as a function of external $B$ and finite $T$. The black dots show the order transition inflection points of the magnetic specific heat, $c_{V}$, obtained by a Lorentzian fitting. The ferromagnetic (fully saturated, FM) and paramagnetic (PM) phases are indicated. The dashed black lines are guides for the eyes.}
\label{fig:pdcoir-phasediagram}
\end{figure}

The obtained $(B,T)$ phase diagram of Pd/Co/Ir(111), based on the time average (1 ps) of the topological charge $Q$, is exhibited in Fig. \ref{fig:pdcoir-phasediagram}a.  In the presence or in the absence of magnetic fields, the Co-based system presents a single-domain FM state. This is corroborated by the average magnetic moments, $\bar{m}_{s}$ (Fig. \ref{fig:pdcoir-phasediagram}b), in which the $\bar{m}_{s}$ value decays regularly with temperature as an effect of time-dependent fluctuations. From our \textit{ab-initio} and spin dynamics calculations, the Pd/Co/Ir(111) FM phase was not modified to a noncollinear texture under the influence of an external magnetic field (evaluated in the in-plane and out-of-plane directions, for field magnitudes up to $B=100$ T). The calculated critical temperature from the peak (divergence) in the magnetic specific heat \cite{Buhrandt2013}, $c_{V}=\frac{\partial E}{\partial T}\Bigr|_{V}$, is about $T_{C}\approx 165$ K (black dots on Fig. \ref{fig:pdcoir-phasediagram}, obtained by a Lorentzian fitting). The calculated $T_{C}$ is far below the Co fcc bulk limit, in qualitative agreement with Ref. \cite{Schneider1990}. For $T>T_{C}$, the random distribution of spin directions over \textit{finite} spaces and times led to arbitrary non-integer $Q$ values, characterizing the paramagnetic phase.

\begin{figure}[htb]
	\centering
\includegraphics[width=1.0\columnwidth]{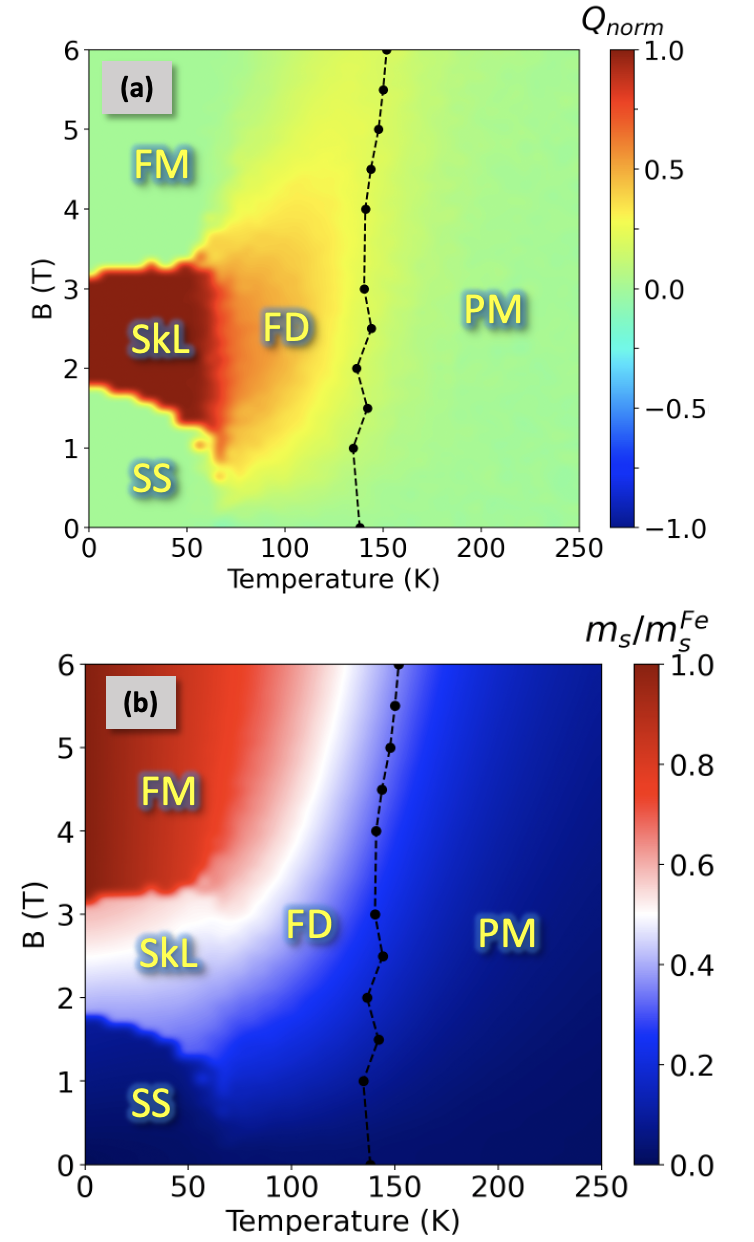}
\caption{(Color online) The same as Fig. \ref{fig:pdcoir-phasediagram}, but for Pd/Fe/Ir(111) with the normalized topological charge, $Q_{norm}$. 
The ferromagnetic (fully saturated, FM), spin-spiral (SS), skyrmion lattice (SkL), flutuation-disordered (FD), and paramagnetic (PM) phases are indicated in yellow. The dashed black lines are guides for the eyes.}
\label{fig:pdfeir-phasediagram}
\end{figure}

\begin{center}
\begin{figure}[hbt]
	\centering
\includegraphics[width=0.9\columnwidth]{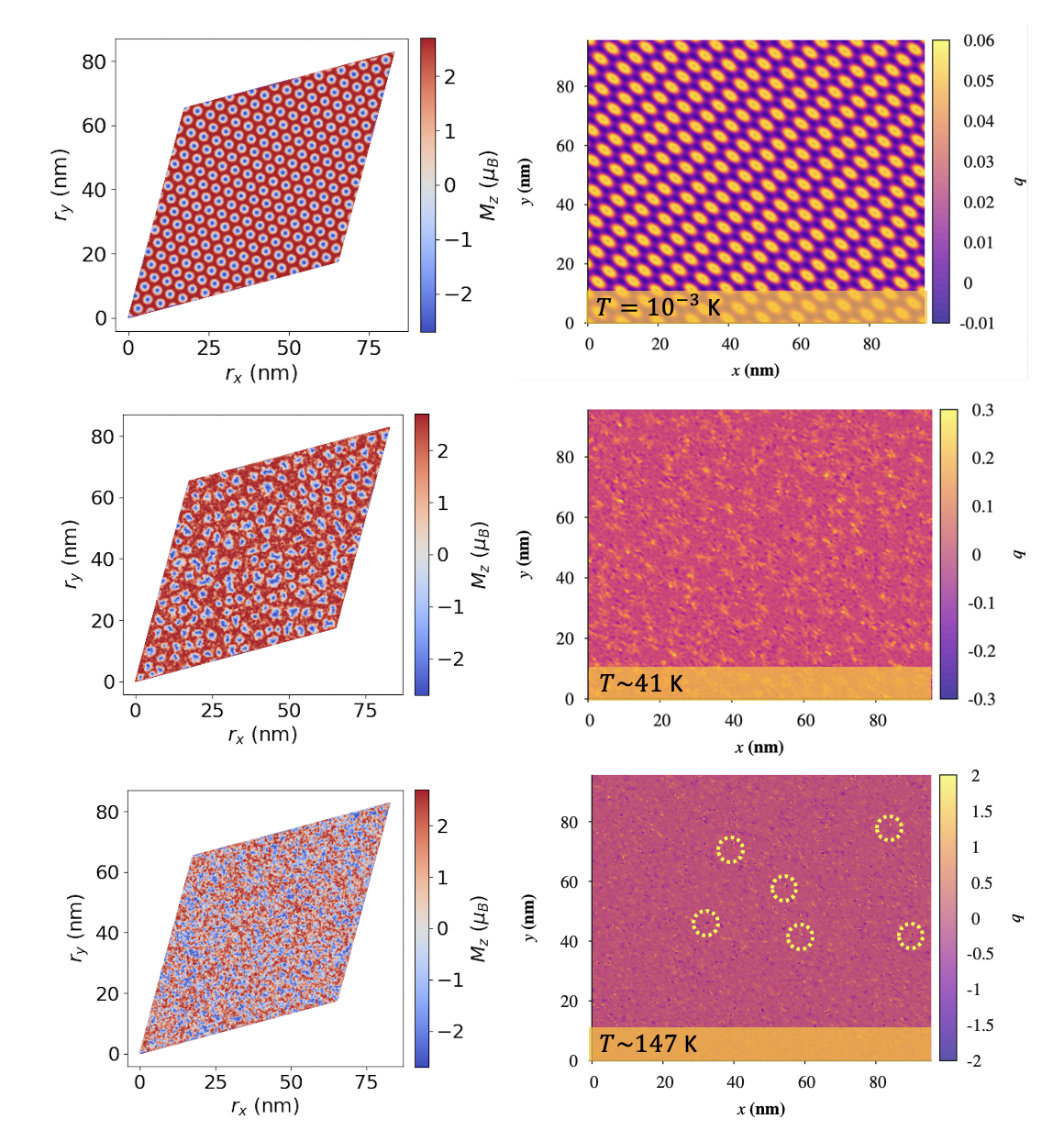}
\vspace*{-0.2cm}
\caption{(Color online) \textit{Left column}: Real-space image of spin configurations of Pd/Fe/Ir(111) for an applied external field of  $B\sim2.6$ T and temperatures of $T=10^{-3}$ K, $T\sim41$ K and $T\sim 147$ K, with the local $z$-component of the magnetic moment, $m_z$, indicated by the colorbars (in $\mu_{B}$). \textit{Right column}: Correspondent topological charge densities with intensity given by the colorbars. The yellow dotted circles indicate the occurrence of positive and negative poles for better visualization. Here, $r_{x}$ and $r_{y}$ are positions in the real-space (given in nm).}
\label{fig:topological-charge-poles}
\end{figure} 
\end{center}

In turn, for Pd/Fe/Ir(111), we obtained a much more complex phase diagram, shown in Fig. \ref{fig:pdfeir-phasediagram}. The most relevant is a SkL phase that occurs when $1.7\,\textnormal{T}\lesssim B\lesssim3\,\textnormal{T}$, excluding an intermediate region, mainly composed by a mixed (SS + skyrmions) state, accessed by SA simulations. The skyrmions vary in diameter and density as the applied magnetic field increases. Our calculations show that, for instance, at $T\sim0$ K the diameter change from $d\sim4.4$ nm for $B=1.8$ T to  $d\sim3.9$ nm for $B=3.3$ T, a reduction of $\sim12\%$ in the average value, in accordance to Refs. \cite{Romming2015,Malottki2019,Wang2018}; their shape, however, switch from the well-known circular shape to an amorphic shape due to thermal fluctuations for finite temperatures.  
Above $B\sim3$ T, the skyrmions gradually vanish, leading to a FM state. Indeed, the skyrmion density is reduced from $\sim 68\times10^{-3}$ skyrmions/nm$^{2}$ for $B\sim1.8$ T (skyrmion lattice) to 
$\sim16\times10^{-3}$ 
skyrmions/nm$^{2}$ for $B\sim3.3$ T. This transition is characterized by the presence of increasingly isolated skyrmions. Again using the criteria of divergence in $c_{V}$, the calculated $T_{C}$ to the PM phase is about $T_{C}\approx145$ K (see the black dots in Fig. \ref{fig:pdfeir-phasediagram}).
Differently from Pd/Co/Ir(111), the transition region to the PM phase is not characterized by an abrupt shift to arbitrary non-integer $Q$ values, but instead by a zone composed by arbitrary \textit{rational} topological charges with the same sign found for $Q$'s in the SkL region; previous studies usually call it a flutuation-disordered (FD) state \cite{Rozsa2016,Lindner2020,Bottcher2018}, where the skyrmion lifetime is finite. In terms of the topological charge densities, states in the FD state are composed by short-lived (positive and negative) poles, as shown in Fig.~\ref{fig:topological-charge-poles} (for $T\sim 147$ K). In contrast, for finite temperatures inside the SkL region (Fig.~\ref{fig:topological-charge-poles}, $T\sim 41$ K), the topological charge densities are composed of amorphic geometries indicating the positions where skyrmions are found in the real-space. The transition to the FD state, at about  $T\sim70$ K, agrees very well with recent experiments \cite{Lindner2020}.

Below  $B\sim1.7$ T (or   $B\sim1$ T, including the intermediate region), we find a SS configuration with a wavelength of   $\lambda\sim4$ nm when $B=0$ T, in excellent agreement with experimental results of $\lambda=5-7$ nm \cite{Romming2013,Kubetzka2017}. This is also corroborated by the average magnetic moment investigation (see Fig.~\ref{fig:pdfeir-phasediagram}b), in which $\bar{m}_{s}\rightarrow0$ for this range of $B$ values. We note that, because of the general FM character of Pd-Fe $J_{ij}$ interactions (see Fig. \ref{fig:exchange-fe-fe}, \textit{Inset}), induced by the high hybridization of the Fe-Pd ($3d$-$4d$) electronic states (Fig. \ref{fig:ldos-pd-fe-ir}), the Pd cover layer replicates the magnetic configuration formed in the Fe layer, however with lower intensity (due to the diminished induced magnetic moment of Pd in comparison to Fe). This result was confimed with atomistic spin dynamics simulations (data not shown), and it is in agreement with the observations in Ref. \cite{Romming2013}, in which the PdFe bilayer behaves as a single magnetic entity.

\onecolumngrid
\twocolumngrid

\vspace*{+25cm}
\subsubsection{Boundaries for the noncollinear textures at low temperatures}
\label{sec:boundaries}


Established the phase diagrams
, natural questions that arise are: which interactions are essential to the emergence of noncollinear textures in Pd/Fe/Ir(111)? Does Pd/Co/Ir(111) ever presents a noncollinear behavior for the consideration of distinct interactions neighborhoods? To answer these questions, we performed several low-temperature SA calculations with different ranges of $J_{ij}$ and $\vec{D}_{ij}$ interactions, for both surfaces. In the case of Pd/Co/Ir(111), considering DMI, $J_{ij}$'s up to \textit{fifth} neighbors and disregarding $B$ and the anisotropy term, we found noncollinear configurations (SS, with $\lambda>40$ nm, and SkL, composed by skyrmions with $d>20$ nm) as almost degenerated lower-energy solutions. In contrast, when the $K_{\textnormal{eff}}^{i}$ is included (Eq. \ref{eq:spinhamiltonian}), we always find a FM (single-domain) solution as the most stable; therefore, the shift of $E_{F}$ to higher energies, turning second neighbors and further $J_{ij}$ interactions to positive (or less negative) values, indeed vanishes the possibility of a noncollinear ordering in the Pd/Co bilayer. 

\begin{figure}[htb]
\centering
\includegraphics[width=\columnwidth]{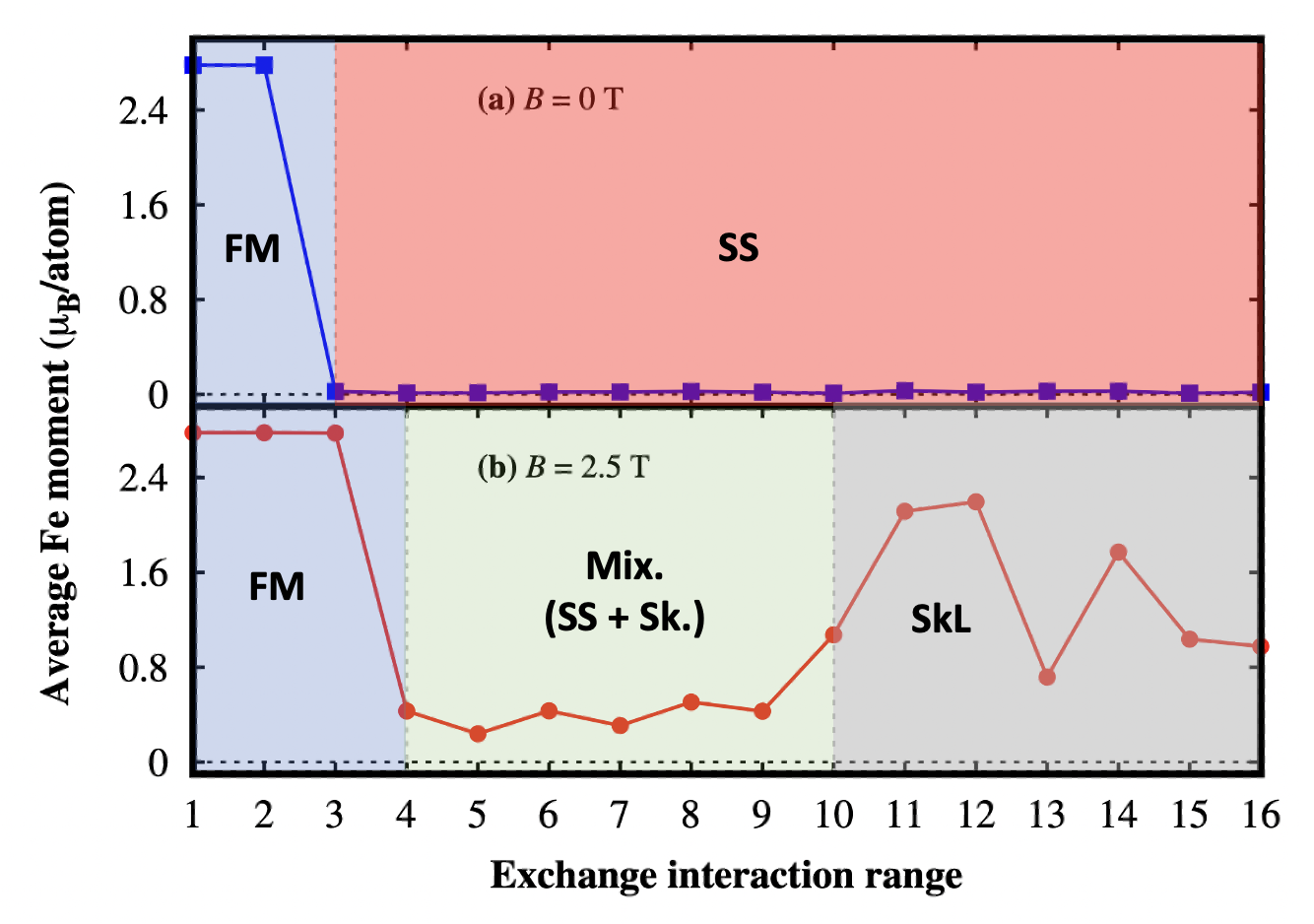}
\vspace*{-0.4cm}
\caption{(Color online) Evolution of the average Fe $m_{s}$ in Pd/Fe/Ir(111) at low temperature ($T=10^{-3}$ K), for (a) $B=0$; and (b) $B=2.5$ T, as further $J_{ij}$ interactions are considered (\textit{e.g.}, ``1'' indicates up to first neighbors), and in the presence of DMI and the experimental anisotropy. Lines are guides for the eyes.}
\label{fig:transition-to-noncollinear}
\end{figure}

For Pd/Fe/Ir(111), however, the case is more interesting. Without considering DMI, the inclusion of $J_{ij}$'s from fourth to seventh neighbors results in isolated  skyrmionic structures as metastable solutions in the presence of external magnetic fields, alternating between   stable SS, in agreement with \cite{Dupe2014}, and mixed states when $B=0$. With the full $J_{ij}$ set and $(B,K_{\textnormal{eff}}^{i})=(0,0)$, we find noncollinear metastable solutions (FM ground-state~\cite{Simon2014}). In turn, in the presence of DMI, $B=0$, and high anisotropies ($\sim1$ meV/Fe), the consideration until \textit{third} neighbors interactions produces chiral skyrmions with arbitrary topological charge \cite{Rybakov2019} as metastable solutions. With  the experimental anisotropy, the inclusion of $J_{ij}$'s further than   third neighbors, we observe the main transition to the SS phase. When $B=2.5$ T, the fourth neighbors $J_{ij}$'s also produces a transition from FM $\rightarrow$ mixed state (SS + skyrmions), while the  SkL (Fig. \ref{fig:pdfeir-phasediagram}) only emerges when considering \textit{tenth} neighbors interactions. Figure \ref{fig:transition-to-noncollinear} summarizes the regions for Pd/Fe/Ir(111) when DMI is present for different neighborhoods of exchange interactions. In brief, the fact that the first neighbors DMI Fe-Fe interaction has a relevant in-plane component is a necessary although not a sufficient condition to drive the complex noncollinear configurations here.

\subsection{Tuning Pd/Co/Ir(111)}

Now, we explore the possibility of tuning the magnetic ($J_{ij}$, $\vec{D}_{ij}$) parameters of Pd/Co/Ir(111), aiming for the emergence of noncollinear textures in this material which has a simple (single-domain) magnetic configuration. As mentioned in the Introduction, recent investigations have shown that these intrinsic interactions can be experimentally modified by the application of external electric fields, strain, or by introducing high-spin orbit defects. Also, considering the demonstrated  rigid-band-filling nature of the interactions, one can modify the $J_{ij}$ and $\vec{D}_{ij}$ by shifting $E_{F}$ through, for instance, FeCo alloying \cite{Levzaic2007,Spethmann2022}. Therefore, in this Section we consider initially the \textit{ab-initio} parameters calculated for Pd/Co/Ir(111), which, after tuned, will be analyzed separately.

\begin{center}
\begin{figure}[!htb]
	\centering
\includegraphics[width=0.9\columnwidth]{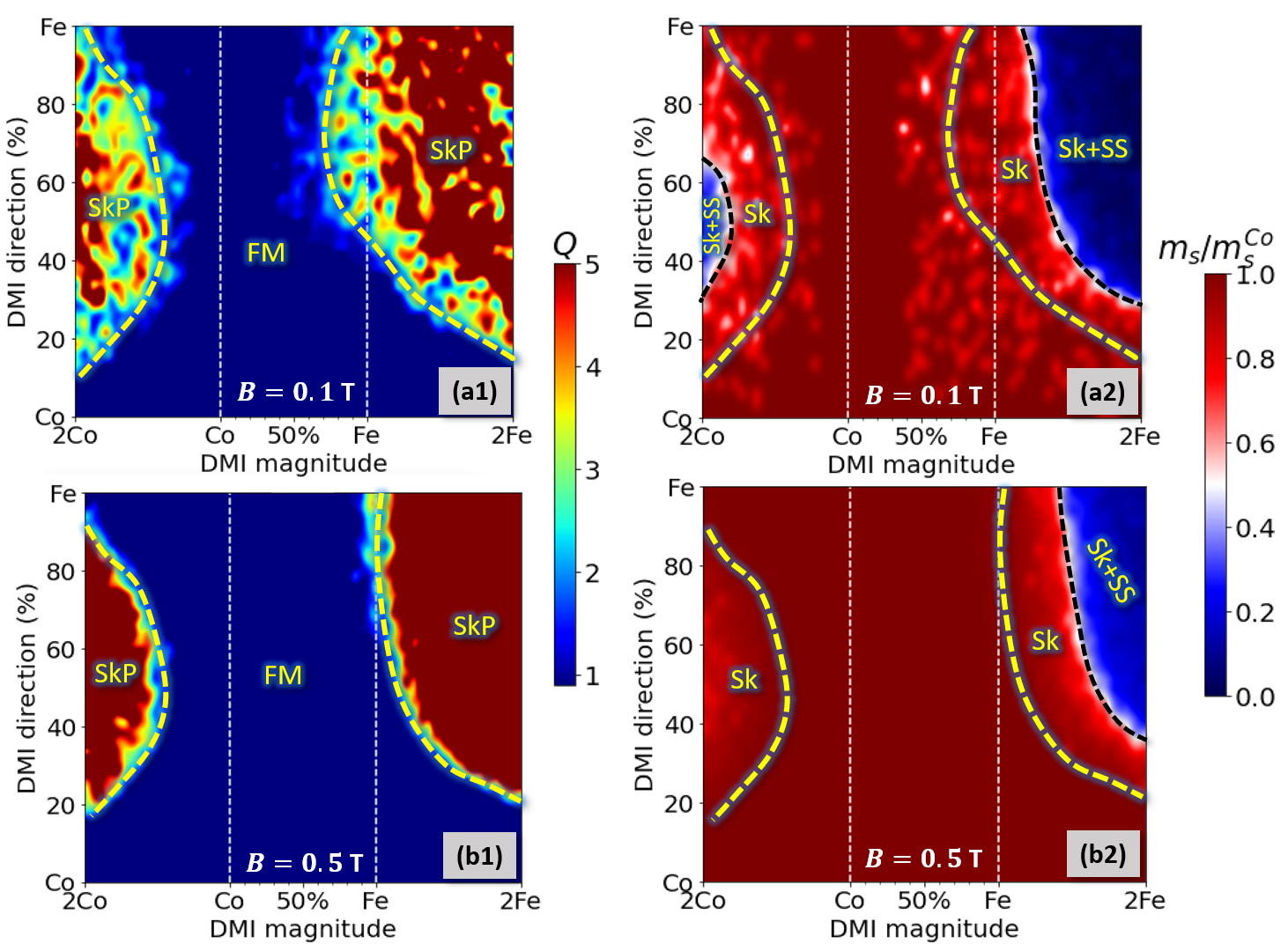}
\vspace*{-0.1cm}
	\caption{(Color online) 
	Skyrmion phase diagrams and the correspondent average relative spin moments $\left(\frac{m_{s}}{m_{s}^{\textnormal{Co}}}\right)$ with respect to the applied magnetic fields of (a1,a2) $B=0.1$ T; and (b1,b2) $B=0.5$ T, as a function of the DMI vector directions and magnitudes. Directions of $0\%$ and $100\%$ represent, respectively, the original Pd/Co/Ir(111) and Pd/Fe/Ir(111) DMI orientations. The $\vec{D}_{ij}$ magnitude, ranging from two times the Pd/Co/Ir(111) DMI (\textit{2Co}) to two times the Pd/Fe/Ir(111) DMI (\textit{2Fe}), passes through an intermediate transition from $|\vec{D}_{\textnormal{Co-Co}}|$ to $|\vec{D}_{\textnormal{Fe-Fe}}|$. The absolute topological charges are indicated in the colorbar, and the ferromagnetic (FM, single-domain), skyrmionic (SkP), isolated-skyrmion (Sk), and mixed (Sk + SS) phases are indicated in yellow. In all simulations, we considered $T=10^{-3}$ K.
	The dashed white, yellow, and black lines are guides for the eyes.}
	\label{fig:dm-diagram}
\end{figure}
\end{center}

\begin{figure}[htb]
	\centering
\hspace*{-0.7cm}
\includegraphics[width=1.2\columnwidth]{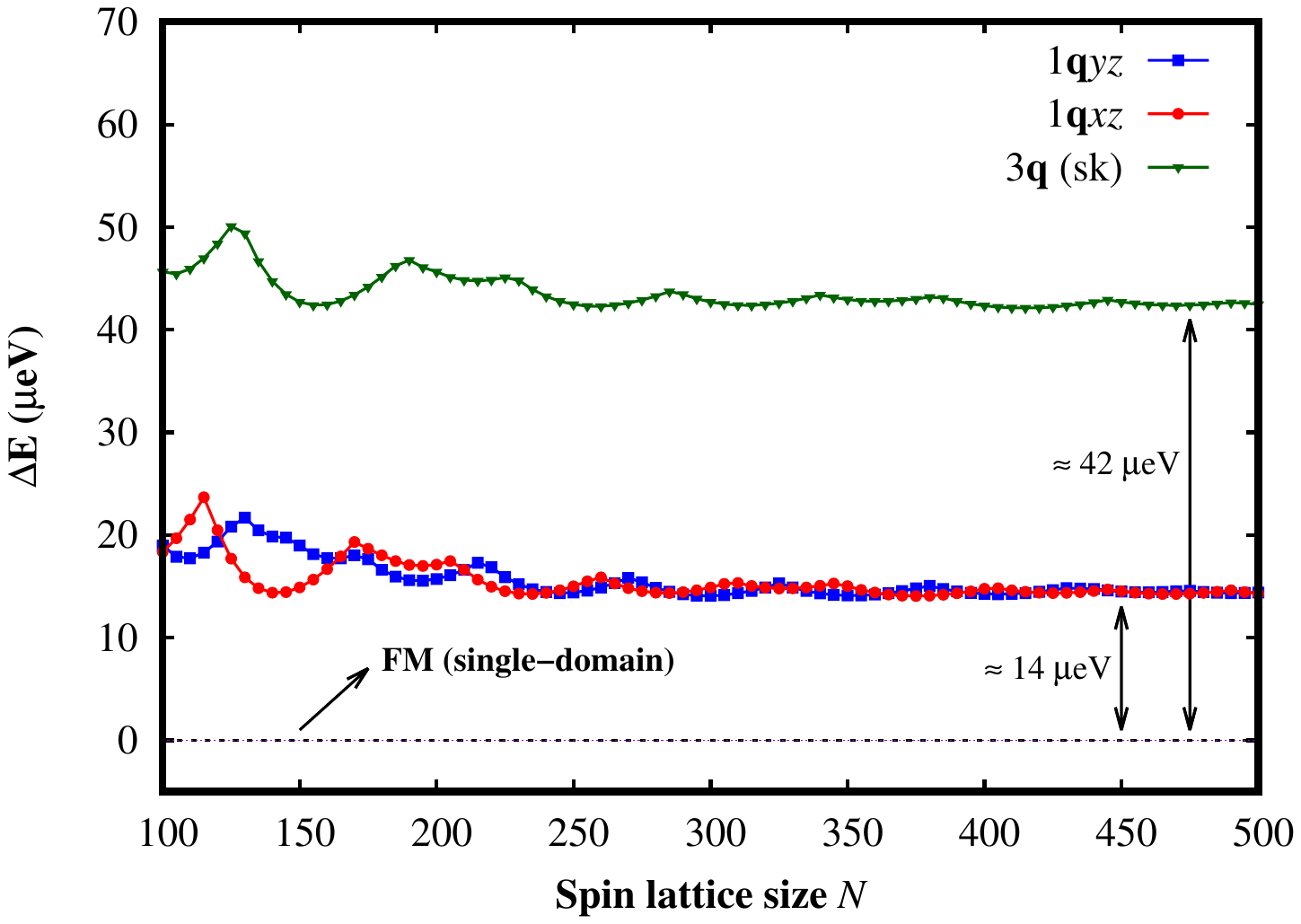}
\vspace*{-0.8cm}
	\caption{(Color online) Energy difference, $\Delta E$, with respect to the FM single-domain configuration (dashed line) as a function of the spin lattice size $N$ for: SS phase in the $\hat{y}$ ($1\textbf{q}\,xz$, red dots) and $\hat{x}$ ($1\textbf{q}\,yz$, blue squares) directions, and SkL phase ($3\textbf{q}$, green triangles). 
	For the simulations, $J_{ij}$ interaction set is kept as the one obtained for Pd/Co/Ir(111). 
	DMI magnitudes are two times the original $\left|\vec{D}_{\textnormal{Co-Co}}\right|$, and rotated to $\sim50\%$ of the Pd/Fe/Ir(111) $\vec{D}_{ij}$ directions. In all cases, we considered $T=10^{-3}$ K and $B=0.1$ T.}
	\label{fig:fm-sk-qx-energy}
\end{figure}

Since the DMI presents an essentially distinct behavior in Fe-based and Co-based bilayers, we inspect how the orientations and magnitudes influence in the emergence of skyrmions. Figure \ref{fig:dm-diagram} shows the absolute topological charge phase diagrams for the applied magnetic fields $B=0.1$ T and $B=0.5$ T as a function of the DMI vector rotations and magnitudes (exchange couplings are not changed). The $\vec{D}_{ij}$ orientations are varied from the obtained for Pd/Co/Ir(111) ($0\%$) to the obtained for Pd/Fe/Ir(111) ($100\%$), considering an interaction distance up to almost $\sim2.7$ nm (35 shells of neighbors, or $\sim7a$). Here, we generalize naming the set of these states as skyrmionic phases (SkP). If Pd/Co/Ir(111) DMI is simply twisted, keeping its strength (white dotted line following \textit{Co} in Fig. \ref{fig:dm-diagram}(a1,b1)), no noncollinear configuration is achieved. On the other hand, if the Co-Co DMI magnitude is increased, isolated skyrmionic structures (Sk) or mixed states (Sk + SS) are found in the rotation interval of $\sim15\%$ to $\sim100\%$, for both analyzed magnetic fields; as $B$ enhances, we also obtain an increase in the skyrmion counting in this region (higher $Q$'s). An example of state considering a rotation of $\sim50\%$ of the DMI orientation of Pd/Fe/Ir(111) and the double of the original $\vec{D}_{\textnormal{Co-Co}}$ magnitude, for $B=0.1$ T, is shown in Fig. \ref{fig:fm-sk-qx-energy}. The energy differences to the FM (single-domain) state as well as a comparison with SS phase in the $\hat{x}$ and $\hat{y}$ directions ($1\textbf{q}\,yz$ and $1\textbf{q}\,xz$, respectively), and with a pure SkL phase ($3\textbf{q}$), are calculated. As can be seen, both SS ($\Delta E\sim 14\,\mu$eV/Co) and SkL ($\Delta E\sim 42\,\mu$eV/Co) present small energy differences with respect to the FM configuration, thus being characterized by metastable solutions in the whole range of spin lattice sizes. The situation is similar to the reported for Co/Ru(0001) \cite{Herve2018}, in which metastable isolated skyrmions are found.

In Fig. \ref{fig:dm-diagram}, for DMI directions between $\sim40\%$ and $\sim60\%$ of $\vec{D}_{\textnormal{Fe-Fe}}$, and $B=0.1$ T, a mixed phase (SS + skyrmions) can occur specially near higher $\vec{D}_{ij}$ magnitudes (near two times $\vec{D}_{\textnormal{Co-Co}}$). This region vanishes when higher external magnetic fields (such as $B=0.5$ T) are applied, making room for an isolated-skyrmions phase. The situation is similar in the transition from Co $\rightarrow$ Fe and higher DMI magnitudes, except that, even for $B=0.5$ T, a region characterized by mixed states still survives. In this case, following the trends shown by Figs. \ref{fig:dm-diagram}(b1,b2), the large DMI magnitudes require even higher external fields to transform this (SS + Sk) region into an isolated-skyrmions zone.

\onecolumngrid
\twocolumngrid

\begin{center}
\begin{figure}[htb]
	\centering
\includegraphics[width=1.0\columnwidth]{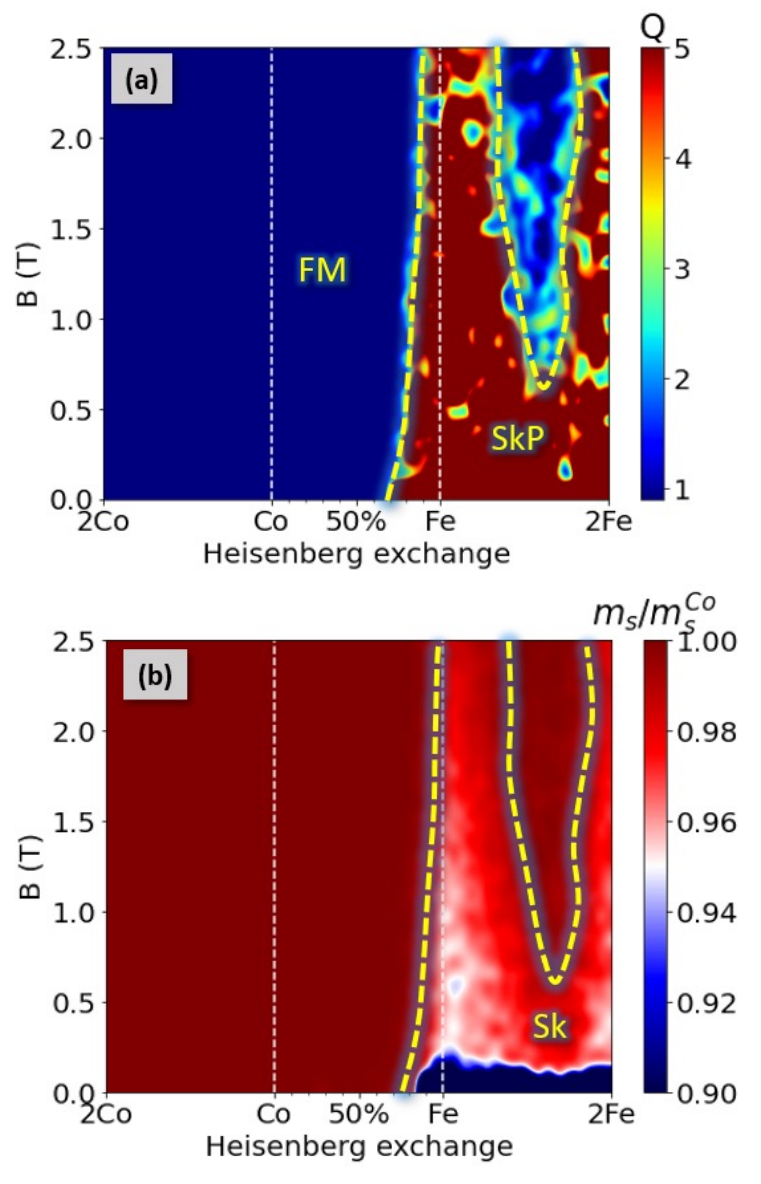}
\vspace*{-0.6cm}
	\caption{(Color online) 
	Skyrmion phase diagram and the correspondent average relative spin moments $\left(\frac{m_{s}}{m_{s}^{\textnormal{Co}}}\right)$ as a function of the exchange coupling, $J_{ij}$, and external magnetic field, $B$. The exchange magnitudes are varied from two times the Pd/Co/Ir(111) $J_{ij}$'s (\textit{2Co}) to two times the Pd/Fe/Ir(111) $J_{ij}$'s (\textit{2Fe}), passing through an intermediate transition from $J_{\textnormal{Co-Co}}$ to $J_{\textnormal{Fe-Fe}}$. The absolute topological charges are indicated in the colorbar, and the ferromagnetic (FM, single-domain) and skyrmionic phases (SkP) are indicated in yellow. In all simulations, we considered $T=10^{-3}$ K.
	}
	\label{fig:j-diagram}
\end{figure}
\end{center}

The effect of exchange coupling frustrations is investigated in Fig. \ref{fig:j-diagram}, which shows the skyrmion phase diagram as a function of the $J_{ij}$ strength and the applied $B$; in this case, the original DMI magnitudes and directions of Pd/Co/Ir(111) are not modified. This transition allows for $J_{\textnormal{Co-Co}}$ to transform into $J_{\textnormal{Fe-Fe}}$ in an interaction distance up to $\sim2.7$ nm (or $\sim7a$), so the FM-AFM exchange coupling competition existing in Pd/Fe/Ir(111) is gradually emerged. Interestingly, we note that a SkP only appears near the $J_{\textnormal{Fe-Fe}}$ line (white dotted line following \textit{Fe} in Figs. \ref{fig:j-diagram}(a,b)), for external magnetic fields smaller than $B=2.5$ T. On $J_{ij}$ strengths higher than $J_{\textnormal{Fe-Fe}}$, the strong frustration induced forces the system to a metastable isolated-skyrmions phase, specially for low external magnetic fields. 
Therefore, the effect of magnetic frustration is sufficient to render skyrmions from about the $J_{\textnormal{Fe-Fe}}$ limit, which are stabilized with the help of magnetic fields, the not completely out-of-plane $\vec{D}_{\textnormal{Co-Co}}$, and an out-of-plane magnetic anisotropy. The latter has been previously identified as a key ingredient for the emergence of complex structures in frustrated magnets \cite{Hayami2016}.


\twocolumngrid

\section{Conclusions}
\label{sec:conclusions}

 We investigated the electronic and magnetic properties of Pd/Fe/Ir(111) and Pd/Co/Ir(111) using a combination of first-principles calculations and spin-dynamics simulations, aiming to explain their contrasting magnetic behaviors. For Pd/Fe/Ir(111), we obtained that the ground-state is characterized by a noncollinear spin arrangement, in which both exchange and DMI have, simultaneously, significant roles. In particular, the emergent FM-AFM magnetic frustration present on the Fe-Fe exchange interactions is not sufficient to drive, alone, the Pd/Fe/Ir(111) bilayer into a noncollinear ground-state configuration, but leads to a metastable one. In turn, the Pd/Fe/Ir(111) DMI first neighbors vectors,  $\vec{D}_{ij}^{1st}$, exhibit an almost contrary behavior from the Pd/Co/Ir(111), although both present a long-range influence, particularly driven by the $D_{y}$ (in-plane) component. 
Due to this long-range nature of $J_{ij}$ and $\vec{D}_{ij}$, the correct Pd/Fe/Ir(111) noncollinear states are only achieved with the consideration of, at least, fourth neighbors interactions.
Nevertheless, in the case of Pd/Co/Ir(111), these further neighbors interactions are not sufficient to change the magnetic behavior from the simple FM (single-domain) configuration. In order to unravel this puzzle, by performing state-of-the-art \textit{in silico} simulations, the original Pd/Co/Ir(111) $\vec{D}_{ij}$ and $J_{ij}$ parameters were separetly tuned into the ones of Pd/Fe/Ir(111), and  higher magnitudes. Interestingly, we found conditions for the emergence of SkP in the originally FM single-domain Pd/Co/Ir(111), by rotating the DMI vectors, or inducing a FM-AFM competition in the exchange interactions.

The \textit{ab-initio} $J_{ij}$ and DMI behaviors have shown to exhibit a simple rigid-band-like model nature. Among the most important effects, by switching the Fermi level to higher energies, the extra minority $3d$ electron in Co (w.r.t. Fe) is responsible for the emergence of a relevant AFM $J^{3rd}_{\textnormal{Fe}-\textnormal{Fe}}$ and a more in-plane $\vec{D}^{1st}_{\textnormal{Fe}-\textnormal{Fe}}$. 

 The present deep investigation contributes to the understanding of the origin of the noncollinear magnetism (and skyrmions) in Pd/Fe/Ir(111) and not in Pd/Co/Ir(111).  Moreover, it also opens avenues to the study of conditions for the emergence of skyrmions in non-skyrmionic magnetic thin films. 

\section{Acknowledgements}

H.M.P. and A.B.K. acknowledge financial
support from CAPES, CNPq and FAPESP, Brazil. A.B. acknowledges eSSENCE. 
I.M. acknowledges financial support from CAPES, Finance Code 001, process n$^{\circ}$ 88882.332894/2018-01, and in the Institutional Program of Overseas Sandwich Doctorate, process n$^{\circ}$ 88881.187258/2018-01. 
The calculations were performed at the computational
facilities of the HPC-USP/CENAPAD-UNICAMP (Brazil), at the National Laboratory for Scientific Computing (LNCC/MCTI, Brazil), and at the Swedish National Infrastructure for Computing (SNIC).

\bibliographystyle{apsrev4-2}
\bibliography{library.bib}

\begin{thebibliography}{97}%
\makeatletter
\providecommand \@ifxundefined [1]{%
 \@ifx{#1\undefined}
}%
\providecommand \@ifnum [1]{%
 \ifnum #1\expandafter \@firstoftwo
 \else \expandafter \@secondoftwo
 \fi
}%
\providecommand \@ifx [1]{%
 \ifx #1\expandafter \@firstoftwo
 \else \expandafter \@secondoftwo
 \fi
}%
\providecommand \natexlab [1]{#1}%
\providecommand \enquote  [1]{``#1''}%
\providecommand \bibnamefont  [1]{#1}%
\providecommand \bibfnamefont [1]{#1}%
\providecommand \citenamefont [1]{#1}%
\providecommand \href@noop [0]{\@secondoftwo}%
\providecommand \href [0]{\begingroup \@sanitize@url \@href}%
\providecommand \@href[1]{\@@startlink{#1}\@@href}%
\providecommand \@@href[1]{\endgroup#1\@@endlink}%
\providecommand \@sanitize@url [0]{\catcode `\\12\catcode `\$12\catcode
  `\&12\catcode `\#12\catcode `\^12\catcode `\_12\catcode `\%12\relax}%
\providecommand \@@startlink[1]{}%
\providecommand \@@endlink[0]{}%
\providecommand \url  [0]{\begingroup\@sanitize@url \@url }%
\providecommand \@url [1]{\endgroup\@href {#1}{\urlprefix }}%
\providecommand \urlprefix  [0]{URL }%
\providecommand \Eprint [0]{\href }%
\providecommand \doibase [0]{https://doi.org/}%
\providecommand \selectlanguage [0]{\@gobble}%
\providecommand \bibinfo  [0]{\@secondoftwo}%
\providecommand \bibfield  [0]{\@secondoftwo}%
\providecommand \translation [1]{[#1]}%
\providecommand \BibitemOpen [0]{}%
\providecommand \bibitemStop [0]{}%
\providecommand \bibitemNoStop [0]{.\EOS\space}%
\providecommand \EOS [0]{\spacefactor3000\relax}%
\providecommand \BibitemShut  [1]{\csname bibitem#1\endcsname}%
\let\auto@bib@innerbib\@empty
\bibitem [{\citenamefont {Holzberger}\ \emph {et~al.}(2013)\citenamefont
  {Holzberger}, \citenamefont {Schuh}, \citenamefont {Bl{\"{u}}gel},
  \citenamefont {Lounis},\ and\ \citenamefont {Wulfhekel}}]{Holzberger2013}%
  \BibitemOpen
  \bibfield  {author} {\bibinfo {author} {\bibfnamefont {S.}~\bibnamefont
  {Holzberger}}, \bibinfo {author} {\bibfnamefont {T.}~\bibnamefont {Schuh}},
  \bibinfo {author} {\bibfnamefont {S.}~\bibnamefont {Bl{\"{u}}gel}}, \bibinfo
  {author} {\bibfnamefont {S.}~\bibnamefont {Lounis}},\ and\ \bibinfo {author}
  {\bibfnamefont {W.}~\bibnamefont {Wulfhekel}},\ }\href
  {https://doi.org/10.1103/PhysRevLett.110.157206} {\bibfield  {journal}
  {\bibinfo  {journal} {Phys. Rev. Lett.}\ }\textbf {\bibinfo {volume} {110}},\
  \bibinfo {pages} {157206} (\bibinfo {year} {2013})}\BibitemShut {NoStop}%
\bibitem [{\citenamefont {Lounis}\ \emph {et~al.}(2008)\citenamefont {Lounis},
  \citenamefont {Dederichs},\ and\ \citenamefont {Bl{\"{u}}gel}}]{Lounis2008}%
  \BibitemOpen
  \bibfield  {author} {\bibinfo {author} {\bibfnamefont {S.}~\bibnamefont
  {Lounis}}, \bibinfo {author} {\bibfnamefont {P.~H.}\ \bibnamefont
  {Dederichs}},\ and\ \bibinfo {author} {\bibfnamefont {S.}~\bibnamefont
  {Bl{\"{u}}gel}},\ }\href {https://doi.org/10.1103/PhysRevLett.101.107204}
  {\bibfield  {journal} {\bibinfo  {journal} {Phys. Rev. Lett.}\ }\textbf
  {\bibinfo {volume} {101}},\ \bibinfo {pages} {107204} (\bibinfo {year}
  {2008})}\BibitemShut {NoStop}%
\bibitem [{\citenamefont {Igarashi}\ \emph {et~al.}(2012)\citenamefont
  {Igarashi}, \citenamefont {Klautau}, \citenamefont {Muniz}, \citenamefont
  {Sanyal},\ and\ \citenamefont {Petrilli}}]{Igarashi2012b}%
  \BibitemOpen
  \bibfield  {author} {\bibinfo {author} {\bibfnamefont {R.~N.}\ \bibnamefont
  {Igarashi}}, \bibinfo {author} {\bibfnamefont {A.~B.}\ \bibnamefont
  {Klautau}}, \bibinfo {author} {\bibfnamefont {R.~B.}\ \bibnamefont {Muniz}},
  \bibinfo {author} {\bibfnamefont {B.}~\bibnamefont {Sanyal}},\ and\ \bibinfo
  {author} {\bibfnamefont {H.~M.}\ \bibnamefont {Petrilli}},\ }\href
  {https://doi.org/10.1103/PhysRevB.85.014436} {\bibfield  {journal} {\bibinfo
  {journal} {Phys. Rev. B}\ }\textbf {\bibinfo {volume} {85}},\ \bibinfo
  {pages} {014436} (\bibinfo {year} {2012})}\BibitemShut {NoStop}%
\bibitem [{\citenamefont {Ribeiro}\ \emph
  {et~al.}(2011{\natexlab{a}})\citenamefont {Ribeiro}, \citenamefont
  {Corr{\^{e}}a}, \citenamefont {Bergman}, \citenamefont {Nordstr{\"{o}}m},
  \citenamefont {Eriksson},\ and\ \citenamefont {Klautau}}]{Ribeiro2011}%
  \BibitemOpen
  \bibfield  {author} {\bibinfo {author} {\bibfnamefont {M.~S.}\ \bibnamefont
  {Ribeiro}}, \bibinfo {author} {\bibfnamefont {G.~B.}\ \bibnamefont
  {Corr{\^{e}}a}}, \bibinfo {author} {\bibfnamefont {A.}~\bibnamefont
  {Bergman}}, \bibinfo {author} {\bibfnamefont {L.}~\bibnamefont
  {Nordstr{\"{o}}m}}, \bibinfo {author} {\bibfnamefont {O.}~\bibnamefont
  {Eriksson}},\ and\ \bibinfo {author} {\bibfnamefont {A.~B.}\ \bibnamefont
  {Klautau}},\ }\href {https://doi.org/10.1103/PhysRevB.83.014406} {\bibfield
  {journal} {\bibinfo  {journal} {Phys. Rev. B}\ }\textbf {\bibinfo {volume}
  {83}},\ \bibinfo {pages} {014406} (\bibinfo {year}
  {2011}{\natexlab{a}})}\BibitemShut {NoStop}%
\bibitem [{\citenamefont {Igarashi}\ \emph {et~al.}(2016)\citenamefont
  {Igarashi}, \citenamefont {Miranda}, \citenamefont {Eleno}, \citenamefont
  {Klautau},\ and\ \citenamefont {Petrilli}}]{Igarashi2016}%
  \BibitemOpen
  \bibfield  {author} {\bibinfo {author} {\bibfnamefont {R.~N.}\ \bibnamefont
  {Igarashi}}, \bibinfo {author} {\bibfnamefont {I.~P.}\ \bibnamefont
  {Miranda}}, \bibinfo {author} {\bibfnamefont {L.~T.~F.}\ \bibnamefont
  {Eleno}}, \bibinfo {author} {\bibfnamefont {A.~B.}\ \bibnamefont {Klautau}},\
  and\ \bibinfo {author} {\bibfnamefont {H.~M.}\ \bibnamefont {Petrilli}},\
  }\href {https://doi.org/10.1088/0953-8984/28/32/326001} {\bibfield  {journal}
  {\bibinfo  {journal} {J. Phys.: Condens. Matter}\ }\textbf {\bibinfo {volume}
  {28}},\ \bibinfo {pages} {326001} (\bibinfo {year} {2016})}\BibitemShut
  {NoStop}%
\bibitem [{\citenamefont {Lounis}(2014)}]{Lounis2014}%
  \BibitemOpen
  \bibfield  {author} {\bibinfo {author} {\bibfnamefont {S.}~\bibnamefont
  {Lounis}},\ }\href {https://doi.org/10.1088/0953-8984/26/27/273201}
  {\bibfield  {journal} {\bibinfo  {journal} {J. Phys.: Condens. Matter}\
  }\textbf {\bibinfo {volume} {26}},\ \bibinfo {pages} {273201} (\bibinfo
  {year} {2014})}\BibitemShut {NoStop}%
\bibitem [{\citenamefont {Bezerra-Neto}\ \emph {et~al.}(2013)\citenamefont
  {Bezerra-Neto}, \citenamefont {Ribeiro}, \citenamefont {Sanyal},
  \citenamefont {Bergman}, \citenamefont {Muniz}, \citenamefont {Eriksson},\
  and\ \citenamefont {Klautau}}]{Bezerra-Neto2013}%
  \BibitemOpen
  \bibfield  {author} {\bibinfo {author} {\bibfnamefont {M.~M.}\ \bibnamefont
  {Bezerra-Neto}}, \bibinfo {author} {\bibfnamefont {M.~S.}\ \bibnamefont
  {Ribeiro}}, \bibinfo {author} {\bibfnamefont {B.}~\bibnamefont {Sanyal}},
  \bibinfo {author} {\bibfnamefont {A.}~\bibnamefont {Bergman}}, \bibinfo
  {author} {\bibfnamefont {R.~B.}\ \bibnamefont {Muniz}}, \bibinfo {author}
  {\bibfnamefont {O.}~\bibnamefont {Eriksson}},\ and\ \bibinfo {author}
  {\bibfnamefont {A.~B.}\ \bibnamefont {Klautau}},\ }\href
  {https://doi.org/10.1038/srep03054} {\bibfield  {journal} {\bibinfo
  {journal} {Sci. Rep.}\ }\textbf {\bibinfo {volume} {3}},\ \bibinfo {pages}
  {3054} (\bibinfo {year} {2013})}\BibitemShut {NoStop}%
\bibitem [{\citenamefont {Phark}\ \emph {et~al.}(2014)\citenamefont {Phark},
  \citenamefont {Fischer}, \citenamefont {Corbetta}, \citenamefont {Sander},
  \citenamefont {Nakamura},\ and\ \citenamefont {Kirschner}}]{Phark2014}%
  \BibitemOpen
  \bibfield  {author} {\bibinfo {author} {\bibfnamefont {S.~H.}\ \bibnamefont
  {Phark}}, \bibinfo {author} {\bibfnamefont {J.~A.}\ \bibnamefont {Fischer}},
  \bibinfo {author} {\bibfnamefont {M.}~\bibnamefont {Corbetta}}, \bibinfo
  {author} {\bibfnamefont {D.}~\bibnamefont {Sander}}, \bibinfo {author}
  {\bibfnamefont {K.}~\bibnamefont {Nakamura}},\ and\ \bibinfo {author}
  {\bibfnamefont {J.}~\bibnamefont {Kirschner}},\ }\href
  {https://doi.org/10.1038/ncomms6183} {\bibfield  {journal} {\bibinfo
  {journal} {Nat. Commun.}\ }\textbf {\bibinfo {volume} {5}},\ \bibinfo {pages}
  {5183} (\bibinfo {year} {2014})}\BibitemShut {NoStop}%
\bibitem [{\citenamefont {Cardias}\ \emph
  {et~al.}(2016{\natexlab{a}})\citenamefont {Cardias}, \citenamefont
  {Bezerra-Neto}, \citenamefont {Ribeiro}, \citenamefont {Bergman},
  \citenamefont {Szilva}, \citenamefont {Eriksson},\ and\ \citenamefont
  {Klautau}}]{Cardias2016}%
  \BibitemOpen
  \bibfield  {author} {\bibinfo {author} {\bibfnamefont {R.}~\bibnamefont
  {Cardias}}, \bibinfo {author} {\bibfnamefont {M.~M.}\ \bibnamefont
  {Bezerra-Neto}}, \bibinfo {author} {\bibfnamefont {M.~S.}\ \bibnamefont
  {Ribeiro}}, \bibinfo {author} {\bibfnamefont {A.}~\bibnamefont {Bergman}},
  \bibinfo {author} {\bibfnamefont {A.}~\bibnamefont {Szilva}}, \bibinfo
  {author} {\bibfnamefont {O.}~\bibnamefont {Eriksson}},\ and\ \bibinfo
  {author} {\bibfnamefont {A.~B.}\ \bibnamefont {Klautau}},\ }\href
  {https://doi.org/10.1103/PhysRevB.93.014438} {\bibfield  {journal} {\bibinfo
  {journal} {Phys. Rev. B}\ }\textbf {\bibinfo {volume} {93}},\ \bibinfo
  {pages} {014438} (\bibinfo {year} {2016}{\natexlab{a}})}\BibitemShut
  {NoStop}%
\bibitem [{\citenamefont {Fert}\ \emph {et~al.}(2017)\citenamefont {Fert},
  \citenamefont {Reyren},\ and\ \citenamefont {Cros}}]{Fert2017}%
  \BibitemOpen
  \bibfield  {author} {\bibinfo {author} {\bibfnamefont {A.}~\bibnamefont
  {Fert}}, \bibinfo {author} {\bibfnamefont {N.}~\bibnamefont {Reyren}},\ and\
  \bibinfo {author} {\bibfnamefont {V.}~\bibnamefont {Cros}},\ }\href
  {https://doi.org/10.1038/natrevmats.2017.31} {\bibfield  {journal} {\bibinfo
  {journal} {Nat. Rev. Mater.}\ }\textbf {\bibinfo {volume} {2}},\ \bibinfo
  {pages} {17031} (\bibinfo {year} {2017})}\BibitemShut {NoStop}%
\bibitem [{\citenamefont {Wiesendanger}(2016)}]{Wiesendanger2016}%
  \BibitemOpen
  \bibfield  {author} {\bibinfo {author} {\bibfnamefont {R.}~\bibnamefont
  {Wiesendanger}},\ }\href {https://doi.org/10.1038/natrevmats.2016.44}
  {\bibfield  {journal} {\bibinfo  {journal} {Nat. Rev. Mater.}\ }\textbf
  {\bibinfo {volume} {1}},\ \bibinfo {pages} {16044} (\bibinfo {year}
  {2016})}\BibitemShut {NoStop}%
\bibitem [{\citenamefont {Romming}\ \emph {et~al.}(2013)\citenamefont
  {Romming}, \citenamefont {Hanneken}, \citenamefont {Menzel}, \citenamefont
  {Bickel}, \citenamefont {Wolter}, \citenamefont {von Bergmann}, \citenamefont
  {Kubetzka},\ and\ \citenamefont {Wiesendanger}}]{Romming2013}%
  \BibitemOpen
  \bibfield  {author} {\bibinfo {author} {\bibfnamefont {N.}~\bibnamefont
  {Romming}}, \bibinfo {author} {\bibfnamefont {C.}~\bibnamefont {Hanneken}},
  \bibinfo {author} {\bibfnamefont {M.}~\bibnamefont {Menzel}}, \bibinfo
  {author} {\bibfnamefont {J.~E.}\ \bibnamefont {Bickel}}, \bibinfo {author}
  {\bibfnamefont {B.}~\bibnamefont {Wolter}}, \bibinfo {author} {\bibfnamefont
  {K.}~\bibnamefont {von Bergmann}}, \bibinfo {author} {\bibfnamefont
  {A.}~\bibnamefont {Kubetzka}},\ and\ \bibinfo {author} {\bibfnamefont
  {R.}~\bibnamefont {Wiesendanger}},\ }\href
  {https://doi.org/10.1126/science.1240573} {\bibfield  {journal} {\bibinfo
  {journal} {Science}\ }\textbf {\bibinfo {volume} {341}},\ \bibinfo {pages}
  {636} (\bibinfo {year} {2013})}\BibitemShut {NoStop}%
\bibitem [{\citenamefont {M{\"{u}}hlbauer}\ \emph {et~al.}(2009)\citenamefont
  {M{\"{u}}hlbauer}, \citenamefont {Binz}, \citenamefont {Jonietz},
  \citenamefont {Pfleiderer}, \citenamefont {Rosch}, \citenamefont {Neubauer},
  \citenamefont {Georgii},\ and\ \citenamefont {B{\"{o}}ni}}]{Mulhbauer2009}%
  \BibitemOpen
  \bibfield  {author} {\bibinfo {author} {\bibfnamefont {S.}~\bibnamefont
  {M{\"{u}}hlbauer}}, \bibinfo {author} {\bibfnamefont {B.}~\bibnamefont
  {Binz}}, \bibinfo {author} {\bibfnamefont {F.}~\bibnamefont {Jonietz}},
  \bibinfo {author} {\bibfnamefont {C.}~\bibnamefont {Pfleiderer}}, \bibinfo
  {author} {\bibfnamefont {A.}~\bibnamefont {Rosch}}, \bibinfo {author}
  {\bibfnamefont {A.}~\bibnamefont {Neubauer}}, \bibinfo {author}
  {\bibfnamefont {R.}~\bibnamefont {Georgii}},\ and\ \bibinfo {author}
  {\bibfnamefont {P.}~\bibnamefont {B{\"{o}}ni}},\ }\href
  {https://doi.org/10.1126/science.1166767} {\bibfield  {journal} {\bibinfo
  {journal} {Science}\ }\textbf {\bibinfo {volume} {323}},\ \bibinfo {pages}
  {915} (\bibinfo {year} {2009})}\BibitemShut {NoStop}%
\bibitem [{\citenamefont {Sampaio}\ \emph {et~al.}(2013)\citenamefont
  {Sampaio}, \citenamefont {Cros}, \citenamefont {Rohart}, \citenamefont
  {Thiaville},\ and\ \citenamefont {Fert}}]{Sampaio2013}%
  \BibitemOpen
  \bibfield  {author} {\bibinfo {author} {\bibfnamefont {J.}~\bibnamefont
  {Sampaio}}, \bibinfo {author} {\bibfnamefont {V.}~\bibnamefont {Cros}},
  \bibinfo {author} {\bibfnamefont {S.}~\bibnamefont {Rohart}}, \bibinfo
  {author} {\bibfnamefont {A.}~\bibnamefont {Thiaville}},\ and\ \bibinfo
  {author} {\bibfnamefont {A.}~\bibnamefont {Fert}},\ }\href
  {https://doi.org/10.1038/nnano.2013.210} {\bibfield  {journal} {\bibinfo
  {journal} {Nat. Nanotechnol.}\ }\textbf {\bibinfo {volume} {8}},\ \bibinfo
  {pages} {839} (\bibinfo {year} {2013})}\BibitemShut {NoStop}%
\bibitem [{\citenamefont {Zhang}\ \emph {et~al.}(2015)\citenamefont {Zhang},
  \citenamefont {Ezawa},\ and\ \citenamefont {Zhou}}]{Zhang2015a}%
  \BibitemOpen
  \bibfield  {author} {\bibinfo {author} {\bibfnamefont {X.}~\bibnamefont
  {Zhang}}, \bibinfo {author} {\bibfnamefont {M.}~\bibnamefont {Ezawa}},\ and\
  \bibinfo {author} {\bibfnamefont {Y.}~\bibnamefont {Zhou}},\ }\href
  {https://doi.org/10.1038/srep09400} {\bibfield  {journal} {\bibinfo
  {journal} {Sci. Rep.}\ }\textbf {\bibinfo {volume} {5}},\ \bibinfo {pages}
  {9400} (\bibinfo {year} {2015})}\BibitemShut {NoStop}%
\bibitem [{\citenamefont {Kiselev}\ \emph {et~al.}(2011)\citenamefont
  {Kiselev}, \citenamefont {Bogdanov}, \citenamefont {Sch{\"{a}}fer},\ and\
  \citenamefont {Ler}}]{Kiselev2011}%
  \BibitemOpen
  \bibfield  {author} {\bibinfo {author} {\bibfnamefont {N.~S.}\ \bibnamefont
  {Kiselev}}, \bibinfo {author} {\bibfnamefont {a.~N.}\ \bibnamefont
  {Bogdanov}}, \bibinfo {author} {\bibfnamefont {R.}~\bibnamefont
  {Sch{\"{a}}fer}},\ and\ \bibinfo {author} {\bibfnamefont {U.~K.~R.}\
  \bibnamefont {Ler}},\ }\href {https://doi.org/10.1088/0022-3727/44/39/392001}
  {\bibfield  {journal} {\bibinfo  {journal} {J. Phys. D: Appl. Phys.}\
  }\textbf {\bibinfo {volume} {44}},\ \bibinfo {pages} {392001} (\bibinfo
  {year} {2011})}\BibitemShut {NoStop}%
\bibitem [{\citenamefont {Heinze}\ \emph {et~al.}(2011)\citenamefont {Heinze},
  \citenamefont {von Bergmann}, \citenamefont {Menzel}, \citenamefont {Brede},
  \citenamefont {Kubetzka}, \citenamefont {Wiesendanger}, \citenamefont
  {Bihlmayer},\ and\ \citenamefont {Bl{\"{u}}gel}}]{Heinze2011}%
  \BibitemOpen
  \bibfield  {author} {\bibinfo {author} {\bibfnamefont {S.}~\bibnamefont
  {Heinze}}, \bibinfo {author} {\bibfnamefont {K.}~\bibnamefont {von
  Bergmann}}, \bibinfo {author} {\bibfnamefont {M.}~\bibnamefont {Menzel}},
  \bibinfo {author} {\bibfnamefont {J.}~\bibnamefont {Brede}}, \bibinfo
  {author} {\bibfnamefont {A.}~\bibnamefont {Kubetzka}}, \bibinfo {author}
  {\bibfnamefont {R.}~\bibnamefont {Wiesendanger}}, \bibinfo {author}
  {\bibfnamefont {G.}~\bibnamefont {Bihlmayer}},\ and\ \bibinfo {author}
  {\bibfnamefont {S.}~\bibnamefont {Bl{\"{u}}gel}},\ }\href
  {https://doi.org/10.1038/nphys2045} {\bibfield  {journal} {\bibinfo
  {journal} {Nat. Phys.}\ }\textbf {\bibinfo {volume} {7}},\ \bibinfo {pages}
  {713} (\bibinfo {year} {2011})}\BibitemShut {NoStop}%
\bibitem [{\citenamefont {Herv{\'{e}}}\ \emph {et~al.}(2018)\citenamefont
  {Herv{\'{e}}}, \citenamefont {Dup{\'{e}}}, \citenamefont {Lopes},
  \citenamefont {B{\"{o}}ttcher}, \citenamefont {Martins}, \citenamefont
  {Balashov}, \citenamefont {Gerhard}, \citenamefont {Sinova},\ and\
  \citenamefont {Wulfhekel}}]{Herve2018}%
  \BibitemOpen
  \bibfield  {author} {\bibinfo {author} {\bibfnamefont {M.}~\bibnamefont
  {Herv{\'{e}}}}, \bibinfo {author} {\bibfnamefont {B.}~\bibnamefont
  {Dup{\'{e}}}}, \bibinfo {author} {\bibfnamefont {R.}~\bibnamefont {Lopes}},
  \bibinfo {author} {\bibfnamefont {M.}~\bibnamefont {B{\"{o}}ttcher}},
  \bibinfo {author} {\bibfnamefont {M.~D.}\ \bibnamefont {Martins}}, \bibinfo
  {author} {\bibfnamefont {T.}~\bibnamefont {Balashov}}, \bibinfo {author}
  {\bibfnamefont {L.}~\bibnamefont {Gerhard}}, \bibinfo {author} {\bibfnamefont
  {J.}~\bibnamefont {Sinova}},\ and\ \bibinfo {author} {\bibfnamefont
  {W.}~\bibnamefont {Wulfhekel}},\ }\href
  {https://doi.org/10.1038/s41467-018-03240-w} {\bibfield  {journal} {\bibinfo
  {journal} {Nat. Commun.}\ }\textbf {\bibinfo {volume} {9}},\ \bibinfo {pages}
  {1015} (\bibinfo {year} {2018})}\BibitemShut {NoStop}%
\bibitem [{\citenamefont {Dzyaloshinskii}(1958)}]{Dzyaloshinskii1958}%
  \BibitemOpen
  \bibfield  {author} {\bibinfo {author} {\bibfnamefont {I.~E.}\ \bibnamefont
  {Dzyaloshinskii}},\ }\href {https://doi.org/10.1016/0022-3697(58)90076-3}
  {\bibfield  {journal} {\bibinfo  {journal} {J. Phys. Chem. Solids}\ }\textbf
  {\bibinfo {volume} {4}},\ \bibinfo {pages} {241} (\bibinfo {year}
  {1958})}\BibitemShut {NoStop}%
\bibitem [{\citenamefont {Moriya}(1960)}]{Moriya1960}%
  \BibitemOpen
  \bibfield  {author} {\bibinfo {author} {\bibfnamefont {T.}~\bibnamefont
  {Moriya}},\ }\href {https://doi.org/10.1103/PhysRev.120.91} {\bibfield
  {journal} {\bibinfo  {journal} {Phys. Rev.}\ }\textbf {\bibinfo {volume}
  {120}},\ \bibinfo {pages} {91} (\bibinfo {year} {1960})}\BibitemShut
  {NoStop}%
\bibitem [{\citenamefont {Bergman}(2006)}]{Bergman2006}%
  \BibitemOpen
  \bibfield  {author} {\bibinfo {author} {\bibfnamefont {A.}~\bibnamefont
  {Bergman}},\ }\emph {\bibinfo {title} {{A theoretical study of magnetism in
  nanostructured materials}}},\ \href@noop {} {Ph.D. thesis},\ \bibinfo
  {school} {Uppsala University} (\bibinfo {year} {2006})\BibitemShut {NoStop}%
\bibitem [{\citenamefont {Okubo}\ \emph {et~al.}(2012)\citenamefont {Okubo},
  \citenamefont {Chung},\ and\ \citenamefont {Kawamura}}]{Okubo2012}%
  \BibitemOpen
  \bibfield  {author} {\bibinfo {author} {\bibfnamefont {T.}~\bibnamefont
  {Okubo}}, \bibinfo {author} {\bibfnamefont {S.}~\bibnamefont {Chung}},\ and\
  \bibinfo {author} {\bibfnamefont {H.}~\bibnamefont {Kawamura}},\ }\href
  {https://doi.org/10.1103/PhysRevLett.108.017206} {\bibfield  {journal}
  {\bibinfo  {journal} {Phys. Rev. Lett.}\ }\textbf {\bibinfo {volume} {108}},\
  \bibinfo {pages} {017206} (\bibinfo {year} {2012})}\BibitemShut {NoStop}%
\bibitem [{\citenamefont {R{\'{o}}zsa}\ \emph
  {et~al.}(2016{\natexlab{a}})\citenamefont {R{\'{o}}zsa}, \citenamefont
  {De{\'{a}}k}, \citenamefont {Simon}, \citenamefont {Yanes}, \citenamefont
  {Udvardi}, \citenamefont {Szunyogh},\ and\ \citenamefont
  {Nowak}}]{Rozsa2016a}%
  \BibitemOpen
  \bibfield  {author} {\bibinfo {author} {\bibfnamefont {L.}~\bibnamefont
  {R{\'{o}}zsa}}, \bibinfo {author} {\bibfnamefont {A.}~\bibnamefont
  {De{\'{a}}k}}, \bibinfo {author} {\bibfnamefont {E.}~\bibnamefont {Simon}},
  \bibinfo {author} {\bibfnamefont {R.}~\bibnamefont {Yanes}}, \bibinfo
  {author} {\bibfnamefont {L.}~\bibnamefont {Udvardi}}, \bibinfo {author}
  {\bibfnamefont {L.}~\bibnamefont {Szunyogh}},\ and\ \bibinfo {author}
  {\bibfnamefont {U.}~\bibnamefont {Nowak}},\ }\href
  {https://doi.org/10.1103/PhysRevLett.117.157205} {\bibfield  {journal}
  {\bibinfo  {journal} {Phys. Rev. Lett.}\ }\textbf {\bibinfo {volume} {117}},\
  \bibinfo {pages} {157205} (\bibinfo {year} {2016}{\natexlab{a}})}\BibitemShut
  {NoStop}%
\bibitem [{\citenamefont {Jadaun}\ \emph {et~al.}(2020)\citenamefont {Jadaun},
  \citenamefont {Register},\ and\ \citenamefont {Banerjee}}]{Jadaun2020}%
  \BibitemOpen
  \bibfield  {author} {\bibinfo {author} {\bibfnamefont {P.}~\bibnamefont
  {Jadaun}}, \bibinfo {author} {\bibfnamefont {L.~F.}\ \bibnamefont
  {Register}},\ and\ \bibinfo {author} {\bibfnamefont {S.~K.}\ \bibnamefont
  {Banerjee}},\ }\href {https://doi.org/10.1038/s41524-020-00351-1} {\bibfield
  {journal} {\bibinfo  {journal} {npj Comput. Mater.}\ }\textbf {\bibinfo
  {volume} {6}},\ \bibinfo {pages} {1} (\bibinfo {year} {2020})}\BibitemShut
  {NoStop}%
\bibitem [{\citenamefont {Romming}\ \emph {et~al.}(2015)\citenamefont
  {Romming}, \citenamefont {Kubetzka}, \citenamefont {Hanneken}, \citenamefont
  {von Bergmann},\ and\ \citenamefont {Wiesendanger}}]{Romming2015}%
  \BibitemOpen
  \bibfield  {author} {\bibinfo {author} {\bibfnamefont {N.}~\bibnamefont
  {Romming}}, \bibinfo {author} {\bibfnamefont {A.}~\bibnamefont {Kubetzka}},
  \bibinfo {author} {\bibfnamefont {C.}~\bibnamefont {Hanneken}}, \bibinfo
  {author} {\bibfnamefont {K.}~\bibnamefont {von Bergmann}},\ and\ \bibinfo
  {author} {\bibfnamefont {R.}~\bibnamefont {Wiesendanger}},\ }\href
  {https://doi.org/10.1103/PhysRevLett.114.177203} {\bibfield  {journal}
  {\bibinfo  {journal} {Phys. Rev. Lett.}\ }\textbf {\bibinfo {volume} {114}},\
  \bibinfo {pages} {177203} (\bibinfo {year} {2015})}\BibitemShut {NoStop}%
\bibitem [{\citenamefont {Spethmann}\ \emph {et~al.}(2022)\citenamefont
  {Spethmann}, \citenamefont {Vedmedenko}, \citenamefont {Wiesendanger},
  \citenamefont {Kubetzka},\ and\ \citenamefont {von
  Bergmann}}]{Spethmann2022}%
  \BibitemOpen
  \bibfield  {author} {\bibinfo {author} {\bibfnamefont {J.}~\bibnamefont
  {Spethmann}}, \bibinfo {author} {\bibfnamefont {E.~Y.}\ \bibnamefont
  {Vedmedenko}}, \bibinfo {author} {\bibfnamefont {R.}~\bibnamefont
  {Wiesendanger}}, \bibinfo {author} {\bibfnamefont {A.}~\bibnamefont
  {Kubetzka}},\ and\ \bibinfo {author} {\bibfnamefont {K.}~\bibnamefont {von
  Bergmann}},\ }\href {https://doi.org/10.1038/s42005-021-00796-w} {\bibfield
  {journal} {\bibinfo  {journal} {Commun. Phys.}\ }\textbf {\bibinfo {volume}
  {5}},\ \bibinfo {pages} {1} (\bibinfo {year} {2022})}\BibitemShut {NoStop}%
\bibitem [{\citenamefont {R{\'{o}}zsa}\ \emph
  {et~al.}(2016{\natexlab{b}})\citenamefont {R{\'{o}}zsa}, \citenamefont
  {Simon}, \citenamefont {Palot{\'{a}}s}, \citenamefont {Udvardi},\ and\
  \citenamefont {Szunyogh}}]{Rozsa2016}%
  \BibitemOpen
  \bibfield  {author} {\bibinfo {author} {\bibfnamefont {L.}~\bibnamefont
  {R{\'{o}}zsa}}, \bibinfo {author} {\bibfnamefont {E.}~\bibnamefont {Simon}},
  \bibinfo {author} {\bibfnamefont {K.}~\bibnamefont {Palot{\'{a}}s}}, \bibinfo
  {author} {\bibfnamefont {L.}~\bibnamefont {Udvardi}},\ and\ \bibinfo {author}
  {\bibfnamefont {L.}~\bibnamefont {Szunyogh}},\ }\href
  {https://doi.org/10.1103/PhysRevB.93.024417} {\bibfield  {journal} {\bibinfo
  {journal} {Phys. Rev. B}\ }\textbf {\bibinfo {volume} {93}},\ \bibinfo
  {pages} {024417} (\bibinfo {year} {2016}{\natexlab{b}})}\BibitemShut
  {NoStop}%
\bibitem [{\citenamefont {Simon}\ \emph {et~al.}(2014)\citenamefont {Simon},
  \citenamefont {Palot{\'{a}}s}, \citenamefont {R{\'{o}}zsa}, \citenamefont
  {Udvardi},\ and\ \citenamefont {Szunyogh}}]{Simon2014}%
  \BibitemOpen
  \bibfield  {author} {\bibinfo {author} {\bibfnamefont {E.}~\bibnamefont
  {Simon}}, \bibinfo {author} {\bibfnamefont {K.}~\bibnamefont
  {Palot{\'{a}}s}}, \bibinfo {author} {\bibfnamefont {L.}~\bibnamefont
  {R{\'{o}}zsa}}, \bibinfo {author} {\bibfnamefont {L.}~\bibnamefont
  {Udvardi}},\ and\ \bibinfo {author} {\bibfnamefont {L.}~\bibnamefont
  {Szunyogh}},\ }\href {https://doi.org/10.1103/PhysRevB.90.094410} {\bibfield
  {journal} {\bibinfo  {journal} {Phys. Rev. B}\ }\textbf {\bibinfo {volume}
  {90}},\ \bibinfo {pages} {094410} (\bibinfo {year} {2014})}\BibitemShut
  {NoStop}%
\bibitem [{\citenamefont {Hanneken}\ \emph {et~al.}(2015)\citenamefont
  {Hanneken}, \citenamefont {Otte}, \citenamefont {Kubetzka}, \citenamefont
  {Dup{\'{e}}}, \citenamefont {Romming}, \citenamefont {von Bergmann},
  \citenamefont {Wiesendanger},\ and\ \citenamefont {Heinze}}]{Hanneken2015}%
  \BibitemOpen
  \bibfield  {author} {\bibinfo {author} {\bibfnamefont {C.}~\bibnamefont
  {Hanneken}}, \bibinfo {author} {\bibfnamefont {F.}~\bibnamefont {Otte}},
  \bibinfo {author} {\bibfnamefont {A.}~\bibnamefont {Kubetzka}}, \bibinfo
  {author} {\bibfnamefont {B.}~\bibnamefont {Dup{\'{e}}}}, \bibinfo {author}
  {\bibfnamefont {N.}~\bibnamefont {Romming}}, \bibinfo {author} {\bibfnamefont
  {K.}~\bibnamefont {von Bergmann}}, \bibinfo {author} {\bibfnamefont
  {R.}~\bibnamefont {Wiesendanger}},\ and\ \bibinfo {author} {\bibfnamefont
  {S.}~\bibnamefont {Heinze}},\ }\href {https://doi.org/10.1038/nnano.2015.218}
  {\bibfield  {journal} {\bibinfo  {journal} {Nat. Nanotechnol}\ }\textbf
  {\bibinfo {volume} {10}},\ \bibinfo {pages} {1039} (\bibinfo {year}
  {2015})}\BibitemShut {NoStop}%
\bibitem [{\citenamefont {B{\"o}ttcher}\ \emph {et~al.}(2018)\citenamefont
  {B{\"o}ttcher}, \citenamefont {Heinze}, \citenamefont {Egorov}, \citenamefont
  {Sinova},\ and\ \citenamefont {Dup{\'e}}}]{Bottcher2018}%
  \BibitemOpen
  \bibfield  {author} {\bibinfo {author} {\bibfnamefont {M.}~\bibnamefont
  {B{\"o}ttcher}}, \bibinfo {author} {\bibfnamefont {S.}~\bibnamefont
  {Heinze}}, \bibinfo {author} {\bibfnamefont {S.}~\bibnamefont {Egorov}},
  \bibinfo {author} {\bibfnamefont {J.}~\bibnamefont {Sinova}},\ and\ \bibinfo
  {author} {\bibfnamefont {B.}~\bibnamefont {Dup{\'e}}},\ }\href
  {https://doi.org/10.1088/1367-2630/aae282} {\bibfield  {journal} {\bibinfo
  {journal} {New J. Phys.}\ }\textbf {\bibinfo {volume} {20}},\ \bibinfo
  {pages} {103014} (\bibinfo {year} {2018})}\BibitemShut {NoStop}%
\bibitem [{\citenamefont {Dup{\'{e}}}\ \emph {et~al.}(2014)\citenamefont
  {Dup{\'{e}}}, \citenamefont {Hoffmann}, \citenamefont {Paillard},\ and\
  \citenamefont {Heinze}}]{Dupe2014}%
  \BibitemOpen
  \bibfield  {author} {\bibinfo {author} {\bibfnamefont {B.}~\bibnamefont
  {Dup{\'{e}}}}, \bibinfo {author} {\bibfnamefont {M.}~\bibnamefont
  {Hoffmann}}, \bibinfo {author} {\bibfnamefont {C.}~\bibnamefont {Paillard}},\
  and\ \bibinfo {author} {\bibfnamefont {S.}~\bibnamefont {Heinze}},\ }\href
  {https://doi.org/10.1038/ncomms5030} {\bibfield  {journal} {\bibinfo
  {journal} {Nat. Commun.}\ }\textbf {\bibinfo {volume} {5}},\ \bibinfo {pages}
  {4030} (\bibinfo {year} {2014})}\BibitemShut {NoStop}%
\bibitem [{\citenamefont {Dzemiantsova}\ \emph {et~al.}(2012)\citenamefont
  {Dzemiantsova}, \citenamefont {Hortamani}, \citenamefont {Hanneken},
  \citenamefont {Kubetzka}, \citenamefont {von Bergmann},\ and\ \citenamefont
  {Wiesendanger}}]{Dzemiantsova2012}%
  \BibitemOpen
  \bibfield  {author} {\bibinfo {author} {\bibfnamefont {L.~V.}\ \bibnamefont
  {Dzemiantsova}}, \bibinfo {author} {\bibfnamefont {M.}~\bibnamefont
  {Hortamani}}, \bibinfo {author} {\bibfnamefont {C.}~\bibnamefont {Hanneken}},
  \bibinfo {author} {\bibfnamefont {A.}~\bibnamefont {Kubetzka}}, \bibinfo
  {author} {\bibfnamefont {K.}~\bibnamefont {von Bergmann}},\ and\ \bibinfo
  {author} {\bibfnamefont {R.}~\bibnamefont {Wiesendanger}},\ }\href
  {https://doi.org/10.1103/PhysRevB.86.094427} {\bibfield  {journal} {\bibinfo
  {journal} {Phys. Rev. B}\ }\textbf {\bibinfo {volume} {86}},\ \bibinfo
  {pages} {094427} (\bibinfo {year} {2012})}\BibitemShut {NoStop}%
\bibitem [{\citenamefont {Back}\ \emph {et~al.}(2020)\citenamefont {Back},
  \citenamefont {Cros}, \citenamefont {Ebert}, \citenamefont {Everschor-Sitte},
  \citenamefont {Fert}, \citenamefont {Garst}, \citenamefont {Ma},
  \citenamefont {Mankovsky}, \citenamefont {Monchesky}, \citenamefont
  {Mostovoy} \emph {et~al.}}]{Back2020}%
  \BibitemOpen
  \bibfield  {author} {\bibinfo {author} {\bibfnamefont {C.}~\bibnamefont
  {Back}}, \bibinfo {author} {\bibfnamefont {V.}~\bibnamefont {Cros}}, \bibinfo
  {author} {\bibfnamefont {H.}~\bibnamefont {Ebert}}, \bibinfo {author}
  {\bibfnamefont {K.}~\bibnamefont {Everschor-Sitte}}, \bibinfo {author}
  {\bibfnamefont {A.}~\bibnamefont {Fert}}, \bibinfo {author} {\bibfnamefont
  {M.}~\bibnamefont {Garst}}, \bibinfo {author} {\bibfnamefont
  {T.}~\bibnamefont {Ma}}, \bibinfo {author} {\bibfnamefont {S.}~\bibnamefont
  {Mankovsky}}, \bibinfo {author} {\bibfnamefont {T.}~\bibnamefont
  {Monchesky}}, \bibinfo {author} {\bibfnamefont {M.}~\bibnamefont {Mostovoy}},
  \emph {et~al.},\ }\href {https://doi.org/10.1088/1361-6463/ab8418} {\bibfield
   {journal} {\bibinfo  {journal} {J. Phys. D: Appl. Phys.}\ }\textbf {\bibinfo
  {volume} {53}},\ \bibinfo {pages} {363001} (\bibinfo {year}
  {2020})}\BibitemShut {NoStop}%
\bibitem [{\citenamefont {Mankovsky}\ \emph {et~al.}(2021)\citenamefont
  {Mankovsky}, \citenamefont {Simon}, \citenamefont {Polesya}, \citenamefont
  {Marmodoro},\ and\ \citenamefont {Ebert}}]{Mankovsky2021}%
  \BibitemOpen
  \bibfield  {author} {\bibinfo {author} {\bibfnamefont {S.}~\bibnamefont
  {Mankovsky}}, \bibinfo {author} {\bibfnamefont {E.}~\bibnamefont {Simon}},
  \bibinfo {author} {\bibfnamefont {S.}~\bibnamefont {Polesya}}, \bibinfo
  {author} {\bibfnamefont {A.}~\bibnamefont {Marmodoro}},\ and\ \bibinfo
  {author} {\bibfnamefont {H.}~\bibnamefont {Ebert}},\ }\href
  {https://doi.org/10.1103/PhysRevB.104.174443} {\bibfield  {journal} {\bibinfo
   {journal} {Phys. Rev. B}\ }\textbf {\bibinfo {volume} {104}},\ \bibinfo
  {pages} {174443} (\bibinfo {year} {2021})}\BibitemShut {NoStop}%
\bibitem [{\citenamefont {Yang}\ \emph {et~al.}(2018)\citenamefont {Yang},
  \citenamefont {Boulle}, \citenamefont {Cros}, \citenamefont {Fert},\ and\
  \citenamefont {Chshiev}}]{Yang2018}%
  \BibitemOpen
  \bibfield  {author} {\bibinfo {author} {\bibfnamefont {H.}~\bibnamefont
  {Yang}}, \bibinfo {author} {\bibfnamefont {O.}~\bibnamefont {Boulle}},
  \bibinfo {author} {\bibfnamefont {V.}~\bibnamefont {Cros}}, \bibinfo {author}
  {\bibfnamefont {A.}~\bibnamefont {Fert}},\ and\ \bibinfo {author}
  {\bibfnamefont {M.}~\bibnamefont {Chshiev}},\ }\href
  {https://doi.org/10.1038/s41598-018-30063-y} {\bibfield  {journal} {\bibinfo
  {journal} {Sci. Rep.}\ }\textbf {\bibinfo {volume} {8}},\ \bibinfo {pages}
  {1} (\bibinfo {year} {2018})}\BibitemShut {NoStop}%
\bibitem [{\citenamefont {Gusev}\ \emph {et~al.}(2020)\citenamefont {Gusev},
  \citenamefont {Sadovnikov}, \citenamefont {Nikitov}, \citenamefont
  {Sapozhnikov},\ and\ \citenamefont {Udalov}}]{Gusev2020}%
  \BibitemOpen
  \bibfield  {author} {\bibinfo {author} {\bibfnamefont {N.~S.}\ \bibnamefont
  {Gusev}}, \bibinfo {author} {\bibfnamefont {A.~V.}\ \bibnamefont
  {Sadovnikov}}, \bibinfo {author} {\bibfnamefont {S.~A.}\ \bibnamefont
  {Nikitov}}, \bibinfo {author} {\bibfnamefont {M.~V.}\ \bibnamefont
  {Sapozhnikov}},\ and\ \bibinfo {author} {\bibfnamefont {O.~G.}\ \bibnamefont
  {Udalov}},\ }\href {https://doi.org/10.1103/PhysRevLett.124.157202}
  {\bibfield  {journal} {\bibinfo  {journal} {Phys. Rev. Lett.}\ }\textbf
  {\bibinfo {volume} {124}},\ \bibinfo {pages} {157202} (\bibinfo {year}
  {2020})}\BibitemShut {NoStop}%
\bibitem [{\citenamefont {Desplat}\ \emph {et~al.}(2021)\citenamefont
  {Desplat}, \citenamefont {Meyer}, \citenamefont {Bouaziz}, \citenamefont
  {Buhl}, \citenamefont {Lounis}, \citenamefont {Dup\'e},\ and\ \citenamefont
  {Hervieux}}]{Desplat2021}%
  \BibitemOpen
  \bibfield  {author} {\bibinfo {author} {\bibfnamefont {L.}~\bibnamefont
  {Desplat}}, \bibinfo {author} {\bibfnamefont {S.}~\bibnamefont {Meyer}},
  \bibinfo {author} {\bibfnamefont {J.}~\bibnamefont {Bouaziz}}, \bibinfo
  {author} {\bibfnamefont {P.~M.}\ \bibnamefont {Buhl}}, \bibinfo {author}
  {\bibfnamefont {S.}~\bibnamefont {Lounis}}, \bibinfo {author} {\bibfnamefont
  {B.}~\bibnamefont {Dup\'e}},\ and\ \bibinfo {author} {\bibfnamefont {P.-A.}\
  \bibnamefont {Hervieux}},\ }\href
  {https://doi.org/10.1103/PhysRevB.104.L060409} {\bibfield  {journal}
  {\bibinfo  {journal} {Phys. Rev. B}\ }\textbf {\bibinfo {volume} {104}},\
  \bibinfo {pages} {L060409} (\bibinfo {year} {2021})}\BibitemShut {NoStop}%
\bibitem [{\citenamefont {Srivastava}\ \emph {et~al.}(2018)\citenamefont
  {Srivastava}, \citenamefont {Schott}, \citenamefont {Juge}, \citenamefont
  {Krizakova}, \citenamefont {Belmeguenai}, \citenamefont {Roussign{\'e}},
  \citenamefont {Bernand-Mantel}, \citenamefont {Ranno}, \citenamefont
  {Pizzini}, \citenamefont {Ch{\'e}rif} \emph {et~al.}}]{Srivastava2018}%
  \BibitemOpen
  \bibfield  {author} {\bibinfo {author} {\bibfnamefont {T.}~\bibnamefont
  {Srivastava}}, \bibinfo {author} {\bibfnamefont {M.}~\bibnamefont {Schott}},
  \bibinfo {author} {\bibfnamefont {R.}~\bibnamefont {Juge}}, \bibinfo {author}
  {\bibfnamefont {V.}~\bibnamefont {Krizakova}}, \bibinfo {author}
  {\bibfnamefont {M.}~\bibnamefont {Belmeguenai}}, \bibinfo {author}
  {\bibfnamefont {Y.}~\bibnamefont {Roussign{\'e}}}, \bibinfo {author}
  {\bibfnamefont {A.}~\bibnamefont {Bernand-Mantel}}, \bibinfo {author}
  {\bibfnamefont {L.}~\bibnamefont {Ranno}}, \bibinfo {author} {\bibfnamefont
  {S.}~\bibnamefont {Pizzini}}, \bibinfo {author} {\bibfnamefont {S.-M.}\
  \bibnamefont {Ch{\'e}rif}}, \emph {et~al.},\ }\href
  {https://doi.org/10.1021/acs.nanolett.8b01502} {\bibfield  {journal}
  {\bibinfo  {journal} {Nano Lett.}\ }\textbf {\bibinfo {volume} {18}},\
  \bibinfo {pages} {4871} (\bibinfo {year} {2018})}\BibitemShut {NoStop}%
\bibitem [{\citenamefont {Paul}\ and\ \citenamefont {Heinze}(2021)}]{Paul2021}%
  \BibitemOpen
  \bibfield  {author} {\bibinfo {author} {\bibfnamefont {S.}~\bibnamefont
  {Paul}}\ and\ \bibinfo {author} {\bibfnamefont {S.}~\bibnamefont {Heinze}},\
  }\href {https://arxiv.org/pdf/2104.11986.pdf} {\bibfield  {journal} {\bibinfo
   {journal} {arXiv:2104.11986}\ } (\bibinfo {year} {2021})}\BibitemShut
  {NoStop}%
\bibitem [{\citenamefont {Peduto}\ \emph {et~al.}(1991)\citenamefont {Peduto},
  \citenamefont {Frota-Pess{\^{o}}a},\ and\ \citenamefont
  {Methfessel}}]{Peduto1991}%
  \BibitemOpen
  \bibfield  {author} {\bibinfo {author} {\bibfnamefont {P.~R.}\ \bibnamefont
  {Peduto}}, \bibinfo {author} {\bibfnamefont {S.}~\bibnamefont
  {Frota-Pess{\^{o}}a}},\ and\ \bibinfo {author} {\bibfnamefont {M.~S.}\
  \bibnamefont {Methfessel}},\ }\href
  {https://doi.org/10.1103/PhysRevB.44.13283} {\bibfield  {journal} {\bibinfo
  {journal} {Phys. Rev. B}\ }\textbf {\bibinfo {volume} {44}},\ \bibinfo
  {pages} {13283} (\bibinfo {year} {1991})}\BibitemShut {NoStop}%
\bibitem [{\citenamefont {Frota-Pess{\^{o}}a}(1992)}]{Frota-Pessoa1992}%
  \BibitemOpen
  \bibfield  {author} {\bibinfo {author} {\bibfnamefont {S.}~\bibnamefont
  {Frota-Pess{\^{o}}a}},\ }\href {https://doi.org/10.1103/PhysRevB.46.14570}
  {\bibfield  {journal} {\bibinfo  {journal} {Phys. Rev. B}\ }\textbf {\bibinfo
  {volume} {46}},\ \bibinfo {pages} {14570} (\bibinfo {year}
  {1992})}\BibitemShut {NoStop}%
\bibitem [{\citenamefont {Bergman}\ \emph
  {et~al.}(2006{\natexlab{a}})\citenamefont {Bergman}, \citenamefont
  {Nordstr\"om}, \citenamefont {Klautau}, \citenamefont {Frota-Pess\^oa},\ and\
  \citenamefont {Eriksson}}]{Bergman2006a}%
  \BibitemOpen
  \bibfield  {author} {\bibinfo {author} {\bibfnamefont {A.}~\bibnamefont
  {Bergman}}, \bibinfo {author} {\bibfnamefont {L.}~\bibnamefont
  {Nordstr\"om}}, \bibinfo {author} {\bibfnamefont {A.~B.}\ \bibnamefont
  {Klautau}}, \bibinfo {author} {\bibfnamefont {S.}~\bibnamefont
  {Frota-Pess\^oa}},\ and\ \bibinfo {author} {\bibfnamefont {O.}~\bibnamefont
  {Eriksson}},\ }\href {https://doi.org/10.1103/PhysRevB.73.174434} {\bibfield
  {journal} {\bibinfo  {journal} {Phys. Rev. B}\ }\textbf {\bibinfo {volume}
  {73}},\ \bibinfo {pages} {174434} (\bibinfo {year}
  {2006}{\natexlab{a}})}\BibitemShut {NoStop}%
\bibitem [{\citenamefont {Bergman}\ \emph {et~al.}(2007)\citenamefont
  {Bergman}, \citenamefont {Nordstr{\"{o}}m}, \citenamefont {{Burlamaqui
  Klautau}}, \citenamefont {Frota-Pess{\^{o}}a},\ and\ \citenamefont
  {Eriksson}}]{Bergman2007}%
  \BibitemOpen
  \bibfield  {author} {\bibinfo {author} {\bibfnamefont {A.}~\bibnamefont
  {Bergman}}, \bibinfo {author} {\bibfnamefont {L.}~\bibnamefont
  {Nordstr{\"{o}}m}}, \bibinfo {author} {\bibfnamefont {A.}~\bibnamefont
  {{Burlamaqui Klautau}}}, \bibinfo {author} {\bibfnamefont {S.}~\bibnamefont
  {Frota-Pess{\^{o}}a}},\ and\ \bibinfo {author} {\bibfnamefont
  {O.}~\bibnamefont {Eriksson}},\ }\href
  {https://doi.org/10.1103/PhysRevB.75.224425} {\bibfield  {journal} {\bibinfo
  {journal} {Phys. Rev. B}\ }\textbf {\bibinfo {volume} {75}},\ \bibinfo
  {pages} {224425} (\bibinfo {year} {2007})}\BibitemShut {NoStop}%
\bibitem [{\citenamefont {Bergman}\ \emph
  {et~al.}(2006{\natexlab{b}})\citenamefont {Bergman}, \citenamefont
  {Nordström}, \citenamefont {Klautau}, \citenamefont {Frota-Pessôa},\ and\
  \citenamefont {Eriksson}}]{BERGMAN20064838}%
  \BibitemOpen
  \bibfield  {author} {\bibinfo {author} {\bibfnamefont {A.}~\bibnamefont
  {Bergman}}, \bibinfo {author} {\bibfnamefont {L.}~\bibnamefont {Nordström}},
  \bibinfo {author} {\bibfnamefont {A.~B.}\ \bibnamefont {Klautau}}, \bibinfo
  {author} {\bibfnamefont {S.}~\bibnamefont {Frota-Pessôa}},\ and\ \bibinfo
  {author} {\bibfnamefont {O.}~\bibnamefont {Eriksson}},\ }\href
  {https://doi.org/https://doi.org/10.1016/j.susc.2006.08.004} {\bibfield
  {journal} {\bibinfo  {journal} {Surf. Sci.}\ }\textbf {\bibinfo {volume}
  {600}},\ \bibinfo {pages} {4838} (\bibinfo {year}
  {2006}{\natexlab{b}})}\BibitemShut {NoStop}%
\bibitem [{\citenamefont {Cardias}\ \emph
  {et~al.}(2016{\natexlab{b}})\citenamefont {Cardias}, \citenamefont
  {Bezerra-Neto}, \citenamefont {Ribeiro}, \citenamefont {Bergman},
  \citenamefont {Szilva}, \citenamefont {Eriksson},\ and\ \citenamefont
  {Klautau}}]{PhysRevB.93.014438}%
  \BibitemOpen
  \bibfield  {author} {\bibinfo {author} {\bibfnamefont {R.}~\bibnamefont
  {Cardias}}, \bibinfo {author} {\bibfnamefont {M.~M.}\ \bibnamefont
  {Bezerra-Neto}}, \bibinfo {author} {\bibfnamefont {M.~S.}\ \bibnamefont
  {Ribeiro}}, \bibinfo {author} {\bibfnamefont {A.}~\bibnamefont {Bergman}},
  \bibinfo {author} {\bibfnamefont {A.}~\bibnamefont {Szilva}}, \bibinfo
  {author} {\bibfnamefont {O.}~\bibnamefont {Eriksson}},\ and\ \bibinfo
  {author} {\bibfnamefont {A.~B.}\ \bibnamefont {Klautau}},\ }\href
  {https://doi.org/10.1103/PhysRevB.93.014438} {\bibfield  {journal} {\bibinfo
  {journal} {Phys. Rev. B}\ }\textbf {\bibinfo {volume} {93}},\ \bibinfo
  {pages} {014438} (\bibinfo {year} {2016}{\natexlab{b}})}\BibitemShut
  {NoStop}%
\bibitem [{\citenamefont {Ribeiro}\ \emph
  {et~al.}(2011{\natexlab{b}})\citenamefont {Ribeiro}, \citenamefont
  {Corr\^ea}, \citenamefont {Bergman}, \citenamefont {Nordstr\"om},
  \citenamefont {Eriksson},\ and\ \citenamefont
  {Klautau}}]{PhysRevB.83.014406}%
  \BibitemOpen
  \bibfield  {author} {\bibinfo {author} {\bibfnamefont {M.~S.}\ \bibnamefont
  {Ribeiro}}, \bibinfo {author} {\bibfnamefont {G.~B.}\ \bibnamefont
  {Corr\^ea}}, \bibinfo {author} {\bibfnamefont {A.}~\bibnamefont {Bergman}},
  \bibinfo {author} {\bibfnamefont {L.}~\bibnamefont {Nordstr\"om}}, \bibinfo
  {author} {\bibfnamefont {O.}~\bibnamefont {Eriksson}},\ and\ \bibinfo
  {author} {\bibfnamefont {A.~B.}\ \bibnamefont {Klautau}},\ }\href
  {https://doi.org/10.1103/PhysRevB.83.014406} {\bibfield  {journal} {\bibinfo
  {journal} {Phys. Rev. B}\ }\textbf {\bibinfo {volume} {83}},\ \bibinfo
  {pages} {014406} (\bibinfo {year} {2011}{\natexlab{b}})}\BibitemShut
  {NoStop}%
\bibitem [{\citenamefont {Andersen}(1975)}]{Andersen1975}%
  \BibitemOpen
  \bibfield  {author} {\bibinfo {author} {\bibfnamefont {O.~K.}\ \bibnamefont
  {Andersen}},\ }\href {https://doi.org/10.1103/PhysRevB.12.3060} {\bibfield
  {journal} {\bibinfo  {journal} {Phys. Rev. B}\ }\textbf {\bibinfo {volume}
  {12}},\ \bibinfo {pages} {3060} (\bibinfo {year} {1975})}\BibitemShut
  {NoStop}%
\bibitem [{\citenamefont {Haydock}(1980)}]{Haydock1980}%
  \BibitemOpen
  \bibfield  {author} {\bibinfo {author} {\bibfnamefont {R.}~\bibnamefont
  {Haydock}},\ }\href@noop {} {\emph {\bibinfo {title} {{Solid State
  Physics}}}},\ edited by\ \bibinfo {editor} {\bibfnamefont {H.}~\bibnamefont
  {Ehrenreich}}, \bibinfo {editor} {\bibfnamefont {F.}~\bibnamefont {Seitz}},\
  and\ \bibinfo {editor} {\bibfnamefont {D.}~\bibnamefont {Turnbull}}\
  (\bibinfo  {publisher} {Academic Press},\ \bibinfo {address} {New York},\
  \bibinfo {year} {1980})\BibitemShut {NoStop}%
\bibitem [{\citenamefont {von Barth}\ \emph {et~al.}(1972)\citenamefont {von
  Barth}, \citenamefont {Hedin}, \citenamefont {von Barth},\ and\ \citenamefont
  {Hedin}}]{Barth1972}%
  \BibitemOpen
  \bibfield  {author} {\bibinfo {author} {\bibfnamefont {U.}~\bibnamefont {von
  Barth}}, \bibinfo {author} {\bibfnamefont {L.}~\bibnamefont {Hedin}},
  \bibinfo {author} {\bibfnamefont {U.}~\bibnamefont {von Barth}},\ and\
  \bibinfo {author} {\bibfnamefont {L.}~\bibnamefont {Hedin}},\ }\href
  {https://doi.org/10.1088/0022-3719/5/13/012} {\bibfield  {journal} {\bibinfo
  {journal} {J. Phys. C: Solid State Phys.}\ }\textbf {\bibinfo {volume} {5}},\
  \bibinfo {pages} {1629} (\bibinfo {year} {1972})}\BibitemShut {NoStop}%
\bibitem [{\citenamefont {Beer}\ and\ \citenamefont
  {Pettifor}(1984)}]{Beer1984}%
  \BibitemOpen
  \bibfield  {author} {\bibinfo {author} {\bibfnamefont {N.}~\bibnamefont
  {Beer}}\ and\ \bibinfo {author} {\bibfnamefont {D.~G.}\ \bibnamefont
  {Pettifor}},\ }\href@noop {} {\emph {\bibinfo {title} {{The Electronic
  Structure of Complex Systems}}}},\ edited by\ \bibinfo {editor}
  {\bibfnamefont {W.}~\bibnamefont {Temmermann}}\ and\ \bibinfo {editor}
  {\bibfnamefont {P.}~\bibnamefont {Phariseau}}\ (\bibinfo  {publisher} {Plenum
  Press},\ \bibinfo {address} {New York},\ \bibinfo {year} {1984})\BibitemShut
  {NoStop}%
\bibitem [{\citenamefont {Liechtenstein}\ \emph {et~al.}(1987)\citenamefont
  {Liechtenstein}, \citenamefont {Katsnelson}, \citenamefont {Antropov},\ and\
  \citenamefont {Gubanov}}]{Liechtenstein1987}%
  \BibitemOpen
  \bibfield  {author} {\bibinfo {author} {\bibfnamefont {A.}~\bibnamefont
  {Liechtenstein}}, \bibinfo {author} {\bibfnamefont {M.}~\bibnamefont
  {Katsnelson}}, \bibinfo {author} {\bibfnamefont {V.}~\bibnamefont
  {Antropov}},\ and\ \bibinfo {author} {\bibfnamefont {V.}~\bibnamefont
  {Gubanov}},\ }\href {https://doi.org/10.1016/0304-8853(87)90721-9} {\bibfield
   {journal} {\bibinfo  {journal} {J. Magn. Magn. Mater.}\ }\textbf {\bibinfo
  {volume} {67}},\ \bibinfo {pages} {65} (\bibinfo {year} {1987})}\BibitemShut
  {NoStop}%
\bibitem [{\citenamefont {Frota-Pess{\^{o}}a}\ \emph
  {et~al.}(2000)\citenamefont {Frota-Pess{\^{o}}a}, \citenamefont {Muniz},\
  and\ \citenamefont {Kudrnovsk{\'{y}}}}]{Frota-Pessoa2000}%
  \BibitemOpen
  \bibfield  {author} {\bibinfo {author} {\bibfnamefont {S.}~\bibnamefont
  {Frota-Pess{\^{o}}a}}, \bibinfo {author} {\bibfnamefont {R.~B.}\ \bibnamefont
  {Muniz}},\ and\ \bibinfo {author} {\bibfnamefont {J.}~\bibnamefont
  {Kudrnovsk{\'{y}}}},\ }\href {https://doi.org/10.1103/PhysRevB.62.5293}
  {\bibfield  {journal} {\bibinfo  {journal} {Phys. Rev. B}\ }\textbf {\bibinfo
  {volume} {62}},\ \bibinfo {pages} {5293} (\bibinfo {year}
  {2000})}\BibitemShut {NoStop}%
\bibitem [{\citenamefont {Szilva}\ \emph {et~al.}(2017)\citenamefont {Szilva},
  \citenamefont {Thonig}, \citenamefont {Bessarab}, \citenamefont {Kvashnin},
  \citenamefont {Rodrigues}, \citenamefont {Cardias}, \citenamefont {Pereiro},
  \citenamefont {Nordstr\"om}, \citenamefont {Bergman}, \citenamefont
  {Klautau},\ and\ \citenamefont {Eriksson}}]{PhysRevB.96.144413}%
  \BibitemOpen
  \bibfield  {author} {\bibinfo {author} {\bibfnamefont {A.}~\bibnamefont
  {Szilva}}, \bibinfo {author} {\bibfnamefont {D.}~\bibnamefont {Thonig}},
  \bibinfo {author} {\bibfnamefont {P.~F.}\ \bibnamefont {Bessarab}}, \bibinfo
  {author} {\bibfnamefont {Y.~O.}\ \bibnamefont {Kvashnin}}, \bibinfo {author}
  {\bibfnamefont {D.~C.~M.}\ \bibnamefont {Rodrigues}}, \bibinfo {author}
  {\bibfnamefont {R.}~\bibnamefont {Cardias}}, \bibinfo {author} {\bibfnamefont
  {M.}~\bibnamefont {Pereiro}}, \bibinfo {author} {\bibfnamefont
  {L.}~\bibnamefont {Nordstr\"om}}, \bibinfo {author} {\bibfnamefont
  {A.}~\bibnamefont {Bergman}}, \bibinfo {author} {\bibfnamefont {A.~B.}\
  \bibnamefont {Klautau}},\ and\ \bibinfo {author} {\bibfnamefont
  {O.}~\bibnamefont {Eriksson}},\ }\href
  {https://doi.org/10.1103/PhysRevB.96.144413} {\bibfield  {journal} {\bibinfo
  {journal} {Phys. Rev. B}\ }\textbf {\bibinfo {volume} {96}},\ \bibinfo
  {pages} {144413} (\bibinfo {year} {2017})}\BibitemShut {NoStop}%
\bibitem [{\citenamefont {Cardias}\ \emph {et~al.}(2020)\citenamefont
  {Cardias}, \citenamefont {Szilva}, \citenamefont {Bezerra-Neto},
  \citenamefont {Ribeiro}, \citenamefont {Bergman}, \citenamefont {Kvashnin},
  \citenamefont {Fransson}, \citenamefont {Klautau}, \citenamefont {Eriksson},\
  and\ \citenamefont {Nordstr{\"o}m}}]{Cardias2020}%
  \BibitemOpen
  \bibfield  {author} {\bibinfo {author} {\bibfnamefont {R.}~\bibnamefont
  {Cardias}}, \bibinfo {author} {\bibfnamefont {A.}~\bibnamefont {Szilva}},
  \bibinfo {author} {\bibfnamefont {M.}~\bibnamefont {Bezerra-Neto}}, \bibinfo
  {author} {\bibfnamefont {M.}~\bibnamefont {Ribeiro}}, \bibinfo {author}
  {\bibfnamefont {A.}~\bibnamefont {Bergman}}, \bibinfo {author} {\bibfnamefont
  {Y.~O.}\ \bibnamefont {Kvashnin}}, \bibinfo {author} {\bibfnamefont
  {J.}~\bibnamefont {Fransson}}, \bibinfo {author} {\bibfnamefont
  {A.}~\bibnamefont {Klautau}}, \bibinfo {author} {\bibfnamefont
  {O.}~\bibnamefont {Eriksson}},\ and\ \bibinfo {author} {\bibfnamefont
  {L.}~\bibnamefont {Nordstr{\"o}m}},\ }\href
  {https://doi.org/10.1038/s41598-020-77219-3} {\bibfield  {journal} {\bibinfo
  {journal} {Sci. Rep.}\ }\textbf {\bibinfo {volume} {10}},\ \bibinfo {pages}
  {1} (\bibinfo {year} {2020})}\BibitemShut {NoStop}%
\bibitem [{\citenamefont {Gutzeit}\ \emph {et~al.}(2021)\citenamefont
  {Gutzeit}, \citenamefont {Haldar}, \citenamefont {Meyer},\ and\ \citenamefont
  {Heinze}}]{Gutzeit2021}%
  \BibitemOpen
  \bibfield  {author} {\bibinfo {author} {\bibfnamefont {M.}~\bibnamefont
  {Gutzeit}}, \bibinfo {author} {\bibfnamefont {S.}~\bibnamefont {Haldar}},
  \bibinfo {author} {\bibfnamefont {S.}~\bibnamefont {Meyer}},\ and\ \bibinfo
  {author} {\bibfnamefont {S.}~\bibnamefont {Heinze}},\ }\href
  {https://doi.org/10.1103/PhysRevB.104.024420} {\bibfield  {journal} {\bibinfo
   {journal} {Phys. Rev. B}\ }\textbf {\bibinfo {volume} {104}},\ \bibinfo
  {pages} {024420} (\bibinfo {year} {2021})}\BibitemShut {NoStop}%
\bibitem [{\citenamefont {Ashcroft}\ and\ \citenamefont
  {Mermin}(1976)}]{Ashcroft1976}%
  \BibitemOpen
  \bibfield  {author} {\bibinfo {author} {\bibfnamefont {N.~W.}\ \bibnamefont
  {Ashcroft}}\ and\ \bibinfo {author} {\bibfnamefont {N.~D.}\ \bibnamefont
  {Mermin}},\ }\href@noop {} {\emph {\bibinfo {title} {{Solid State
  Physics}}}}\ (\bibinfo  {publisher} {Saunders College},\ \bibinfo {address}
  {Florida},\ \bibinfo {year} {1976})\BibitemShut {NoStop}%
\bibitem [{\citenamefont {Skubic}\ \emph {et~al.}(2008)\citenamefont {Skubic},
  \citenamefont {Hellsvik}, \citenamefont {Nordstr{\"{o}}m},\ and\
  \citenamefont {Eriksson}}]{Skubic2008}%
  \BibitemOpen
  \bibfield  {author} {\bibinfo {author} {\bibfnamefont {B.}~\bibnamefont
  {Skubic}}, \bibinfo {author} {\bibfnamefont {J.}~\bibnamefont {Hellsvik}},
  \bibinfo {author} {\bibfnamefont {L.}~\bibnamefont {Nordstr{\"{o}}m}},\ and\
  \bibinfo {author} {\bibfnamefont {O.}~\bibnamefont {Eriksson}},\ }\href
  {https://doi.org/10.1088/0953-8984/20/31/315203} {\bibfield  {journal}
  {\bibinfo  {journal} {J. Phys.: Condens. Matter}\ }\textbf {\bibinfo {volume}
  {20}},\ \bibinfo {pages} {315203} (\bibinfo {year} {2008})}\BibitemShut
  {NoStop}%
\bibitem [{\citenamefont {Eriksson}\ \emph {et~al.}(2017)\citenamefont
  {Eriksson}, \citenamefont {Bergman}, \citenamefont {Bergqvist},\ and\
  \citenamefont {Hellsvik}}]{Eriksson2017}%
  \BibitemOpen
  \bibfield  {author} {\bibinfo {author} {\bibfnamefont {O.}~\bibnamefont
  {Eriksson}}, \bibinfo {author} {\bibfnamefont {A.}~\bibnamefont {Bergman}},
  \bibinfo {author} {\bibfnamefont {L.}~\bibnamefont {Bergqvist}},\ and\
  \bibinfo {author} {\bibfnamefont {J.}~\bibnamefont {Hellsvik}},\ }\href@noop
  {} {\emph {\bibinfo {title} {{Atomistic Spin Dynamics: Foundations and
  Applications}}}}\ (\bibinfo  {publisher} {Oxford University Press},\ \bibinfo
  {year} {2017})\BibitemShut {NoStop}%
\bibitem [{\citenamefont {Rohart}\ \emph {et~al.}(2016)\citenamefont {Rohart},
  \citenamefont {Miltat},\ and\ \citenamefont {Thiaville}}]{Rohart2016}%
  \BibitemOpen
  \bibfield  {author} {\bibinfo {author} {\bibfnamefont {S.}~\bibnamefont
  {Rohart}}, \bibinfo {author} {\bibfnamefont {J.}~\bibnamefont {Miltat}},\
  and\ \bibinfo {author} {\bibfnamefont {A.}~\bibnamefont {Thiaville}},\ }\href
  {https://doi.org/10.1103/PhysRevB.93.214412} {\bibfield  {journal} {\bibinfo
  {journal} {Phys. Rev. B}\ }\textbf {\bibinfo {volume} {93}},\ \bibinfo
  {pages} {214412} (\bibinfo {year} {2016})}\BibitemShut {NoStop}%
\bibitem [{\citenamefont {Lobanov}\ \emph {et~al.}(2016)\citenamefont
  {Lobanov}, \citenamefont {J\'onsson},\ and\ \citenamefont
  {Uzdin}}]{Lobanov2016}%
  \BibitemOpen
  \bibfield  {author} {\bibinfo {author} {\bibfnamefont {I.~S.}\ \bibnamefont
  {Lobanov}}, \bibinfo {author} {\bibfnamefont {H.}~\bibnamefont {J\'onsson}},\
  and\ \bibinfo {author} {\bibfnamefont {V.~M.}\ \bibnamefont {Uzdin}},\ }\href
  {https://doi.org/10.1103/PhysRevB.94.174418} {\bibfield  {journal} {\bibinfo
  {journal} {Phys. Rev. B}\ }\textbf {\bibinfo {volume} {94}},\ \bibinfo
  {pages} {174418} (\bibinfo {year} {2016})}\BibitemShut {NoStop}%
\bibitem [{\citenamefont {Guo}\ \emph {et~al.}(1991)\citenamefont {Guo},
  \citenamefont {Temmerman},\ and\ \citenamefont {Ebert}}]{Guo1991}%
  \BibitemOpen
  \bibfield  {author} {\bibinfo {author} {\bibfnamefont {G.}~\bibnamefont
  {Guo}}, \bibinfo {author} {\bibfnamefont {W.}~\bibnamefont {Temmerman}},\
  and\ \bibinfo {author} {\bibfnamefont {H.}~\bibnamefont {Ebert}},\ }\href
  {https://doi.org/10.1088/0953-8984/3/42/015} {\bibfield  {journal} {\bibinfo
  {journal} {J. Phys.: Condens. Matter}\ }\textbf {\bibinfo {volume} {3}},\
  \bibinfo {pages} {8205} (\bibinfo {year} {1991})}\BibitemShut {NoStop}%
\bibitem [{\citenamefont {Giannozzi}\ \emph {et~al.}(2017)\citenamefont
  {Giannozzi}, \citenamefont {Andreussi}, \citenamefont {Brumme}, \citenamefont
  {Bunau}, \citenamefont {Nardelli}, \citenamefont {Calandra}, \citenamefont
  {Car}, \citenamefont {Cavazzoni}, \citenamefont {Ceresoli}, \citenamefont
  {Cococcioni} \emph {et~al.}}]{Giannozzi2017}%
  \BibitemOpen
  \bibfield  {author} {\bibinfo {author} {\bibfnamefont {P.}~\bibnamefont
  {Giannozzi}}, \bibinfo {author} {\bibfnamefont {O.}~\bibnamefont
  {Andreussi}}, \bibinfo {author} {\bibfnamefont {T.}~\bibnamefont {Brumme}},
  \bibinfo {author} {\bibfnamefont {O.}~\bibnamefont {Bunau}}, \bibinfo
  {author} {\bibfnamefont {M.~B.}\ \bibnamefont {Nardelli}}, \bibinfo {author}
  {\bibfnamefont {M.}~\bibnamefont {Calandra}}, \bibinfo {author}
  {\bibfnamefont {R.}~\bibnamefont {Car}}, \bibinfo {author} {\bibfnamefont
  {C.}~\bibnamefont {Cavazzoni}}, \bibinfo {author} {\bibfnamefont
  {D.}~\bibnamefont {Ceresoli}}, \bibinfo {author} {\bibfnamefont
  {M.}~\bibnamefont {Cococcioni}}, \emph {et~al.},\ }\href
  {https://doi.org/10.1088/1361-648X/aa8f79} {\bibfield  {journal} {\bibinfo
  {journal} {J. Phys.: Condens. Matter}\ }\textbf {\bibinfo {volume} {29}},\
  \bibinfo {pages} {465901} (\bibinfo {year} {2017})}\BibitemShut {NoStop}%
\bibitem [{\citenamefont {Giannozzi}\ \emph {et~al.}(2009)\citenamefont
  {Giannozzi}, \citenamefont {Baroni}, \citenamefont {Bonini}, \citenamefont
  {Calandra}, \citenamefont {Car}, \citenamefont {Cavazzoni}, \citenamefont
  {Ceresoli}, \citenamefont {Chiarotti}, \citenamefont {Cococcioni},
  \citenamefont {Dabo}, \citenamefont {{Dal Corso}}, \citenamefont
  {de~Gironcoli}, \citenamefont {Fabris}, \citenamefont {Fratesi},
  \citenamefont {Gebauer}, \citenamefont {Gerstmann}, \citenamefont
  {Gougoussis}, \citenamefont {Kokalj}, \citenamefont {Lazzeri}, \citenamefont
  {Martin-Samos}, \citenamefont {Marzari}, \citenamefont {Mauri}, \citenamefont
  {Mazzarello}, \citenamefont {Paolini}, \citenamefont {Pasquarello},
  \citenamefont {Paulatto}, \citenamefont {Sbraccia}, \citenamefont {Scandolo},
  \citenamefont {Sclauzero}, \citenamefont {Seitsonen}, \citenamefont
  {Smogunov}, \citenamefont {Umari},\ and\ \citenamefont
  {Wentzcovitch}}]{Giannozzi2009}%
  \BibitemOpen
  \bibfield  {author} {\bibinfo {author} {\bibfnamefont {P.}~\bibnamefont
  {Giannozzi}}, \bibinfo {author} {\bibfnamefont {S.}~\bibnamefont {Baroni}},
  \bibinfo {author} {\bibfnamefont {N.}~\bibnamefont {Bonini}}, \bibinfo
  {author} {\bibfnamefont {M.}~\bibnamefont {Calandra}}, \bibinfo {author}
  {\bibfnamefont {R.}~\bibnamefont {Car}}, \bibinfo {author} {\bibfnamefont
  {C.}~\bibnamefont {Cavazzoni}}, \bibinfo {author} {\bibfnamefont
  {D.}~\bibnamefont {Ceresoli}}, \bibinfo {author} {\bibfnamefont {G.~L.}\
  \bibnamefont {Chiarotti}}, \bibinfo {author} {\bibfnamefont {M.}~\bibnamefont
  {Cococcioni}}, \bibinfo {author} {\bibfnamefont {I.}~\bibnamefont {Dabo}},
  \bibinfo {author} {\bibfnamefont {A.}~\bibnamefont {{Dal Corso}}}, \bibinfo
  {author} {\bibfnamefont {S.}~\bibnamefont {de~Gironcoli}}, \bibinfo {author}
  {\bibfnamefont {S.}~\bibnamefont {Fabris}}, \bibinfo {author} {\bibfnamefont
  {G.}~\bibnamefont {Fratesi}}, \bibinfo {author} {\bibfnamefont
  {R.}~\bibnamefont {Gebauer}}, \bibinfo {author} {\bibfnamefont
  {U.}~\bibnamefont {Gerstmann}}, \bibinfo {author} {\bibfnamefont
  {C.}~\bibnamefont {Gougoussis}}, \bibinfo {author} {\bibfnamefont
  {A.}~\bibnamefont {Kokalj}}, \bibinfo {author} {\bibfnamefont
  {M.}~\bibnamefont {Lazzeri}}, \bibinfo {author} {\bibfnamefont
  {L.}~\bibnamefont {Martin-Samos}}, \bibinfo {author} {\bibfnamefont
  {N.}~\bibnamefont {Marzari}}, \bibinfo {author} {\bibfnamefont
  {F.}~\bibnamefont {Mauri}}, \bibinfo {author} {\bibfnamefont
  {R.}~\bibnamefont {Mazzarello}}, \bibinfo {author} {\bibfnamefont
  {S.}~\bibnamefont {Paolini}}, \bibinfo {author} {\bibfnamefont
  {A.}~\bibnamefont {Pasquarello}}, \bibinfo {author} {\bibfnamefont
  {L.}~\bibnamefont {Paulatto}}, \bibinfo {author} {\bibfnamefont
  {C.}~\bibnamefont {Sbraccia}}, \bibinfo {author} {\bibfnamefont
  {S.}~\bibnamefont {Scandolo}}, \bibinfo {author} {\bibfnamefont
  {G.}~\bibnamefont {Sclauzero}}, \bibinfo {author} {\bibfnamefont {A.~P.}\
  \bibnamefont {Seitsonen}}, \bibinfo {author} {\bibfnamefont {A.}~\bibnamefont
  {Smogunov}}, \bibinfo {author} {\bibfnamefont {P.}~\bibnamefont {Umari}},\
  and\ \bibinfo {author} {\bibfnamefont {R.~M.}\ \bibnamefont {Wentzcovitch}},\
  }\href {https://doi.org/10.1088/0953-8984/21/39/395502} {\bibfield  {journal}
  {\bibinfo  {journal} {J. Phys.: Condens. Matter}\ }\textbf {\bibinfo {volume}
  {21}},\ \bibinfo {pages} {395502} (\bibinfo {year} {2009})}\BibitemShut
  {NoStop}%
\bibitem [{\citenamefont {Urru}\ and\ \citenamefont
  {Dal~Corso}(2019)}]{Urru2019}%
  \BibitemOpen
  \bibfield  {author} {\bibinfo {author} {\bibfnamefont {A.}~\bibnamefont
  {Urru}}\ and\ \bibinfo {author} {\bibfnamefont {A.}~\bibnamefont
  {Dal~Corso}},\ }\href {https://doi.org/10.1103/PhysRevB.100.045115}
  {\bibfield  {journal} {\bibinfo  {journal} {Phys. Rev. B}\ }\textbf {\bibinfo
  {volume} {100}},\ \bibinfo {pages} {045115} (\bibinfo {year}
  {2019})}\BibitemShut {NoStop}%
\bibitem [{\citenamefont {Monkhorst}\ and\ \citenamefont
  {Pack}(1976)}]{Monkhorst1976}%
  \BibitemOpen
  \bibfield  {author} {\bibinfo {author} {\bibfnamefont {H.~J.}\ \bibnamefont
  {Monkhorst}}\ and\ \bibinfo {author} {\bibfnamefont {J.~D.}\ \bibnamefont
  {Pack}},\ }\href {https://doi.org/10.1103/PhysRevB.13.5188} {\bibfield
  {journal} {\bibinfo  {journal} {Phys. Rev. B}\ }\textbf {\bibinfo {volume}
  {13}},\ \bibinfo {pages} {5188} (\bibinfo {year} {1976})}\BibitemShut
  {NoStop}%
\bibitem [{\citenamefont {Kirkpatrick}\ \emph {et~al.}(1983)\citenamefont
  {Kirkpatrick}, \citenamefont {Gelatt},\ and\ \citenamefont
  {Vecchi}}]{Kirkpatrick1983}%
  \BibitemOpen
  \bibfield  {author} {\bibinfo {author} {\bibfnamefont {S.}~\bibnamefont
  {Kirkpatrick}}, \bibinfo {author} {\bibfnamefont {C.~D.}\ \bibnamefont
  {Gelatt}},\ and\ \bibinfo {author} {\bibfnamefont {M.~P.}\ \bibnamefont
  {Vecchi}},\ }\href@noop {} {\bibfield  {journal} {\bibinfo  {journal}
  {science}\ }\textbf {\bibinfo {volume} {220}},\ \bibinfo {pages} {671}
  (\bibinfo {year} {1983})}\BibitemShut {NoStop}%
\bibitem [{\citenamefont {Berg}\ and\ \citenamefont
  {L{\"u}scher}(1981)}]{Berg1981}%
  \BibitemOpen
  \bibfield  {author} {\bibinfo {author} {\bibfnamefont {B.}~\bibnamefont
  {Berg}}\ and\ \bibinfo {author} {\bibfnamefont {M.}~\bibnamefont
  {L{\"u}scher}},\ }\href {https://doi.org/10.1016/0550-3213(81)90568-X}
  {\bibfield  {journal} {\bibinfo  {journal} {Nucl. Phys. B.}\ }\textbf
  {\bibinfo {volume} {190}},\ \bibinfo {pages} {412} (\bibinfo {year}
  {1981})}\BibitemShut {NoStop}%
\bibitem [{\citenamefont {Weber}\ \emph {et~al.}(1991)\citenamefont {Weber},
  \citenamefont {Wesner}, \citenamefont {G{\"{u}}ntherodt},\ and\ \citenamefont
  {Linke}}]{Weber1991}%
  \BibitemOpen
  \bibfield  {author} {\bibinfo {author} {\bibfnamefont {W.}~\bibnamefont
  {Weber}}, \bibinfo {author} {\bibfnamefont {D.~A.}\ \bibnamefont {Wesner}},
  \bibinfo {author} {\bibfnamefont {G.}~\bibnamefont {G{\"{u}}ntherodt}},\ and\
  \bibinfo {author} {\bibfnamefont {U.}~\bibnamefont {Linke}},\ }\href
  {https://doi.org/10.1103/PhysRevLett.66.942} {\bibfield  {journal} {\bibinfo
  {journal} {Phys. Rev. Lett.}\ }\textbf {\bibinfo {volume} {66}},\ \bibinfo
  {pages} {942} (\bibinfo {year} {1991})}\BibitemShut {NoStop}%
\bibitem [{\citenamefont {Polesya}\ \emph {et~al.}(2010)\citenamefont
  {Polesya}, \citenamefont {Mankovsky}, \citenamefont {Sipr}, \citenamefont
  {Meindl}, \citenamefont {Strunk},\ and\ \citenamefont {Ebert}}]{Polesya2010}%
  \BibitemOpen
  \bibfield  {author} {\bibinfo {author} {\bibfnamefont {S.}~\bibnamefont
  {Polesya}}, \bibinfo {author} {\bibfnamefont {S.}~\bibnamefont {Mankovsky}},
  \bibinfo {author} {\bibfnamefont {O.}~\bibnamefont {Sipr}}, \bibinfo {author}
  {\bibfnamefont {W.}~\bibnamefont {Meindl}}, \bibinfo {author} {\bibfnamefont
  {C.}~\bibnamefont {Strunk}},\ and\ \bibinfo {author} {\bibfnamefont
  {H.}~\bibnamefont {Ebert}},\ }\href
  {https://doi.org/10.1103/PhysRevB.82.214409} {\bibfield  {journal} {\bibinfo
  {journal} {Phys. Rev. B}\ }\textbf {\bibinfo {volume} {82}},\ \bibinfo
  {pages} {214409} (\bibinfo {year} {2010})}\BibitemShut {NoStop}%
\bibitem [{\citenamefont {Dup{\'{e}}}\ \emph {et~al.}(2016)\citenamefont
  {Dup{\'{e}}}, \citenamefont {Bihlmayer}, \citenamefont {B{\"{o}}ttcher},
  \citenamefont {Bl{\"{u}}gel},\ and\ \citenamefont {Heinze}}]{Dupe2016}%
  \BibitemOpen
  \bibfield  {author} {\bibinfo {author} {\bibfnamefont {B.}~\bibnamefont
  {Dup{\'{e}}}}, \bibinfo {author} {\bibfnamefont {G.}~\bibnamefont
  {Bihlmayer}}, \bibinfo {author} {\bibfnamefont {M.}~\bibnamefont
  {B{\"{o}}ttcher}}, \bibinfo {author} {\bibfnamefont {S.}~\bibnamefont
  {Bl{\"{u}}gel}},\ and\ \bibinfo {author} {\bibfnamefont {S.}~\bibnamefont
  {Heinze}},\ }\href {https://doi.org/10.1038/ncomms11779} {\bibfield
  {journal} {\bibinfo  {journal} {Nat. Commun.}\ }\textbf {\bibinfo {volume}
  {7}},\ \bibinfo {pages} {11779} (\bibinfo {year} {2016})}\BibitemShut
  {NoStop}%
\bibitem [{\citenamefont {Sj\"ostedt}\ and\ \citenamefont
  {Nordstr\"om}(2002)}]{Sjostedt2002}%
  \BibitemOpen
  \bibfield  {author} {\bibinfo {author} {\bibfnamefont {E.}~\bibnamefont
  {Sj\"ostedt}}\ and\ \bibinfo {author} {\bibfnamefont {L.}~\bibnamefont
  {Nordstr\"om}},\ }\href {https://doi.org/10.1103/PhysRevB.66.014447}
  {\bibfield  {journal} {\bibinfo  {journal} {Phys. Rev. B}\ }\textbf {\bibinfo
  {volume} {66}},\ \bibinfo {pages} {014447} (\bibinfo {year}
  {2002})}\BibitemShut {NoStop}%
\bibitem [{\citenamefont {Liz{\'{a}}rraga}\ \emph {et~al.}(2004)\citenamefont
  {Liz{\'{a}}rraga}, \citenamefont {Nordstr{\"{o}}m}, \citenamefont
  {Bergqvist}, \citenamefont {Bergman}, \citenamefont {Sj{\"{o}}stedt},
  \citenamefont {Mohn},\ and\ \citenamefont {Eriksson}}]{Liz??rraga2004}%
  \BibitemOpen
  \bibfield  {author} {\bibinfo {author} {\bibfnamefont {R.}~\bibnamefont
  {Liz{\'{a}}rraga}}, \bibinfo {author} {\bibfnamefont {L.}~\bibnamefont
  {Nordstr{\"{o}}m}}, \bibinfo {author} {\bibfnamefont {L.}~\bibnamefont
  {Bergqvist}}, \bibinfo {author} {\bibfnamefont {A.}~\bibnamefont {Bergman}},
  \bibinfo {author} {\bibfnamefont {E.}~\bibnamefont {Sj{\"{o}}stedt}},
  \bibinfo {author} {\bibfnamefont {P.}~\bibnamefont {Mohn}},\ and\ \bibinfo
  {author} {\bibfnamefont {O.}~\bibnamefont {Eriksson}},\ }\href
  {https://doi.org/10.1103/PhysRevLett.93.107205} {\bibfield  {journal}
  {\bibinfo  {journal} {Phys. Rev. Lett.}\ }\textbf {\bibinfo {volume} {93}},\
  \bibinfo {pages} {107205} (\bibinfo {year} {2004})}\BibitemShut {NoStop}%
\bibitem [{\citenamefont {Cardias}\ \emph {et~al.}(2017)\citenamefont
  {Cardias}, \citenamefont {Szilva}, \citenamefont {Bergman}, \citenamefont
  {Di~Marco}, \citenamefont {Katsnelson}, \citenamefont {Lichtenstein},
  \citenamefont {Nordstr{\"o}m}, \citenamefont {Klautau}, \citenamefont
  {Eriksson},\ and\ \citenamefont {Kvashnin}}]{Cardias2017}%
  \BibitemOpen
  \bibfield  {author} {\bibinfo {author} {\bibfnamefont {R.}~\bibnamefont
  {Cardias}}, \bibinfo {author} {\bibfnamefont {A.}~\bibnamefont {Szilva}},
  \bibinfo {author} {\bibfnamefont {A.}~\bibnamefont {Bergman}}, \bibinfo
  {author} {\bibfnamefont {I.}~\bibnamefont {Di~Marco}}, \bibinfo {author}
  {\bibfnamefont {M.}~\bibnamefont {Katsnelson}}, \bibinfo {author}
  {\bibfnamefont {A.}~\bibnamefont {Lichtenstein}}, \bibinfo {author}
  {\bibfnamefont {L.}~\bibnamefont {Nordstr{\"o}m}}, \bibinfo {author}
  {\bibfnamefont {A.}~\bibnamefont {Klautau}}, \bibinfo {author} {\bibfnamefont
  {O.}~\bibnamefont {Eriksson}},\ and\ \bibinfo {author} {\bibfnamefont
  {Y.~O.}\ \bibnamefont {Kvashnin}},\ }\href
  {https://doi.org/10.1038/s41598-017-04427-9} {\bibfield  {journal} {\bibinfo
  {journal} {Sci. Rep.}\ }\textbf {\bibinfo {volume} {7}},\ \bibinfo {pages}
  {1} (\bibinfo {year} {2017})}\BibitemShut {NoStop}%
\bibitem [{\citenamefont {Ruderman}\ and\ \citenamefont
  {Kittel}(1954)}]{Ruderman1954}%
  \BibitemOpen
  \bibfield  {author} {\bibinfo {author} {\bibfnamefont {M.~A.}\ \bibnamefont
  {Ruderman}}\ and\ \bibinfo {author} {\bibfnamefont {C.}~\bibnamefont
  {Kittel}},\ }\href {https://doi.org/10.1103/PhysRev.96.99} {\bibfield
  {journal} {\bibinfo  {journal} {Phys. Rev.}\ }\textbf {\bibinfo {volume}
  {96}},\ \bibinfo {pages} {99} (\bibinfo {year} {1954})}\BibitemShut {NoStop}%
\bibitem [{\citenamefont {Kasuya}(1956)}]{Kasuya1956}%
  \BibitemOpen
  \bibfield  {author} {\bibinfo {author} {\bibfnamefont {T.}~\bibnamefont
  {Kasuya}},\ }\href {https://doi.org/10.1143/PTP.16.45} {\bibfield  {journal}
  {\bibinfo  {journal} {Prog. Theor. Phys.}\ }\textbf {\bibinfo {volume}
  {16}},\ \bibinfo {pages} {45} (\bibinfo {year} {1956})}\BibitemShut {NoStop}%
\bibitem [{\citenamefont {Yosida}(1957)}]{Yosida1957}%
  \BibitemOpen
  \bibfield  {author} {\bibinfo {author} {\bibfnamefont {K.}~\bibnamefont
  {Yosida}},\ }\href {https://doi.org/10.1103/PhysRev.106.893} {\bibfield
  {journal} {\bibinfo  {journal} {Phys. Rev.}\ }\textbf {\bibinfo {volume}
  {106}},\ \bibinfo {pages} {893} (\bibinfo {year} {1957})}\BibitemShut
  {NoStop}%
\bibitem [{\citenamefont {Khajetoorians}\ \emph {et~al.}(2016)\citenamefont
  {Khajetoorians}, \citenamefont {Steinbrecher}, \citenamefont {Ternes},
  \citenamefont {Bouhassoune}, \citenamefont {{dos Santos Dias}}, \citenamefont
  {Lounis}, \citenamefont {Wiebe},\ and\ \citenamefont
  {Wiesendanger}}]{Khajetoorians2016}%
  \BibitemOpen
  \bibfield  {author} {\bibinfo {author} {\bibfnamefont {A.~A.}\ \bibnamefont
  {Khajetoorians}}, \bibinfo {author} {\bibfnamefont {M.}~\bibnamefont
  {Steinbrecher}}, \bibinfo {author} {\bibfnamefont {M.}~\bibnamefont
  {Ternes}}, \bibinfo {author} {\bibfnamefont {M.}~\bibnamefont {Bouhassoune}},
  \bibinfo {author} {\bibfnamefont {M.}~\bibnamefont {{dos Santos Dias}}},
  \bibinfo {author} {\bibfnamefont {S.}~\bibnamefont {Lounis}}, \bibinfo
  {author} {\bibfnamefont {J.}~\bibnamefont {Wiebe}},\ and\ \bibinfo {author}
  {\bibfnamefont {R.}~\bibnamefont {Wiesendanger}},\ }\href
  {https://doi.org/10.1038/ncomms10620} {\bibfield  {journal} {\bibinfo
  {journal} {Nat. Commun.}\ }\textbf {\bibinfo {volume} {7}},\ \bibinfo {pages}
  {10620} (\bibinfo {year} {2016})}\BibitemShut {NoStop}%
\bibitem [{\citenamefont {Fert}\ and\ \citenamefont {Levy}(1980)}]{Fert1980}%
  \BibitemOpen
  \bibfield  {author} {\bibinfo {author} {\bibfnamefont {A.}~\bibnamefont
  {Fert}}\ and\ \bibinfo {author} {\bibfnamefont {P.~M.}\ \bibnamefont
  {Levy}},\ }\href {https://doi.org/10.1103/PhysRevLett.44.1538} {\bibfield
  {journal} {\bibinfo  {journal} {Phys. Rev. Lett.}\ }\textbf {\bibinfo
  {volume} {44}},\ \bibinfo {pages} {1538} (\bibinfo {year}
  {1980})}\BibitemShut {NoStop}%
\bibitem [{\citenamefont {Cr{\'{e}}pieux}\ and\ \citenamefont
  {Lacroix}(1998)}]{Crepieux1998}%
  \BibitemOpen
  \bibfield  {author} {\bibinfo {author} {\bibfnamefont {A.}~\bibnamefont
  {Cr{\'{e}}pieux}}\ and\ \bibinfo {author} {\bibfnamefont {C.}~\bibnamefont
  {Lacroix}},\ }\href {https://doi.org/10.1016/S0304-8853(97)01044-5}
  {\bibfield  {journal} {\bibinfo  {journal} {J. Magn. Magn. Mater.}\ }\textbf
  {\bibinfo {volume} {182}},\ \bibinfo {pages} {341} (\bibinfo {year}
  {1998})}\BibitemShut {NoStop}%
\bibitem [{\citenamefont {Belabbes}\ \emph {et~al.}(2016)\citenamefont
  {Belabbes}, \citenamefont {Bihlmayer}, \citenamefont {Bechstedt},
  \citenamefont {Bl{\"{u}}gel},\ and\ \citenamefont {Manchon}}]{Belabbes2016}%
  \BibitemOpen
  \bibfield  {author} {\bibinfo {author} {\bibfnamefont {A.}~\bibnamefont
  {Belabbes}}, \bibinfo {author} {\bibfnamefont {G.}~\bibnamefont {Bihlmayer}},
  \bibinfo {author} {\bibfnamefont {F.}~\bibnamefont {Bechstedt}}, \bibinfo
  {author} {\bibfnamefont {S.}~\bibnamefont {Bl{\"{u}}gel}},\ and\ \bibinfo
  {author} {\bibfnamefont {A.}~\bibnamefont {Manchon}},\ }\href
  {https://doi.org/10.1103/PhysRevLett.117.247202} {\bibfield  {journal}
  {\bibinfo  {journal} {Phys. Rev. Lett.}\ }\textbf {\bibinfo {volume} {117}},\
  \bibinfo {pages} {247202} (\bibinfo {year} {2016})}\BibitemShut {NoStop}%
\bibitem [{\citenamefont {Carvalho}\ \emph {et~al.}(2021)\citenamefont
  {Carvalho}, \citenamefont {Miranda}, \citenamefont {Klautau}, \citenamefont
  {Bergman},\ and\ \citenamefont {Petrilli}}]{Carvalho2021}%
  \BibitemOpen
  \bibfield  {author} {\bibinfo {author} {\bibfnamefont {P.~C.}\ \bibnamefont
  {Carvalho}}, \bibinfo {author} {\bibfnamefont {I.~P.}\ \bibnamefont
  {Miranda}}, \bibinfo {author} {\bibfnamefont {A.~B.}\ \bibnamefont
  {Klautau}}, \bibinfo {author} {\bibfnamefont {A.}~\bibnamefont {Bergman}},\
  and\ \bibinfo {author} {\bibfnamefont {H.~M.}\ \bibnamefont {Petrilli}},\
  }\href {https://doi.org/10.1103/PhysRevMaterials.5.124406} {\bibfield
  {journal} {\bibinfo  {journal} {Phys. Rev. Materials}\ }\textbf {\bibinfo
  {volume} {5}},\ \bibinfo {pages} {124406} (\bibinfo {year}
  {2021})}\BibitemShut {NoStop}%
\bibitem [{\citenamefont {Rohart}\ and\ \citenamefont
  {Thiaville}(2013)}]{Rohart2013}%
  \BibitemOpen
  \bibfield  {author} {\bibinfo {author} {\bibfnamefont {S.}~\bibnamefont
  {Rohart}}\ and\ \bibinfo {author} {\bibfnamefont {A.}~\bibnamefont
  {Thiaville}},\ }\href {https://doi.org/10.1103/PhysRevB.88.184422} {\bibfield
   {journal} {\bibinfo  {journal} {Phys. Rev. B}\ }\textbf {\bibinfo {volume}
  {88}},\ \bibinfo {pages} {184422} (\bibinfo {year} {2013})}\BibitemShut
  {NoStop}%
\bibitem [{\citenamefont {Polesya}\ \emph {et~al.}(2014)\citenamefont
  {Polesya}, \citenamefont {Mankovsky}, \citenamefont {Bornemann},
  \citenamefont {K{\"{o}}dderitzsch}, \citenamefont {Min{\'{a}}r},\ and\
  \citenamefont {Ebert}}]{Polesya2014}%
  \BibitemOpen
  \bibfield  {author} {\bibinfo {author} {\bibfnamefont {S.}~\bibnamefont
  {Polesya}}, \bibinfo {author} {\bibfnamefont {S.}~\bibnamefont {Mankovsky}},
  \bibinfo {author} {\bibfnamefont {S.}~\bibnamefont {Bornemann}}, \bibinfo
  {author} {\bibfnamefont {D.}~\bibnamefont {K{\"{o}}dderitzsch}}, \bibinfo
  {author} {\bibfnamefont {J.}~\bibnamefont {Min{\'{a}}r}},\ and\ \bibinfo
  {author} {\bibfnamefont {H.}~\bibnamefont {Ebert}},\ }\href
  {https://doi.org/10.1103/PhysRevB.89.184414} {\bibfield  {journal} {\bibinfo
  {journal} {Phys. Rev. B}\ }\textbf {\bibinfo {volume} {89}},\ \bibinfo
  {pages} {184414} (\bibinfo {year} {2014})}\BibitemShut {NoStop}%
\bibitem [{\citenamefont {Fert}\ \emph {et~al.}(2013)\citenamefont {Fert},
  \citenamefont {Cros},\ and\ \citenamefont {Sampaio}}]{Fert2013}%
  \BibitemOpen
  \bibfield  {author} {\bibinfo {author} {\bibfnamefont {A.}~\bibnamefont
  {Fert}}, \bibinfo {author} {\bibfnamefont {V.}~\bibnamefont {Cros}},\ and\
  \bibinfo {author} {\bibfnamefont {J.}~\bibnamefont {Sampaio}},\ }\href
  {https://doi.org/10.1038/nnano.2013.29} {\bibfield  {journal} {\bibinfo
  {journal} {Nat. Nanotechnol}\ }\textbf {\bibinfo {volume} {8}},\ \bibinfo
  {pages} {152} (\bibinfo {year} {2013})}\BibitemShut {NoStop}%
\bibitem [{\citenamefont {Lindner}\ \emph {et~al.}(2020)\citenamefont
  {Lindner}, \citenamefont {Bargsten}, \citenamefont {Kovarik}, \citenamefont
  {Friedlein}, \citenamefont {Harm}, \citenamefont {Krause},\ and\
  \citenamefont {Wiesendanger}}]{Lindner2020}%
  \BibitemOpen
  \bibfield  {author} {\bibinfo {author} {\bibfnamefont {P.}~\bibnamefont
  {Lindner}}, \bibinfo {author} {\bibfnamefont {L.}~\bibnamefont {Bargsten}},
  \bibinfo {author} {\bibfnamefont {S.}~\bibnamefont {Kovarik}}, \bibinfo
  {author} {\bibfnamefont {J.}~\bibnamefont {Friedlein}}, \bibinfo {author}
  {\bibfnamefont {J.}~\bibnamefont {Harm}}, \bibinfo {author} {\bibfnamefont
  {S.}~\bibnamefont {Krause}},\ and\ \bibinfo {author} {\bibfnamefont
  {R.}~\bibnamefont {Wiesendanger}},\ }\href
  {https://doi.org/10.1103/PhysRevB.101.214445} {\bibfield  {journal} {\bibinfo
   {journal} {Phys. Rev. B}\ }\textbf {\bibinfo {volume} {101}},\ \bibinfo
  {pages} {214445} (\bibinfo {year} {2020})}\BibitemShut {NoStop}%
\bibitem [{\citenamefont {Bessarab}\ \emph {et~al.}(2018)\citenamefont
  {Bessarab}, \citenamefont {M{\"u}ller}, \citenamefont {Lobanov},
  \citenamefont {Rybakov}, \citenamefont {Kiselev}, \citenamefont
  {J{\'o}nsson}, \citenamefont {Uzdin}, \citenamefont {Bl{\"u}gel},
  \citenamefont {Bergqvist},\ and\ \citenamefont {Delin}}]{Bessarab2018}%
  \BibitemOpen
  \bibfield  {author} {\bibinfo {author} {\bibfnamefont {P.~F.}\ \bibnamefont
  {Bessarab}}, \bibinfo {author} {\bibfnamefont {G.~P.}\ \bibnamefont
  {M{\"u}ller}}, \bibinfo {author} {\bibfnamefont {I.~S.}\ \bibnamefont
  {Lobanov}}, \bibinfo {author} {\bibfnamefont {F.~N.}\ \bibnamefont
  {Rybakov}}, \bibinfo {author} {\bibfnamefont {N.~S.}\ \bibnamefont
  {Kiselev}}, \bibinfo {author} {\bibfnamefont {H.}~\bibnamefont
  {J{\'o}nsson}}, \bibinfo {author} {\bibfnamefont {V.~M.}\ \bibnamefont
  {Uzdin}}, \bibinfo {author} {\bibfnamefont {S.}~\bibnamefont {Bl{\"u}gel}},
  \bibinfo {author} {\bibfnamefont {L.}~\bibnamefont {Bergqvist}},\ and\
  \bibinfo {author} {\bibfnamefont {A.}~\bibnamefont {Delin}},\ }\href
  {https://doi.org/10.1038/s41598-018-21623-3} {\bibfield  {journal} {\bibinfo
  {journal} {Sci. Rep.}\ }\textbf {\bibinfo {volume} {8}},\ \bibinfo {pages}
  {1} (\bibinfo {year} {2018})}\BibitemShut {NoStop}%
\bibitem [{\citenamefont {Miranda}\ \emph {et~al.}(2021)\citenamefont
  {Miranda}, \citenamefont {Klautau}, \citenamefont {Bergman}, \citenamefont
  {Thonig}, \citenamefont {Petrilli},\ and\ \citenamefont
  {Eriksson}}]{Miranda2021}%
  \BibitemOpen
  \bibfield  {author} {\bibinfo {author} {\bibfnamefont {I.~P.}\ \bibnamefont
  {Miranda}}, \bibinfo {author} {\bibfnamefont {A.~B.}\ \bibnamefont
  {Klautau}}, \bibinfo {author} {\bibfnamefont {A.}~\bibnamefont {Bergman}},
  \bibinfo {author} {\bibfnamefont {D.}~\bibnamefont {Thonig}}, \bibinfo
  {author} {\bibfnamefont {H.~M.}\ \bibnamefont {Petrilli}},\ and\ \bibinfo
  {author} {\bibfnamefont {O.}~\bibnamefont {Eriksson}},\ }\href
  {https://doi.org/10.1103/PhysRevB.103.L220405} {\bibfield  {journal}
  {\bibinfo  {journal} {Phys. Rev. B}\ }\textbf {\bibinfo {volume} {103}},\
  \bibinfo {pages} {L220405} (\bibinfo {year} {2021})}\BibitemShut {NoStop}%
\bibitem [{\citenamefont {Bisig}\ \emph {et~al.}(2016)\citenamefont {Bisig},
  \citenamefont {Akosa}, \citenamefont {Moon}, \citenamefont {Rhensius},
  \citenamefont {Moutafis}, \citenamefont {von Bieren}, \citenamefont
  {Heidler}, \citenamefont {Kiliani}, \citenamefont {Kammerer}, \citenamefont
  {Curcic}, \citenamefont {Weigand}, \citenamefont {Tyliszczak}, \citenamefont
  {Van~Waeyenberge}, \citenamefont {Stoll}, \citenamefont {Sch\"utz},
  \citenamefont {Lee}, \citenamefont {Manchon},\ and\ \citenamefont
  {Kl\"aui}}]{Andre2016}%
  \BibitemOpen
  \bibfield  {author} {\bibinfo {author} {\bibfnamefont {A.}~\bibnamefont
  {Bisig}}, \bibinfo {author} {\bibfnamefont {C.~A.}\ \bibnamefont {Akosa}},
  \bibinfo {author} {\bibfnamefont {J.-H.}\ \bibnamefont {Moon}}, \bibinfo
  {author} {\bibfnamefont {J.}~\bibnamefont {Rhensius}}, \bibinfo {author}
  {\bibfnamefont {C.}~\bibnamefont {Moutafis}}, \bibinfo {author}
  {\bibfnamefont {A.}~\bibnamefont {von Bieren}}, \bibinfo {author}
  {\bibfnamefont {J.}~\bibnamefont {Heidler}}, \bibinfo {author} {\bibfnamefont
  {G.}~\bibnamefont {Kiliani}}, \bibinfo {author} {\bibfnamefont
  {M.}~\bibnamefont {Kammerer}}, \bibinfo {author} {\bibfnamefont
  {M.}~\bibnamefont {Curcic}}, \bibinfo {author} {\bibfnamefont
  {M.}~\bibnamefont {Weigand}}, \bibinfo {author} {\bibfnamefont
  {T.}~\bibnamefont {Tyliszczak}}, \bibinfo {author} {\bibfnamefont
  {B.}~\bibnamefont {Van~Waeyenberge}}, \bibinfo {author} {\bibfnamefont
  {H.}~\bibnamefont {Stoll}}, \bibinfo {author} {\bibfnamefont
  {G.}~\bibnamefont {Sch\"utz}}, \bibinfo {author} {\bibfnamefont {K.-J.}\
  \bibnamefont {Lee}}, \bibinfo {author} {\bibfnamefont {A.}~\bibnamefont
  {Manchon}},\ and\ \bibinfo {author} {\bibfnamefont {M.}~\bibnamefont
  {Kl\"aui}},\ }\href {https://doi.org/10.1103/PhysRevLett.117.277203}
  {\bibfield  {journal} {\bibinfo  {journal} {Phys. Rev. Lett.}\ }\textbf
  {\bibinfo {volume} {117}},\ \bibinfo {pages} {277203} (\bibinfo {year}
  {2016})}\BibitemShut {NoStop}%
\bibitem [{\citenamefont {R\'ozsa}\ \emph {et~al.}(2018)\citenamefont
  {R\'ozsa}, \citenamefont {Hagemeister}, \citenamefont {Vedmedenko},\ and\
  \citenamefont {Wiesendanger}}]{Rozsa2018}%
  \BibitemOpen
  \bibfield  {author} {\bibinfo {author} {\bibfnamefont {L.}~\bibnamefont
  {R\'ozsa}}, \bibinfo {author} {\bibfnamefont {J.}~\bibnamefont
  {Hagemeister}}, \bibinfo {author} {\bibfnamefont {E.~Y.}\ \bibnamefont
  {Vedmedenko}},\ and\ \bibinfo {author} {\bibfnamefont {R.}~\bibnamefont
  {Wiesendanger}},\ }\href {https://doi.org/10.1103/PhysRevB.98.100404}
  {\bibfield  {journal} {\bibinfo  {journal} {Phys. Rev. B}\ }\textbf {\bibinfo
  {volume} {98}},\ \bibinfo {pages} {100404} (\bibinfo {year}
  {2018})}\BibitemShut {NoStop}%
\bibitem [{\citenamefont {Buhrandt}\ and\ \citenamefont
  {Fritz}(2013)}]{Buhrandt2013}%
  \BibitemOpen
  \bibfield  {author} {\bibinfo {author} {\bibfnamefont {S.}~\bibnamefont
  {Buhrandt}}\ and\ \bibinfo {author} {\bibfnamefont {L.}~\bibnamefont
  {Fritz}},\ }\href {https://doi.org/10.1103/PhysRevB.88.195137} {\bibfield
  {journal} {\bibinfo  {journal} {Phys. Rev. B}\ }\textbf {\bibinfo {volume}
  {88}},\ \bibinfo {pages} {195137} (\bibinfo {year} {2013})}\BibitemShut
  {NoStop}%
\bibitem [{\citenamefont {Schneider}\ \emph {et~al.}(1990)\citenamefont
  {Schneider}, \citenamefont {Bressler}, \citenamefont {Schuster},
  \citenamefont {Kirschner}, \citenamefont {de~Miguel},\ and\ \citenamefont
  {Miranda}}]{Schneider1990}%
  \BibitemOpen
  \bibfield  {author} {\bibinfo {author} {\bibfnamefont {C.~M.}\ \bibnamefont
  {Schneider}}, \bibinfo {author} {\bibfnamefont {P.}~\bibnamefont {Bressler}},
  \bibinfo {author} {\bibfnamefont {P.}~\bibnamefont {Schuster}}, \bibinfo
  {author} {\bibfnamefont {J.}~\bibnamefont {Kirschner}}, \bibinfo {author}
  {\bibfnamefont {J.~J.}\ \bibnamefont {de~Miguel}},\ and\ \bibinfo {author}
  {\bibfnamefont {R.}~\bibnamefont {Miranda}},\ }\href
  {https://doi.org/10.1103/PhysRevLett.64.1059} {\bibfield  {journal} {\bibinfo
   {journal} {Phys. Rev. Lett.}\ }\textbf {\bibinfo {volume} {64}},\ \bibinfo
  {pages} {1059} (\bibinfo {year} {1990})}\BibitemShut {NoStop}%
\bibitem [{\citenamefont {von Malottki}\ \emph {et~al.}(2019)\citenamefont {von
  Malottki}, \citenamefont {Bessarab}, \citenamefont {Haldar}, \citenamefont
  {Delin},\ and\ \citenamefont {Heinze}}]{Malottki2019}%
  \BibitemOpen
  \bibfield  {author} {\bibinfo {author} {\bibfnamefont {S.}~\bibnamefont {von
  Malottki}}, \bibinfo {author} {\bibfnamefont {P.~F.}\ \bibnamefont
  {Bessarab}}, \bibinfo {author} {\bibfnamefont {S.}~\bibnamefont {Haldar}},
  \bibinfo {author} {\bibfnamefont {A.}~\bibnamefont {Delin}},\ and\ \bibinfo
  {author} {\bibfnamefont {S.}~\bibnamefont {Heinze}},\ }\href
  {https://doi.org/10.1103/PhysRevB.99.060409} {\bibfield  {journal} {\bibinfo
  {journal} {Phys. Rev. B}\ }\textbf {\bibinfo {volume} {99}},\ \bibinfo
  {pages} {060409(R)} (\bibinfo {year} {2019})}\BibitemShut {NoStop}%
\bibitem [{\citenamefont {Wang}\ \emph {et~al.}(2018)\citenamefont {Wang},
  \citenamefont {Yuan},\ and\ \citenamefont {Wang}}]{Wang2018}%
  \BibitemOpen
  \bibfield  {author} {\bibinfo {author} {\bibfnamefont {X.}~\bibnamefont
  {Wang}}, \bibinfo {author} {\bibfnamefont {H.}~\bibnamefont {Yuan}},\ and\
  \bibinfo {author} {\bibfnamefont {X.}~\bibnamefont {Wang}},\ }\href
  {https://doi.org/10.1038/s42005-018-0029-0} {\bibfield  {journal} {\bibinfo
  {journal} {Commun. Phys.}\ }\textbf {\bibinfo {volume} {1}},\ \bibinfo
  {pages} {1} (\bibinfo {year} {2018})}\BibitemShut {NoStop}%
\bibitem [{\citenamefont {Kubetzka}\ \emph {et~al.}(2017)\citenamefont
  {Kubetzka}, \citenamefont {Hanneken}, \citenamefont {Wiesendanger},\ and\
  \citenamefont {Von~Bergmann}}]{Kubetzka2017}%
  \BibitemOpen
  \bibfield  {author} {\bibinfo {author} {\bibfnamefont {A.}~\bibnamefont
  {Kubetzka}}, \bibinfo {author} {\bibfnamefont {C.}~\bibnamefont {Hanneken}},
  \bibinfo {author} {\bibfnamefont {R.}~\bibnamefont {Wiesendanger}},\ and\
  \bibinfo {author} {\bibfnamefont {K.}~\bibnamefont {Von~Bergmann}},\ }\href
  {https://doi.org/10.1103/PhysRevB.95.104433} {\bibfield  {journal} {\bibinfo
  {journal} {Phys. Rev. B}\ }\textbf {\bibinfo {volume} {95}},\ \bibinfo
  {pages} {104433} (\bibinfo {year} {2017})}\BibitemShut {NoStop}%
\bibitem [{\citenamefont {Rybakov}\ and\ \citenamefont
  {Kiselev}(2019)}]{Rybakov2019}%
  \BibitemOpen
  \bibfield  {author} {\bibinfo {author} {\bibfnamefont {F.~N.}\ \bibnamefont
  {Rybakov}}\ and\ \bibinfo {author} {\bibfnamefont {N.~S.}\ \bibnamefont
  {Kiselev}},\ }\href {https://doi.org/10.1103/PhysRevB.99.064437} {\bibfield
  {journal} {\bibinfo  {journal} {Phys. Rev. B}\ }\textbf {\bibinfo {volume}
  {99}},\ \bibinfo {pages} {064437} (\bibinfo {year} {2019})}\BibitemShut
  {NoStop}%
\bibitem [{\citenamefont {Le{\v{z}}ai{\'c}}\ \emph {et~al.}(2007)\citenamefont
  {Le{\v{z}}ai{\'c}}, \citenamefont {Mavropoulos},\ and\ \citenamefont
  {Bl{\"u}gel}}]{Levzaic2007}%
  \BibitemOpen
  \bibfield  {author} {\bibinfo {author} {\bibfnamefont {M.}~\bibnamefont
  {Le{\v{z}}ai{\'c}}}, \bibinfo {author} {\bibfnamefont {P.}~\bibnamefont
  {Mavropoulos}},\ and\ \bibinfo {author} {\bibfnamefont {S.}~\bibnamefont
  {Bl{\"u}gel}},\ }\href {https://doi.org/10.1063/1.2710181} {\bibfield
  {journal} {\bibinfo  {journal} {Appl. Phys. Lett.}\ }\textbf {\bibinfo
  {volume} {90}},\ \bibinfo {pages} {082504} (\bibinfo {year}
  {2007})}\BibitemShut {NoStop}%
\bibitem [{\citenamefont {Hayami}\ \emph {et~al.}(2016)\citenamefont {Hayami},
  \citenamefont {Lin},\ and\ \citenamefont {Batista}}]{Hayami2016}%
  \BibitemOpen
  \bibfield  {author} {\bibinfo {author} {\bibfnamefont {S.}~\bibnamefont
  {Hayami}}, \bibinfo {author} {\bibfnamefont {S.-Z.}\ \bibnamefont {Lin}},\
  and\ \bibinfo {author} {\bibfnamefont {C.~D.}\ \bibnamefont {Batista}},\
  }\href {https://doi.org/10.1103/PhysRevB.93.184413} {\bibfield  {journal}
  {\bibinfo  {journal} {Phys. Rev. B}\ }\textbf {\bibinfo {volume} {93}},\
  \bibinfo {pages} {184413} (\bibinfo {year} {2016})}\BibitemShut {NoStop}%
\end{thebibliography}%

\end{document}